\documentclass[12pt]{article}
\pdfoutput=1
\usepackage{a4wide}
\usepackage[centertags]{amsmath}
\usepackage{amssymb}
\usepackage[sort&compress,numbers]{natbib}
\usepackage{ifpdf}
\usepackage{setspace}
\usepackage{amsfonts}
\usepackage{color}
\usepackage{threeparttable}
\usepackage{cancel}
\usepackage{tikz}
\usetikzlibrary{decorations.pathmorphing}
\usepackage{graphicx}
\usepackage{subcaption}

\usepackage{sidecap}

\usepackage{yfonts}

\usetikzlibrary{decorations.markings}
\ifpdf
\usepackage[colorlinks=true,
linkcolor=black,
citecolor=black,
urlcolor=blue,
filecolor=blue,
pdfstartview=FitV,
pdftitle={},
pdfauthor={},
pdfsubject={Hydrodynamics},
pdfkeywords={hydrodynamics, AdS/CFT},
pdfpagemode=None,
bookmarksopen=true
]{hyperref}
\else
\usepackage[hypertex]{hyperref}
\fi
\usepackage{rotating}
\usepackage{rotating}
\makeatletter
\@addtoreset{equation}{section}

\newcommand{\qn}{{\textswab{q}}}
\newcommand{\wn}{{\textswab{w}}}

\makeatletter
\renewcommand\section{\@startsection {section}{1}{\z@}%
	{-3.5ex \@plus -1ex \@minus -.2ex}%nn
	{2.3ex \@plus.2ex}%
	{\normalfont\large\bfseries}}
\renewcommand\subsection{\@startsection{subsection}{2}{\z@}%
	{-3.25ex\@plus -1ex \@minus -.2ex}%
	{1.5ex \@plus .2ex}%
	{\normalfont\bfseries}}

\def\sec#1{\S \;\ref{#1}}

\title{{Pole-skipping as order parameter to probe a quantum critical point}}
\author{Navid Abbasi$^{a}$\footnote{abbasi@lzu.edu.cn} and Karl Landsteiner$^{b}$\footnote{karl.landsteiner@csic.es}\\
\small{\emph{$^{a}$School of Nuclear Science and Technology, Lanzhou University,}}\\
\small{\emph{ 222 South Tianshui Road, Lanzhou 730000, China }} \\
\small{\emph{$^{b}$Instituto de F\'\i{}sica Te\'o{}rica, IFT-UAM/CSIC}}\\
\small{\emph{Nicol\'a{}s Cabrerea 13-15, 28049 Campus Cantoblanco, Spain }} \\
}

\begin{document}

\setlength{\baselineskip}{16pt}
\begin{titlepage}
\maketitle

\vspace{-36pt}

%Abstract
\begin{abstract}
The holographic system described by Einstein-Maxwell-Chern-Simons dynamics in the bulk of AdS exhibits a chiral magnetic effect and a quantum critical point. Through numerical calculations, we find that the butterfly velocity can serve as a new identifier for the quantum critical point in this system. We show that the critical point is the point at which the butterfly velocity is equal to the speed of light in the direction of the magnetic field, while in the opposite direction the  butterfly propagation vanishes. Furthermore, by studying the pole-skipping points of the response function of the operator dual to the tensor part of the metric perturbation in the bulk, we discover a set of order parameters that distinguish the two states of the system near the quantum critical point.  Each of these order parameters is the sum of the absolute values of the real parts of momentum at all pole-skipping points associated with a particular frequency.  This quantity vanishes in the disordered state while taking a positive value in the ordered state. In addition, our results confirm the idea that the chiral magnetic effect can manifest macroscopically through quantum chaos.

  \end{abstract}
\thispagestyle{empty}
\setcounter{page}{0}
\end{titlepage}

\renewcommand{\baselinestretch}{1}  %looks better
\tableofcontents
\renewcommand{\baselinestretch}{1.2}  %looks better
%%%%%%%%%%%%%%%%%%%%%%%%%%%%%%%%%%%%
%______________________________________________________________
\section{Introduction}
%______________________________________________________________
In large N holographic systems, quantum chaos is quantified by the so-called \textit{quantum chaos points}, namely $(\omega_c, k_c)\equiv(i\lambda, \frac{i\lambda}{v_B})$, where $\lambda$ is the Lyapunov exponent and $v_B$ is the velocity of butterfly propagation in the system. These two quantities are basically \cite{Shenker:2013pqa, Shenker:2014cwa} encoded in an out-of-time-order correlator (OTOC). However,
OTOC calculations are generally difficult; in holography we should study it via shock wave geometry \cite{Maldacena:2015waa}. In quantum field theory, there are only a few well-known analytical results; for example, in 2d CFT \cite{Roberts:2014ifa}, in the SYK model \cite{Maldacena:2016hyu}, and in the weak coupling \cite{Stanford:2015owe}.
	
	Based on the fact that quantum chaos is related to energy dynamics in holographic systems, another approach is introduced: \textit{pole-skipping}.
	The effective field theory argument \cite{Blake:2017ris} corroborates \cite{Grozdanov:2017ajz}'s interesting holographic result that the dispersion of the energy density correlator, i.e. the line of poles of the energy density response function $G^R_{\mathcal{E}\mathcal{E}}(\omega, k)$, does not exist at the chaos point!
	The latter is equivalent to saying that at the chaos point, the numerator and denominator of $G^R_{\mathcal{E}\mathcal{E}}(\omega, k)$ will disappear; this is the \textit{pole-skipping phenomenon}\footnote{Such points have first been observed for diffusive and shear channels in holographic models in \cite{Amado:2007yr,Amado:2008ji}. They appear at real momenta and are related to causality rather than chaos.}.
	Since computing $G^R_{\mathcal{E}\mathcal{E}}(\omega, k)$ is easier than OTOC in holography, this gives us another way to find $\lambda$ and $v_B$ \cite{Blake:2018leo}.

\begin{figure}
	\centering
	\includegraphics[width=0.8\textwidth]{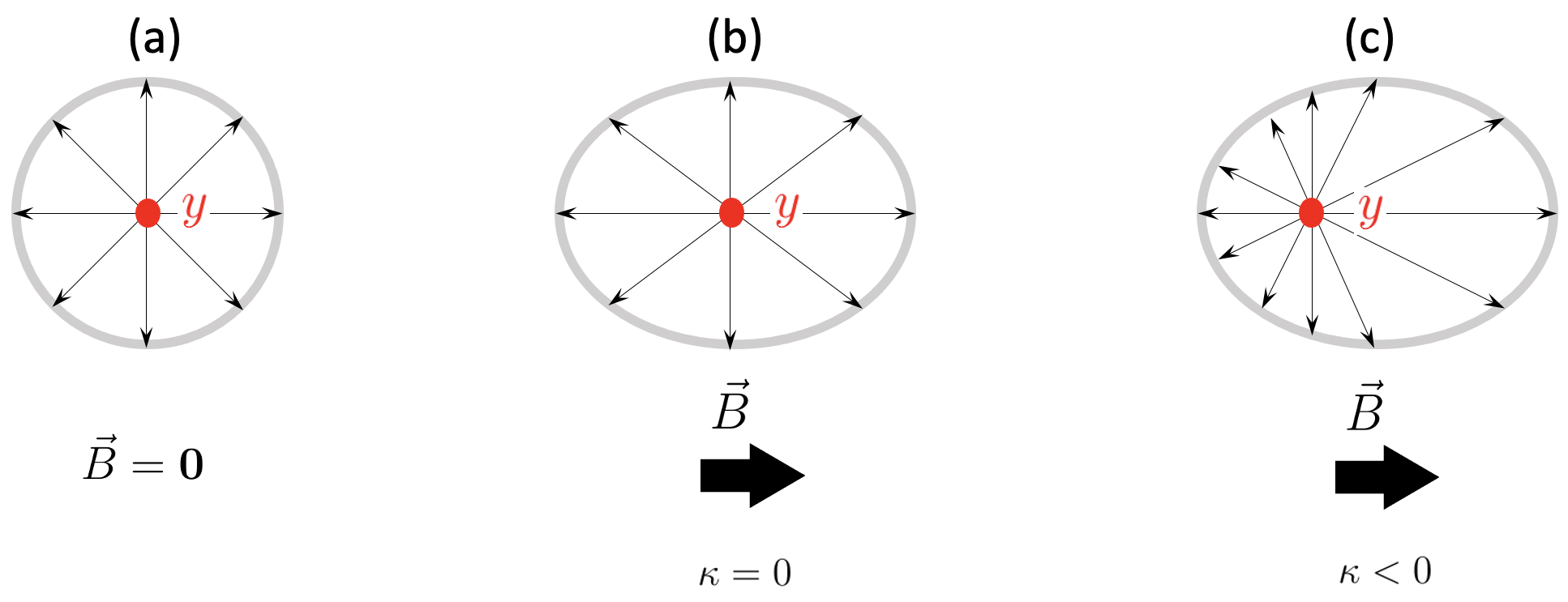}
	%	\,\,\,\includegraphics[width=0.38\textwidth]{Delta_V_B.pdf}
	\caption{Butterfly propagation after perturbing the system at point $y$. (a) Isotropic butterfly propgation in an uncharged  holographic plasma in the absence of magnetic field. (b) Anisotropic butterfly propgation in an electrically charged system in the presence of an externl mgnteic field. (c) The butterfly propagation in the system studied in this paper.  $\kappa$ is proportional to the strenght of the anomaly. When $\kappa=0$, the two butterflies in opposite directions have the same speed (panel b). The anomaly, however, breaks this symmetry. The difference in the speeds of the two butterflies along each axis is a manifestation of the chiral magnetic effect (panel c).}
	\label{intro}
\end{figure}

In this work we want to apply the above idea to a 5-dimensional Einstein-Maxwell theory with Chern-Simons term which is dual to a 4-dimensional gauge theory. The Chern-Simons term is the holographic dual of a 't-Hooft anomaly and the Chern-Simons coupling $\kappa$ determines the strength of the anomaly. At the specific value $\kappa=-2/\sqrt{3}$ this  is the holographic dual of $\mathcal{N}=4$ SYM theory. We study the system in the presence of a constant background magnetic field $B$ and a uniform electric charge density $\rho$ \cite{DHoker:2009ixq,DHoker:2010zpp,DHoker:2012rlj}. Butterfly velocities have been investigated for this system first in \cite{Abbasi:2019rhy} by analytically computing the pole-skipping point for small $B/T^2$ and $\rho/T$.  The main result of \cite{Abbasi:2019rhy} is that the anomaly splits the butterfly propagation velocity along the magnetic field. This is actually different from the anisotropy of the butterfly's velocity due to the magnetic field in the non-anomalous field theory. In the latter case, the butterfly speed depends on the angle between the magnetic field and the measurement axis \footnote{More precisely, it depends on $|\cos\theta|$, where $\theta$ is the mentioned angle.}. Therefore, the same butterfly velocity can be found in both directions parallel to the axis of the magnetic field. However, ref. \cite{Abbasi:2019rhy} shows that in the presence of anomaly, the butterfly speed splits in two directions along the magnetec field; the value in the direction of magnetic field is larger than that in the opposite direction (see figure \ref{intro}). It was noted in  \cite{Abbasi:2019rhy} that the latter is  reminiscent of the chiral magnetic effect \cite{Fukushima:2008xe}.

%%%%%%%%%%%%%%%%%%%%%%%
%%%%%%%%%%%%%%%%%%%%%%%

As discussed in ref. \cite{DHoker:2009ixq}, the Einstein-Maxwell theory with Chern-Simons term exhibits another interesting feature; it has a quantum critical point. In order to investigate how much information about the quantum critical point is encoded in the pole-skipping point(s), in this work, we will extend the study of \cite{Abbasi:2019rhy} to finite magnetic field and finite density. In the language of \cite{Abbasi:2019rhy} we want to numerically study the bulk solution in terms of two parameters $B/\mu^2$ and $T/\mu$ ($\mu$ is the chemical potential associated with $\rho$).  The case of small $B/\mu^2$ and large $T/\mu$ corresponds to \cite{Abbasi:2019rhy}. However, by performing high-precision numerical calculations, we will be able to study the pole-skipping phenomenon in the energy density correlator for arbitrary values of these two parameters. In particular, when $T/\mu$ is small, we vary $B/\mu^2$ to see how the butterfly velocities behave when the system approaches the quantum critical point \footnote{In ref. \cite{Ling:2016ibq}, the butterfly effect has been proposed as the diagnostic of quantum phase transition in an anisotropic holographic model exhibiting	metal-insulator transitions.}.

What we need to calculate is the pole-skipping points of the correlation function of energy momentum tensor, $T^{\mu\nu}$. Such a correlator is related to a certain scalar perturbation in the bulk \footnote{By scalar we mean spin zero under the $SO(2)$ rotations which is defined as the following. Setting the momentum of the correlator in a specific direction, e.g., the third direction, $T^{\mu\nu}$ components can then be classified according to the representation of the $SO(2)$ group of rotations perpendicular to that Fourier momentum.  In the bulk, we take the components of metric perturbation $\delta g_{\mu \nu}$ under the same considerations.}. On the other hand, a full analysis of such perturbations of the  Einstein-Maxwell-Chern-Simons background has some problems \cite{Ammon:2017ded}. Instead, we focus on $E_{vv}$ component of the Einstein's equations, with $v$ being the Eddington-Finkelstein time \cite{Blake:2018leo}. This gives us exclusively the upper-half plane pole-skipping points in the scalar channel \footnote{By upper-half plane we mean upper half of the ``$\text{Im} \,\omega - \text{Im} \,k$'' plane.}, which is sufficient to find the butterfly velocity in the system.

On the other hand, turning on the magnetic field $\hat{B}$ in the bulk corresponds to perturbing the boundary theory by a relevant operator \footnote{In this work, we choose to use the dimensionless magnetic field $\hat{B}$ and temperature $\hat{T}$ \cite{DHoker:2009ixq}, which will be defined in the main text.}.
When $\hat{T}\rightarrow 0$, the boundary theory reduces in the IR to a $(1+1)$ dimensional system, where the operator  dual to $\hat B - \hat B_c$ has the scaling dimension $\Delta=2$. In ref. \cite{DHoker:2010zpp}, the critical magnetic field associated with this relevant perturbation was found to be $\hat{B}_c\approx 0.499$.

The first result of our calculations is that the magnitude of butterfly speeds for $\hat{T}\rightarrow 0$
carries the same information about the phase transition in the system. In this limit, for large $\hat{B}$, we find that the speed of the butterfly in the direction of the magnetic field is equal to the speed of light. For small $\hat{B}$, the butterfly speed in the opposite direction turns out to be zero. We find that the critical point is determined by the magnetic field where the latter two occur simultaneously; it is actually $\hat{B}_c\approx 0.499$.

To gain a deeper understanding of the relationship between pole-skipping and quantum phase transitions in the system, we then investigate the tensor part of the metric perturbation in the bulk. 
%It is clear that the latter is coupled on the boundary to a scalar operator of dimension $\Delta=4$. 
The phenomenon of pole-skipping in correlators for such operators has been extensively discussed in the literature (see \cite{Blake:2019otz} and also  \cite{Natsuume:2019sfp,Natsuume:2019vcv,Wu:2019esr,Ahn:2019rnq,Li:2019bgc,Ceplak:2019ymw,Das:2019tga,Liu:2020yaf,Ahn:2020baf,Kim:2020url,Sil:2020jhr,Yuan:2020fvv,Grozdanov:2019uhi,Abbasi:2020ykq,Jansen:2020hfd,Choi:2020tdj,Abbasi:2020xli,Ceplak:2021efc,Yuan:2021ets,Haehl:2018izb,Ramirez:2020qer,Blake:2021hjj,Mahish:2022xjz,Wang:2022mcq,Amano:2022mlu,Baishya:2023nsb,Yuan:2023tft,Grozdanov:2023txs,Natsuume:2023lzy,Jeong:2023zkf,Natsuume:2023hsz} for an incomplete list of related works). However, none of the previous studies have investigated it near a quantum critical point.

Our results show that phase transitions can also be tracked in this channel; that is, this time pole-skipping introduces an order parameter too! We find that for $\hat{T}\rightarrow 0$ and for large $\hat{B}$, the spectrum of pole-skipping points consists of $2n$ points at $\omega= - n (i\,2 \pi T) $, with pure imaginary frequencies. This result is known for many holographic systems  \cite{Blake:2019otz}. Except for $\hat{T}\gg1$, the spectrum is asymmetric with respect to $\text{Im}\,k$ axis, indicating the chiral anomaly in this system \cite{Abbasi:2019rhy}.

  However, our surprising results arises at finite $\hat{B}$ and low $\hat{T}$. We find that the spectrum develops points with ``complex'' momentum. The ``nonzero'' real part of pole-skipping points is reminiscent of an ordered state below the quantum critical point, identified by a nonzero order parameter. Therefore, we are tempted to consider an order parameter determined by the real part of pole-skipping points. \textit{However, as highlighted in ref. \cite{DHoker:2010zpp}, quantum phase transitions in our system occur with no change in symmetry}. In other words, this is a quantum phase transition that preserves symmetry. Interestingly, it turns out that transitions between these states occur at $\hat{B}_c$.
  
In summary, the structure of pole-skipping points in the ``$\text{Im} \,\omega - \text{Re} \,k$'' plane suggests to define an order parameter;  it determines state of the system; when all points are located at $\text{Re}\,k=0$ on this plane, the system behaves as a Fermi liquid \cite{DHoker:2010zpp}. On the other hand, a point-symmetric distribution about $\text{Re}\,k=0$ on this plane corresponds to a non-Fermi liquid.

In the remainder of this paper, we first present the gravity settings for the boundary system in \sec{section}. In particular we will explain how to find the bulk solution numerically. In \sec{PS_energy_section} we compute the butterfly velocity in the system. The study of pole-skipping in the tensor channel is the subject of \sec{PS_section}. Finally, we conclude in \sec{Conclusion_section} by reviewing and discussing the results.
%______________________________________________________________
\section{Setup}
\label{section}
%______________________________________________________________
The bulk action is given by
%%%%%%%%%%%%%%%%%%%%%%%%%%%%%%%%%%%%%%%% 
\begin{equation}\label{action}
S =  \frac{1}{16 \pi G_5} \int_{\mathcal{M}} d^5 x \,\,\sqrt{-g} \left(R +\frac{12}{L^2} - F^{M N} F_{M N}\right)+S_{CS}+ S_{bdy}
\end{equation}
%%%%%%%%%%%%%%%%%%%%%%%%%%%%%%%%%%
with the Chern-Simons action being as the following
\begin{equation}\label{CS_action}
S_{CS}=\frac{\kappa}{12 \pi G_5}\int A \wedge F \wedge F=\,\frac{\kappa}{48\pi G_5}\int d^5 x \sqrt{-g}\,\,\epsilon^{\rho \mu \nu \alpha \beta}A_{\rho}F_{\mu\nu}F_{\alpha \beta}
\end{equation}
%%%%%%%%%%%%%%%%%%%%%%%%%%%%%%%%%%
and $S_{bdy}$ is the boundary counter term. The equations of motion  are given by:
%%%%%%%%%%%%%%%%%%%%%%%%%%%%%%%%%%
\begin{eqnarray}\label{Gauge_equ}
\nabla_{\nu}F^{\nu\mu }+\frac{\kappa}{4} \epsilon^{\mu \nu \rho  \alpha\beta}F_{\nu \rho}F_{ \alpha \beta}&=&0\\\label{Einstein_equ}
R_{\mu \nu}+4 g_{\mu \nu}+\frac{1}{3}F^{\alpha \beta}F_{\alpha \beta}\,\,g_{\mu \nu}+2 F_{\mu \rho} F^{\rho}_{\,\,\nu}&=&0
\end{eqnarray}
%%%%%%%%%%%%%%%%%%%%%%%%%%%%%%%%%%
from which one can find the following magnetized brane solution in the bulk
%%%%%%%%%%%%%%%%%%%%%%%%%%%%%%%%%%
\begin{equation}\label{metric}
ds^2=\frac{dr^2}{f(r)}-f(r)dt^2+ e^{2W_T(r)}(dx_1^2+dx_2^2)+e^{2W_L(r)}(dx_3+C(r)dt)^2
\end{equation}
%%%%%%%%%%%%%%%%%%%%%%%%%%%%%%%%%%
%%%%%%%%%%%%%%%%%%%%%%%%%%%%%%%%%%
\begin{equation}\label{field_strenght}
F=E(r) dr\wedge dt+B dx_1\wedge dx_2+ P(r) dx_3\wedge dr 
\end{equation}
%%%%%%%%%%%%%%%%%%%%%%%%%%%%%%%%%%
We follow \cite{DHoker:2009ixq} and numerically solve equations \eqref{Gauge_equ} and \eqref{Einstein_equ} to find the bulk fields $\mathcal{G}(r )=\{f(r), w_T(r), w_L(r), C(r), E(r), P(r) \}$.
%______________________________________________________________
\subsection{Near horizon solution}
%______________________________________________________________
 Let us start with the near horizon expansion of $\mathcal{G}$
%%%%%%%%%%%%%%%
\begin{equation}\label{NH}
\mathcal{G}(r)=\,\sum_{n=0} \mathcal{G}^{(n)}\,(r-r_h)^n
\end{equation}
%%%%%%%%%%%%%%%%%
It should be noted that we can always exploit the background symmetry to set $r_h=1$.
\\\\ 
The  boundary conditions for the  the whole bulk solution can be imposed on the leading order coefficients $\mathcal{G}^{(n)}$; to this end we take the horizon data as\footnote{Note that $f^{(1)}= 1$ is equivalent to take $T=1/4\pi$.}
%%%%%%%%%%%%%%%
\begin{equation}\label{horizon_data}
f^{(0)}=0\,,\,\,f^{(1)}= 1\,,\,\,w_T^{(0)}=w_L^{(0)}=0\,,\,\,E^{(0)}=q\,,\,\,C^{(0)}=0\,,\,\,P^{(0)}=p
\end{equation}
%%%%%%%%%%%%%%%%%
This corresponds to taking
%%%%%%%%%%%%%%%%%%%%%%%%%%%%%%%%%%
\begin{equation}\label{mettric_field_strenght_horizon}
\begin{split}
ds_H^2=&\,dx_1^2+dx_2^2+dx_3^2\\
F_H=&\,q \,dr\wedge dt+\,\epsilon_{ijk} b_k dx_i\wedge dx_j+ p_i\, dx_i\wedge dr 
\end{split}
\end{equation}
%%%%%%%%%%%%%%%%%%%%%%%%%%%%%%%%%%
where $\vec{b}=(0,0,b)$ and $\vec{p}=(0,0,p)$.
Using the horizon data, the Einstein and Maxwell equations can be solved perturbatively to find the higher order coefficients in \eqref{NH}. In particular, we find: 
%%%%%%%%%%%%%%%%%
\begin{eqnarray}\nonumber
C^{(1)}&=&\,p+ 2 q \,\kappa\, b \\
w_{T}^{(1)}&=&\frac{2}{3}(6 - 2 b^2 - q^2)\\
w_L^{(1)}&=&\frac{2}{3}\bigg(6+b^2-q^2-3\big(\kappa b +\frac{p}{2q}\big)^2\bigg)
\end{eqnarray}
%%%%%%%%%%%%%%%%%
Then for any set of values of the three parameters $q$, $b$, and $p$, one can numerically integrate the bulk dynamic equations and read the corresponding boundary quantities.
%______________________________________________________________
\subsection{Bulk solution}
%______________________________________________________________
The physical parameters on the boundary are temperature, chemical potential and the magnetic field. However, due to the scale invariance, only dimensionless combinations of quantities have physical meaning. 
We choose to work with normalized dimensionless magnetic field $\hat{B}$ and temperature $\hat{T}$ as  
%%%%%%%%%%%%%%%
\begin{equation}\label{}
\hat{B}=\,\frac{b}{\rho^{2/3}}\,,\,\,\,\,\,\,\,\hat{T}=\,\frac{T}{(b^3+\rho^2)^{1/6}}
\end{equation}
%%%%%%%%%%%%%%%%%
Here $\rho$ is the charge density defined as  
%%%%%%%%%%%%%%%%%%%%%%%%%%%%%%%%%%
\begin{equation}\label{}
	\rho=\,\lim_{r \rightarrow +\infty}\frac{E(r)}{r^2}
\end{equation}
%%%%%%%%%%%%%%%%%%%%%%%%%%%%%%%%%%
In this setting \cite{DHoker:2009ixq}, any specific bulk solution is characterized with two quantities $\hat{B}$ and $\hat{T}$. These are actually the asymptotic data. What we will do is to take a fixed value for $\hat{B}$ and then find the numerical solution to the bulk equations with varying $\hat{T}$. As it was mentioned earlier, any numerical solution itself is specified with the values of the three horizon parameters $q$, $b$, and $p$. For generic values of these horizon parameters we find a solution with a non-vanishing value of the metric function $C(r)$. We chose to implement the additional boundary condition $\lim_{r\rightarrow\infty}C(r)=0$. This give a co-dimension one slice of the horizon data.
In other words, a specific bulk solution maps  a set of $(q, b, p)$ on this co-dimension one slce into a unique set of $(\hat{B}, \hat{T})$ in which the boundary metric is the standard Minkowski metric.
%______________________________________________________________
\section{Pole-skipping in energy correlator and the butterfly velocity}
\label{PS_energy_section}
%______________________________________________________________
In order to find the pole-skipping points of energy density Green's function, it is convenient to work in the ingoing Eddington-Finkelstein coordinates.\footnote{In these coordinates, the regularity of solutions in the future event horizon is automatically satisfied.} It is easy to show that in these coordinates the electromagnetic field strength is given by \eqref{field_strenght}, as before. However, metric transforms to the following form 
%%%%%%%%%%%%%%%
\begin{equation}\label{Eddington}
	ds^2=-\tilde{f}(r)dv^2+2g(r)dr dv+2\left(j(r) dv +\frac{}{}s(r)dr\right) dx_3+e^{2W_{T}(r)}(dx_1^2+dx_2^2)+e^{2W_{L}(r)}dx_3^2
\end{equation}
%%%%%%%%%%%%%%%%%
where $v$ is the Eddington-Finkelstein time coordinate and
%%%%%%%%%%%%%%%%%
\begin{eqnarray}\nonumber
	\tilde{f}(r)&=&f(r)-e^{2W_{L}(r)}C(r)^2\\
	g(r)&=&\left(1-e^{2W_{L}(r)}\frac{C(r)^2}{f(r)}\right)^{1/2}\\\nonumber
	j(r)&=&C(r)e^{2W_{L}(r)}\\\nonumber
	s(r)&=&-\frac{C(r)e^{2W_{L}(r)}}{\sqrt{f(r)\left(f(r)-C(r)^2e^{2W_{L}(r)}\right)}}.
\end{eqnarray}
%%%%%%%%%%%%%%%%%
Turning on $\delta g_{vv}(r,v,x)=\delta g_{vv}(r)e^{-i\omega v+i \vec{k}\cdot \vec{x}}$ and other perturbations that couple it, we need to evaluate $vv$ component of \eqref{Einstein_equ} at $r=r_h=1$  (see Appendix for more details). We find
%%%%%%%%%%%%%%%%%%%%%%%%%%%%%%%%%%
\begin{equation}\label{E_vv}
\begin{split}
\bigg[ k^2-  i  
\vec{k} \cdot (\frac{\vec{p}}{q}+2 \kappa \vec{b})&+\frac{i \omega}{2} \bigg(\big(\frac{\vec{p}}{q}+2 \kappa \vec{b}\big)^2+4 \big(-6+q^2+b^2\big)\bigg)\bigg]\delta g_{vv}^{(0)}\\
+\,&\bigg(\omega- \frac{i}{2}\bigg)\bigg(2 k_i \delta g_{v x_i}^{(0)}+\omega \big(\delta g_{x_1x_1}^{(0)}+\delta g_{x_2x_2}^{(0)}+\delta g_{x_3x_3}^{(0)}\big)\bigg)=\,0
\end{split}
\end{equation}
%%%%%%%%%%%%%%%%%%%%%%%%%%%%%%%%%%
Two comments are in order
\begin{itemize}
	\item We have presented the above equation in a fully $SO(3)$ covariant form. Then this allows us to simply explore both longitudinal and transverse cases. 
	\item At $\omega_p=\frac{i}{2}$, the above equation becomes a decoupled equation for $\delta g_{vv}^{(0)}$. This is in fact the same as the original calculation of  \cite{Blake:2018leo} where the decoupling occurs at $\omega = i 2 \pi T$. Here we are writing the equation in the horizon frame wherein the temperature was set to $T=1/4 \pi$.
	\end{itemize}

%______________________________________________________________
\subsection{Longitudinal}
%______________________________________________________________
 When $\vec{k}\parallel\vec{b},\,\vec{p}$\, we find
%%%%%%%%%%%%%%%%%%%%%%%%%%%%%%%%%%
\begin{equation}\label{}
\bigg[\bigg(k-i\big(\frac{p}{2q}+  \kappa b\big)\bigg)^2+(6-b^2-q^2)\bigg]\delta g_{vv}^{(0)}=\,0
\end{equation}
%%%%%%%%%%%%%%%%%%%%%%%%%%%%%%%%%%
Clearly, the above equation becomes ambiguous at the $k=k_p$ where $k_p$ is the root of the expression in the square brackets.
As a result, we find the butterfly velocity in the horizon frame as 
%%%%%%%%%%%%%%%%%%%%%%%%%%%%%%%%%%
\begin{equation}\label{UHP}
v_B^{(H)}=\,\frac{\omega_p}{k_p}=\,\left(\frac{p}{q}+ 2 \kappa b \pm 2\sqrt{6- b^2 - q^2}\right)^{-1}
\end{equation}
%%%%%%%%%%%%%%%%%%%%%%%%%%%%%%%%%%
To proceed further, let's comment on the homogeneity of \eqref{E_vv} in spatial directions. This is actually a direct consequence of taking the horizon frame as specified by the metric \eqref{mettric_field_strenght_horizon}. On the other hand, we are interested in working in the asymptotic frame given by 
%%%%%%%%%%%%%%%%%%%%%%%%%%%%%%%%%%
\begin{equation}\label{Assymptotic_frame}
\begin{split}
ds^2 \sim&\, \frac{dr^2}{r^2}+r^2(-d\tilde{t}^2+d\tilde{x}_1^2+d\tilde{x}_2^2+d\tilde{x}_3^2)\\
F=&\,E dr \wedge d\tilde{t}+ B d \tilde{x}_1 \wedge d \tilde{x}_2 + P d\tilde{x}_3 \wedge dr
\end{split}
\end{equation}
%%%%%%%%%%%%%%%%%%%%%%%%%%%%%%%%%%
This is actually  found by rescaling the solution \eqref{metric} as 
\begin{equation}\label{}
\tilde{t}=t\,,\,\,\,\,\,\,\tilde{x}_{1,2}= \sqrt{w_T}\,\, x_{12}\,,\,\,\,\,\,\,\tilde{x}_{13}= \sqrt{w_L}\,x_3
\end{equation}
%%%%%%%%%%%%%%%%%%%%%%%%%%%%%%%%%%
where
%%%%%%%%%%%%%%%%%%%%%%%%%%%%%%%%%%
\begin{equation}\label{w_L_T}
w_{T,L}=\,\lim_{r \rightarrow +\infty}\frac{e^{2 W_{T,L}(r)}}{r^2}
\end{equation}
%%%%%%%%%%%%%%%%%%%%%%%%%%%%%%%%%%
In terms of boundary frame coordinates, the horizon metric takes the following form
%%%%%%%%%%%%%%%%%%%%%%%%%%%%%%%%%%
\begin{equation}\label{}
ds_H^2=\,\frac{1}{w_T}\big(d\tilde{x}_1^2+d\tilde{x}_2^2\big)+\frac{1}{w_L}d\tilde{x}_3^2
\end{equation}
%%%%%%%%%%%%%%%%%%%%%%%%%%%%%%%%%%
Therefore a perturbation of the form $e^{-i \omega t+ i k x_3}$ in the horizon frame, takes the form $e^{-i \omega \tilde{t}+ i \frac{k}{\sqrt{w_L}} \tilde{x}_3}$ in the boundary frame. 
As a result, the butterfly velocity in the boundary frame reads:
%%%%%%%%%%%%%%%%%%%%%%%%%%%%%%%%%%
\begin{equation}\label{v_B_L}
v^L_{B}=\, \frac{\omega}{k/\sqrt{w_L}}=\,\sqrt{w_L}\,v_B^{(H)}\,\,\,\,\rightarrow \,\,\,\,\,\boxed{v_{B\pm}^L=\,
\frac{\sqrt{w_L}}{\frac{p}{q}+ 2 \kappa b \pm 2\sqrt{6- b^2 - q^2}}}
\end{equation}
%%%%%%%%%%%%%%%%%%%%%%%%%%%%%%%%%%
For super-symmetric theories the value of $\kappa$ is known: $\kappa=-2/\sqrt{3}$ \cite{Cvetic:1999ne,Chamblin:1999tk}. On the other hand, any given set of $(\hat{B},\hat{T})$ on the boundary is associated with a unique set of $(q,b, p )$ near the horizon. The latter is sufficient to find the whole bulk solution and consequently to read off $w_L$ via \eqref{w_L_T}. Thus \eqref{v_B_L} can be evaluated once $(\hat{B},\hat{T})$ is determined.

The important point of \eqref{v_B_L} is that we found two different butterfly velocities along the magnetic field. This is exactly the same finding in \cite{Abbasi:2019rhy} for small magnetic field and small chemical potential. Thus, the split between butterfly velocities is a general effect across the range of quantities in the system.
%______________________________________________________________
\subsection{Transverse}
%______________________________________________________________
 When $\vec{k}\perp\vec{b},\,\vec{p}$, \eqref{E_vv} reduces to
 %%%%%%%%%%%%%%%%%%%%%%%%%%%%%%%%%%
 \begin{equation}\label{}
 \bigg[k^2-\bigg(\frac{p}{2q}+  \kappa b\bigg)^2+(6-b^2-q^2)\bigg]\delta g_{vv}^{(0)}=\,0
 \end{equation}
 %%%%%%%%%%%%%%%%%%%%%%%%%%%%%%%%%%
Finding the root of the square bracket, $k_p$, and 
taking similar steps as what were done in the longitudinal case, we arrive at 
%%%%%%%%%%%%%%%%%%%%%%%%%%%%%%%%%%
\begin{equation}\label{}
\boxed{v_B^T=\,\pm
\frac{1}{2}\left(\frac{w_T}{6-b^2-q^2-\big(\frac{p}{2q}+  \kappa b\big)^2}\right)^{1/2}}
\end{equation}
%%%%%%%%%%%%%%%%%%%%%%%%%%%%%%%%%%
Again, once  $(\hat{B},\hat{T})$ is given, $v_B^T$ can be calculated from the above formula upon performing numerical calculation to find the bulk solution and consequently $w_T$.

Again, in complete agreeent with the result of \cite{Abbasi:2019rhy}, here we find non-split speeds in the direction perpendicular to the magnteic field. This is another way of saying the anomaly is not detected in the transverse direction for the whole range of physical quntities in the system.
%______________________________________________________________
\subsection{Numerical results}
\begin{figure}[h]
	\centering
	\includegraphics[width=0.3\textwidth]{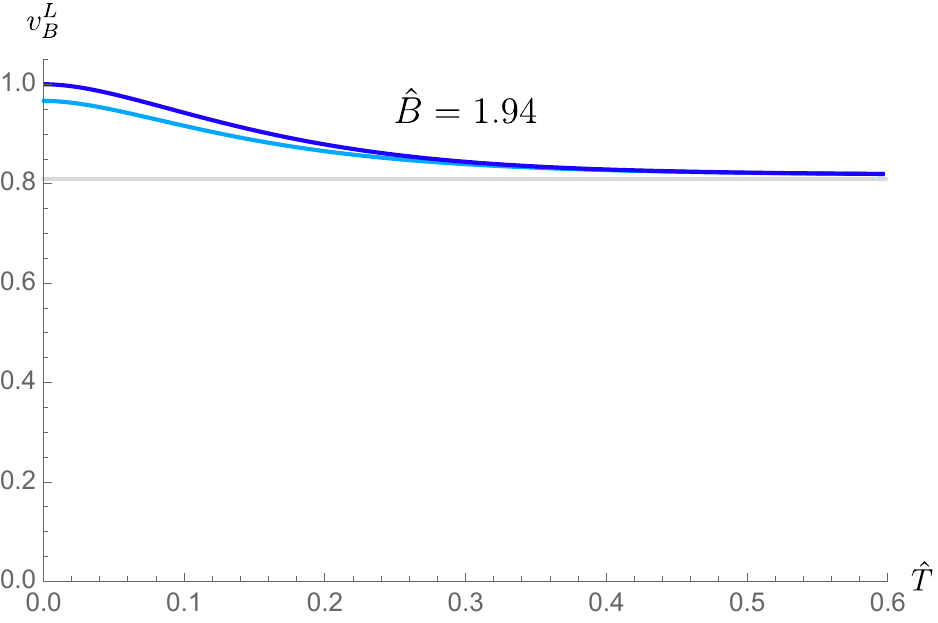}	\includegraphics[width=0.3\textwidth]{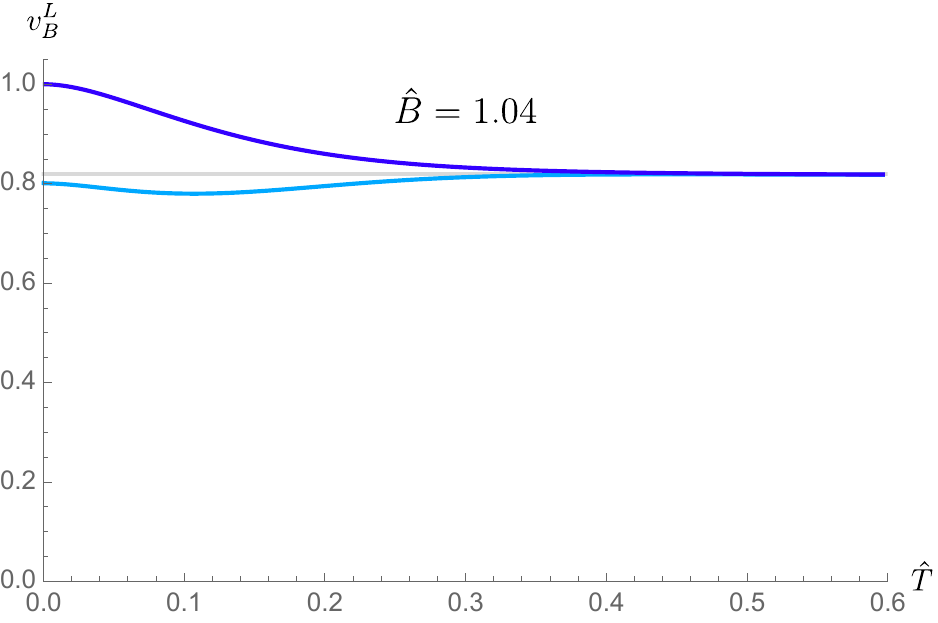}	\includegraphics[width=0.3\textwidth]{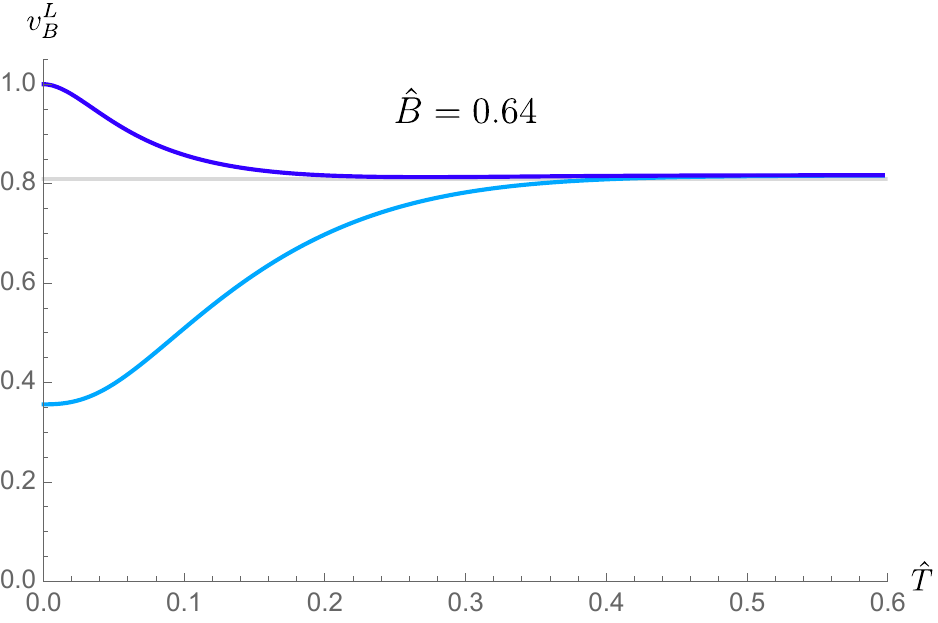}
	\includegraphics[width=0.3\textwidth]{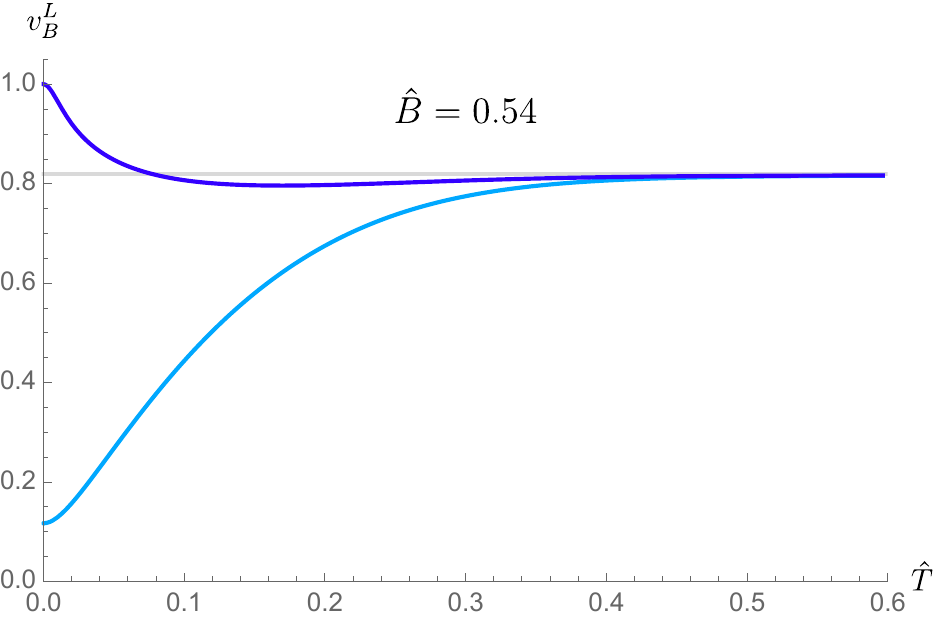}	\includegraphics[width=0.3\textwidth]{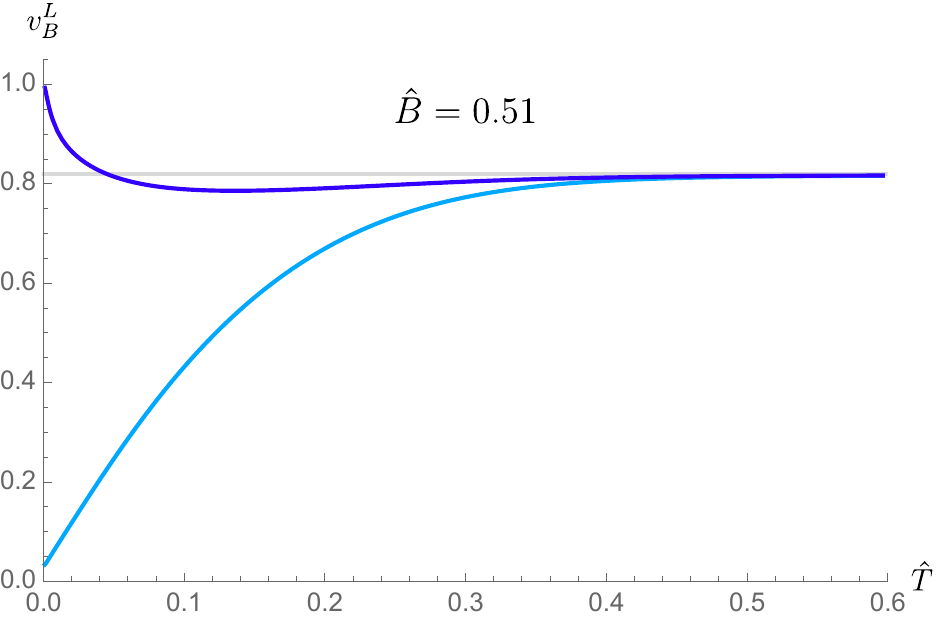}	\includegraphics[width=0.3\textwidth]{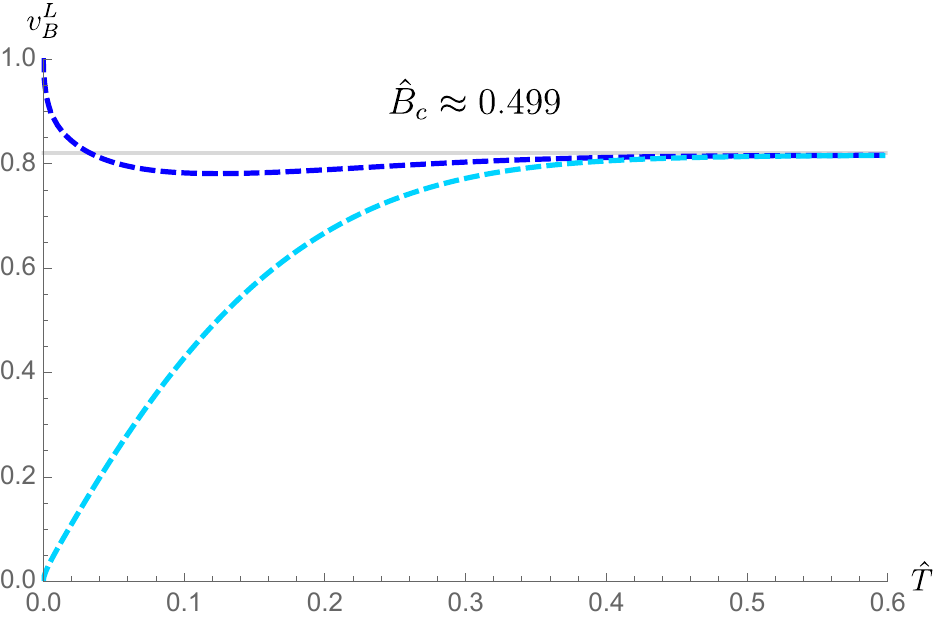}
	\includegraphics[width=0.3\textwidth]{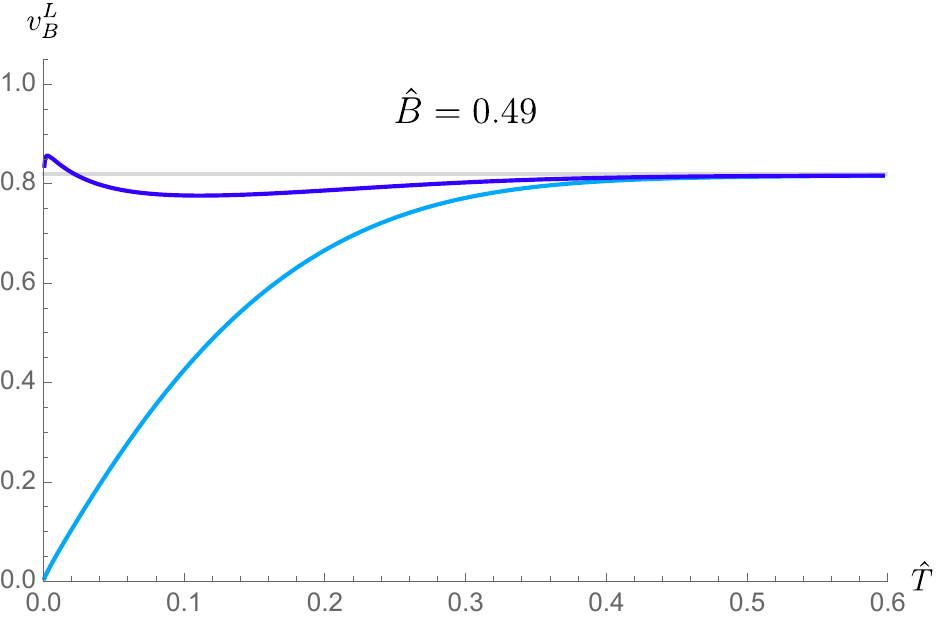}	\includegraphics[width=0.3\textwidth]{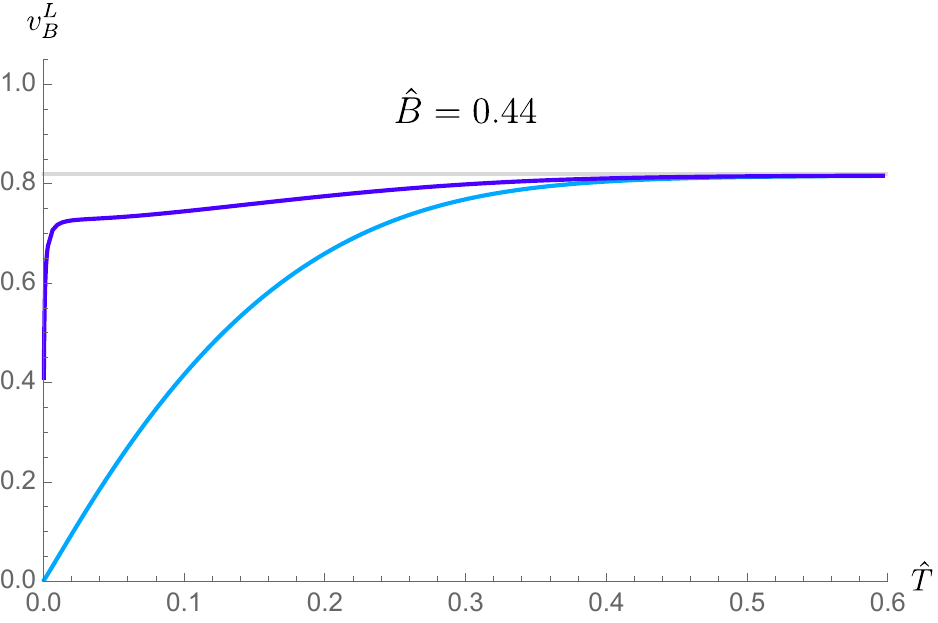}	\includegraphics[width=0.3\textwidth]{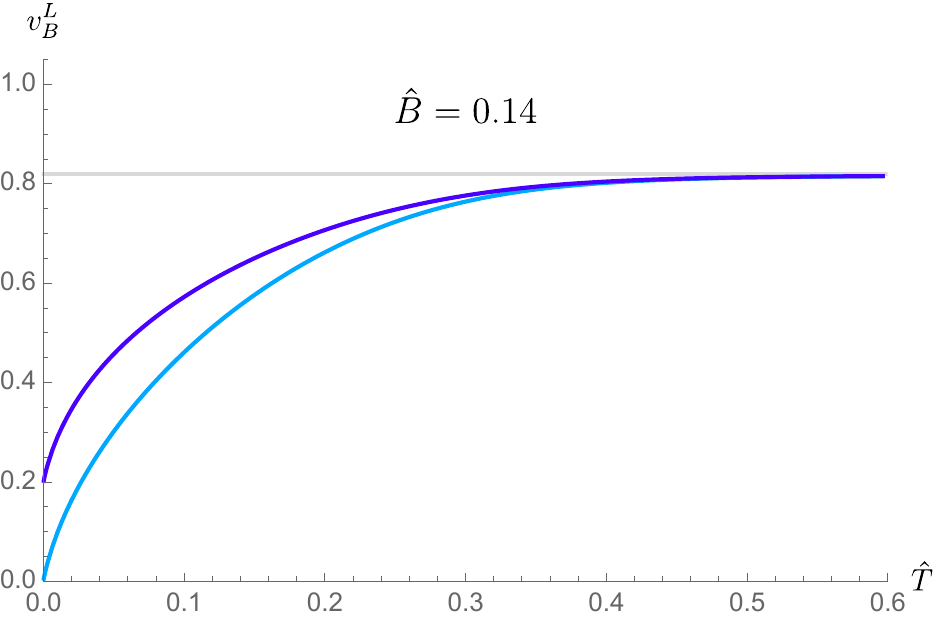}
	%	\,\,\,\includegraphics[width=0.38\textwidth]{Delta_V_B.pdf}
	\caption{Longitudinal butterfly speeds as functions of the dimensionless temperature in a holographic chiral system for a wide range of dimensionless magnetic field ($0.14 < \hat{B} < 1.94$). The dark blue shows the butterfly propagation parallel to the magnetic field, while the light blue curve is associated with the opposite direction.\textbf{ The dashed plot corresponds to critical magnetic field associated with the quantum phase transition found in \cite{DHoker:2009ixq}.} At the critical point, the butterfly propagates at the speed of light in the direction of magnetic while it is frozen in the opposite direction.  The gray line points out to the Schwarzschild result; i.e., $v_B=\sqrt{2/3}$. }
	\label{results}
\end{figure}

\begin{SCfigure}
%	\centering
	\includegraphics[width=0.45\textwidth]{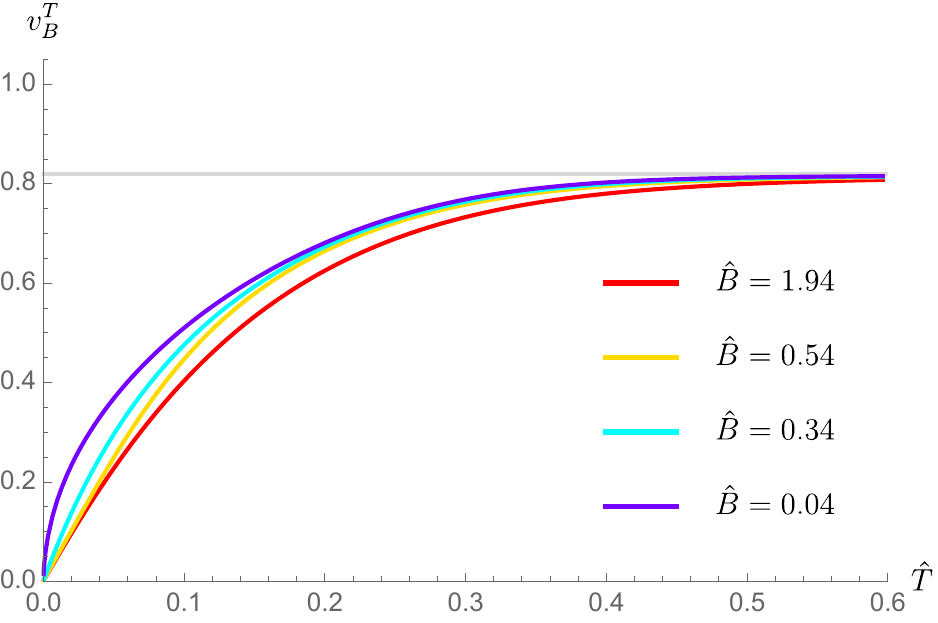}
	%	\,\,\,\includegraphics[width=0.38\textwidth]{Delta_V_B.pdf}
	\caption{Transverse butterfly speed in terms of the dimensionless temperature in a holographic chiral system for a wide range of dimensionless magnetic field ($0.04 < \hat{B} < 1.94$). 
		\\The splitting of butterfly speeds in the longitudinal directions is not the case here. }
	\label{results_2}
\end{SCfigure}

%______________________________________________________________
The results have been given in figures \ref{results} and \ref{results_2}. Let us elaborate on it in the following. In figure \ref{results}, we have shown how $v_{B,+}^L$ (dark blue) and $|v_{B,-}^L|$ (light blue) depend on $\hat{T}$, when the external magnetic field $\hat{B}$ is varied.
%%%%%%%%%%%%%%%%%%%%%%%
%%%%%%%%%%%%%%%%%%%%%%%
\begin{itemize}
	\item At high $\hat{T}$, regardless of the value of $\hat{B}$, the two velocities become the same as the Schwarzschild result, i.e., $v_B=\sqrt{2/3}$ \cite{Shenker:2013pqa}, that is to say the anomaly effects totally disappear.
	\item At finite values of $\hat{T}$ and $\hat{B}$, the two longitudinal butterfly velocities are always split. This is actually a direct consequence of the anomaly in the system. The velocity in the direction of the magnetic field is always greater than the velocity in the opposite direction. The same result was first found in ref. \cite{Abbasi:2019rhy} for small magnetic field and small chemical potential. (See Appendix for details and a summary of \cite{Abbasi:2019rhy} results).	
	\item Interestingly, the splitting at $\hat{T}=0$ encodes important information about the quantum critical point of the theory:
	\begin{itemize}
		\item At $\hat{B}>\hat{B}_c$, $v_{B,+}^L$ is always equal  to $1$ at $\hat{T}=0$. However, by reducing $\hat{B}$ from $\infty$ to $\hat{B}_c$, $|v_{B,-}^L|$ is reduced from $1$ to $0$.
		\item It is known that $\hat{B}_c\approx 0.499$ determines the quantum phase transition \cite{DHoker:2009ixq} in the system when the system is perturbed by the relevant operator of dimension $\Delta=2$. Here we find that, under the same $\hat{B}$, $v_{B,+}^L=1$ and $v_{B,-}^L=0$ (see the right panel of the middle row in the figure). This can be seen as another special feature by which the quantum critical point in this system can be identified.
		\item At $\hat{B}<\hat{B}_c$, $v_{B,-}^L$ remains $0$, and $v_{B,+}^L$ goes from $1$ to $0$ by reducing $\hat{B}$ from $\hat{B}_c$ to $0$.
		\end{itemize}
	\item In only two cases, the longitudinal velocity degenerates at $\hat{T}=0$. First, $\hat{B}=0$, which is actually expected; since in this case anomaly is not excited. Second, at large $\hat{B}$. We comment on this point below.
	\end{itemize}
%%%%%%%%%%%%%%%%%%%%%%%
%%%%%%%%%%%%%%%%%%%%%%%
Figure \ref{results_2} illustrates how the transverse butterfly speed depends on temperature. Interestingly we find that at any value of magnetic field, it vanishes at $\hat{T}=0$. This is reminiscent of weakly coupled field theory under strong magnetic fields; the system reduces to a $(1+1)d$ system at $\hat{T}=0$, and the two transverse directions are completely decoupled. Here we see that although our system is strongly coupled, it exhibits similar behavior. Furthermore, due to the symmetry of the background, the system becomes $2d$ CFT. This confirms why when $\hat{T}=0$, we find two butterflies traveling at the speed of light. 
%______________________________________________________________
\section{Pole-skipping in the tensor channel}
\label{PS_section}
%______________________________________________________________
So far we focused on the scalar part of the energy momentum perturbations. Considering only $E_{vv}=0$, we were able to find the upper-half plane pole-skipping points of the energy density correlation function.  However, in order to find the full spectrum of pole-skipping points in that case the set of all metric perturbations in the scalar channel have to be considered.  We do not elaborate on this issue anymore here. Instead, we choose to investigate the pole-skipping point in the tensor channel. In the latter case, the only involved dynamical fields are $\delta g_{xy}$, $\delta g_{xx}$, $\delta g_{yy}$, and $\delta F_{xy}$.
It is easy to show that $H_{xy}=e^{-2W_T}\,\delta g_{xy}$ behaves like a decoupled scalar field in this channel. Moreover, it sources an operator of weight $\Delta=4$, say $\mathcal{O}$, on the boundary. Our goal in this section is to find the pole-skipping points of the $G^{R}_{\mathcal{O}\mathcal{O}}$. To this end we start with studying the dynamics $H_{xy}(r,t,x_i)$ near the horizon.

Similar to our earlier discussions in the scalar channel, we take  $H_{xy}(r,t,x_i)\equiv H_{xy}(r)e^{- i \omega t + i k x_3}$ \footnote{It should be pointed out that here we only consider the case where the momentum is parallel to the magnetic field. It is easy to show that in the tranverse case, there is no pole-skipping phenomenon.}.  It turns out that $H_{xy}(r)$ obeys the following ordinary differential equation in the bulk
%%%%%%%%%%%%%%%%%%%%%%%%%%%%%%%%%%
\begin{equation}\label{H_xy_eq}
	H''_{xy}(r)+ a(\omega, \textbf{k})H'_{xy}(r)+ b(\omega, \textbf{k}) H_{xy}(r)=\,0
\end{equation}
%%%%%%%%%%%%%%%%%%%%%%%%%%%%%%%%%%
We omit the explicit expression of the functions $a(\omega, \textbf{k})$ and $b(\omega, \textbf{k})$.
In order to analyze this equation near the horizon, we take 
%%%%%%%%%%}%%%%%%%%%%%%%%%%%%%%%%%%
\begin{equation}\label{H_xy}
	H_{xy}(r)=\,\sum_{n=0} \varphi_n\,(r-r_h)^n
\end{equation}
%%%%%%%%%%%%%%%%%%%%%%%%%%%%%%%%%%
and plug it into \eqref{H_xy_eq}. The result is a set of coupled algebraic equations for the coefficients $\varphi_n$. Defining $\wn=\frac{\omega}{2\pi T}$ and $\qn=\frac{\textbf{k}}{2\pi T}$, the first four of these equations can be formally written as 
%%%%%%%%%%%%%%	

%%%%%%%%%%%%%%%
\begin{eqnarray}\label{M_11}
	0&=&M_{11}\varphi_{0}+\,(i \wn-1)\varphi_{1}\, ,\\\label{M_21}
	0&=&M_{21}\varphi_{0}+M_{22}\varphi_{1}+\,(i\wn-2)\varphi_{2}\, ,\\\label{M_31}
	0&=&M_{31}\varphi_{0}+M_{32}\varphi_{1}+M_{33}\varphi_{2}+\,(i \wn-3)\varphi_{3}\, ,\\\label{M_41}
	0&=&M_{41}\varphi_{0}+M_{42}\varphi_{1}+M_{43}\varphi_{2}+M_{44}\varphi_{3}+\,(i \wn-4)\varphi_{4} \, ,
\end{eqnarray}
%%%%%%%%%%%%%%%
where the coefficients, $M_{rs}$, are in fact functions of $\wn$  and $\qn$. In Appendix , we present the explicit expression for the first three of these coefficients. Higher order coefficients have very complicated expressions and we avoid writing them in the paper.    
%As is obvious, at  $\ell^{th}$ order  of the near-horizonn expansion, one finds a linear equation relating $\varphi_0$, $\varphi_1$, $\cdots$ and $\varphi_{\ell+1}$, with the coefficient of $\varphi_{\ell+1}$ vanishing at the pole-skipping frequency $\wn=-i  (\ell+1)$. Here $T$ is the horizon temperature.
As it is obvious from the above equations that just at the frequency $\wn_{\ell}=-i  \ell$, the first $\ell$ equations decouple from the rest of them and take the following form
%%%%%%%%%%%%%%%
\begin{equation}
	0	=\,\mathcal{M}_{\ell\times \ell}(\wn=-i \ell,\tilde{\qn}^2)\begin{pmatrix}
		\phi_0\\
		\phi_1\\
		.\\
		.\\
		\phi_{\ell-1}\\
	\end{pmatrix}.
\end{equation} 
%%%%%%%%%%%%%%%
The roots of the equation $\det \mathcal{M}_{\ell\times \ell}(\wn=-i\ell, \tilde{\qn})=0$,  are then those wave-numbers at which,  for a given UV normalization constant, the ingoing boundary condition at the horizon is not sufficient to uniquely fix  a solution for $H_{xy}$ in the bulk. Let us call the roots $\tilde{\qn}_1, \tilde{\qn}_2, \cdots, \tilde{\qn}_{2\ell}$.  At these $2\ell$ points, the response function of the boundary operator dual to $H_{xy}$, namely $G^R_{\mathcal{O}\mathcal{O}}$, is multi-valued at $\wn=- i \ell$. These points are the so-called \textbf{\textit{level-$\ell$ pole-skipping points}}~\cite{Blake:2017ris,Blake:2018leo}.

Let us for the sake of clarity illustrate the expression of the pole-skipping points associated with $\ell=1$.  Solving the equitation $\det \mathcal{M}_{1\times 1}(\wn=-i, \tilde{\qn}^2)=0$ gives (in the horizon frame)
	%%%%%%%%%%}%%%%%%%%%%%%%%%%%%%%%%%%
	\begin{equation}
\ell=1:\,\,\,\,\,\,\,\,\,\,\,\tilde{\qn}_{1,2}=\,\pm 2 i \sqrt{6-b^2-q^2} - i \left(\frac{2 p}{q} + \kappa b\right)
	\end{equation}
	%%%%%%%%%%%%%%%%%%%%%%%%%%%%%%%%%%
	which is the same as the momentum of the upper-half plane pole-skipping  in \eqref{UHP}. 	
	
	It is clear that in the absence of anomaly effects, i.e. when $p=b=0$, the pole-skipping points are symmetrically located along $\text{Im}\,\qn$ axis. However, as was first found in ref. \cite{Abbasi:2019rhy}, anomaly makes this symmetry break. This was actually observed for the pole-skipping points of the lowest four levels, $\ell=1,2,3,4$, and in the limit of small charge and small magnetic field. Here by use of numerical analysis we will discuss the first seven levels at finite value of magnetic field in a wide range of temperatures down to $T=0$.  Our goal is to investigate the behavior of pole-skipping points when the system is in the vicinity of the quantum critical point. Let us emphasize that the analytical expressions for higher level  pole-skipping points, i.e., $\ell=2,3,4,5,6,7$ are either extremely complicated. Therefore we omit their explicit expressions here and proceed with illustrating their numerical values. 
	
	We present the numerical results in three parts, in the following three subsections.
%______________________________________________________________
\subsection{High temperature limit}
\begin{SCfigure}
		\includegraphics[width=0.48\textwidth]{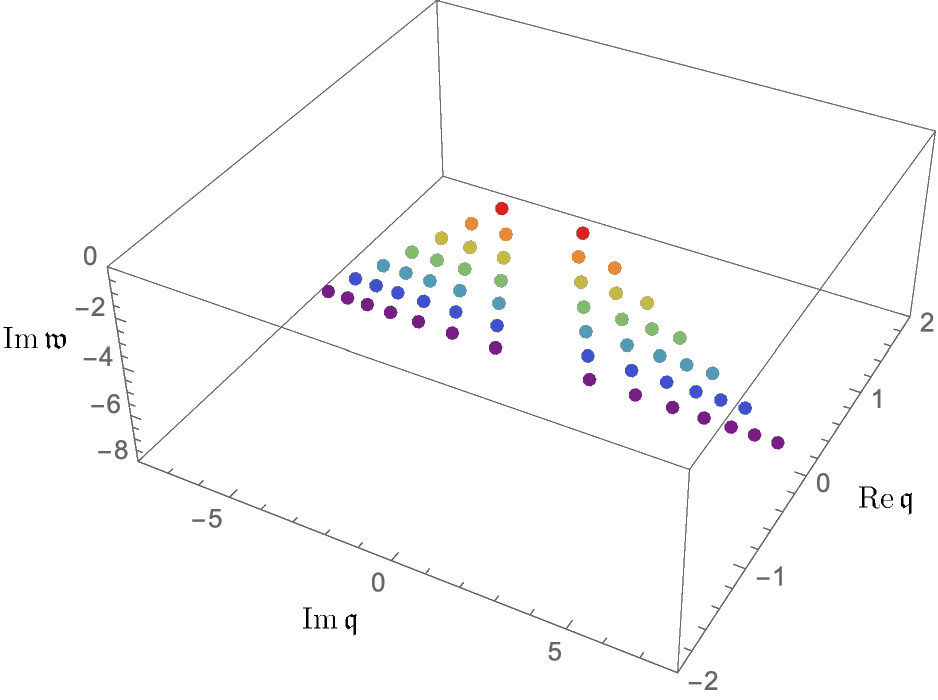}
		\caption{High temperature limit $(\hat{T}\gtrsim 0.6)$: The pole-skipping occurs at Matsubara frequencies and at special purely imaginary values of momenta. We have illustrated the pole-skipping points associated with first seven Matsubara frequencies. As many other studies in the literature in this limit, the spectrum is symmetric, in the sense that anomaly is not detected. Moreover, varying the magnetic field within the range $0.01 < \hat{B}< 1.94$, the spectrum remains unchanged; as if there is no any magnetic field. }
		\label{PS_High}
	\end{SCfigure}

%______________________________________________________________
	From the observation in the scalar channel, we refer to $\hat{T}\gtrsim 0.6$ as the high temperature. As we found there, at such temperatures the magnitudes of the two butterfly speeds approaches the known value associated with an uncharged holographic system.
		Here we find a similar qualitative behavior. 
For all the magnetic field values we probed numerically, the distribution of pole-skipping points  in $\text{Im}\, \wn - \text{Im}\, \qn$ remains unchanged by varying the temperature within the range  $\hat{T}\gtrsim 0.6$. The results is illustrated in figure  \ref{PS_High}. Here are a few comments about the figure:
	\begin{itemize}
		\item For the reasons that are to be mentioned in next section, we have represented the pole-skipping points in a three dimensional plot. However, as in the uncharged system \cite{Blake:2019otz}, all of them lie in   $\text{Im}\, \wn - \text{Im} \,\qn$ plane.
		\item There are two pole-skipping points at $\wn_1=-i$ (red dots), four of them at $\wn_{2}=-2 i$ (orange dots), and $\cdots$.
		\item The distribution is symmetric with respect to $\qn=0$; as a result, no anomaly effect is detected. 
		\end{itemize}
The situation becomes interesting when the temperature is decreased. 
	%%%%%%%%%%%%%%%%%%%%%%%
	%%%%%%%%%%%%%%%%%%%%%%%
	
%______________________________________________________________
\subsection{Scanning the pole-skipping points at lower temperatures}
%______________________________________________________________
In this subsection, we scan the spectrum of pole-skipping points at two different values of $\hat{B}$ on either side of $\hat{B}_c$. In each case, we account for nine different values of $\hat{T}$, illustrating pole-skipping points in both planes, starting at $\hat{T}=0.6$ and approaching $\hat{T}=0$.

\begin{figure}[]
	\centering
\fbox{
	\begin{minipage}{0.28\textwidth}
		\centering
		\begin{subfigure}{1.05\linewidth}
			\centering
			\includegraphics[width=\linewidth]{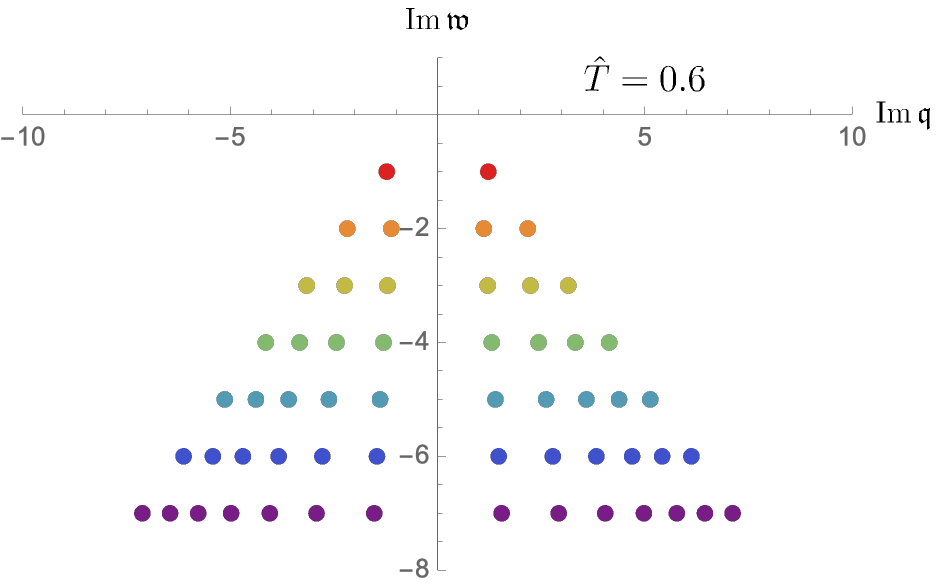}
		\end{subfigure}
	\\
		\begin{subfigure}{1.05\linewidth}
			\centering
			\includegraphics[width=\linewidth]{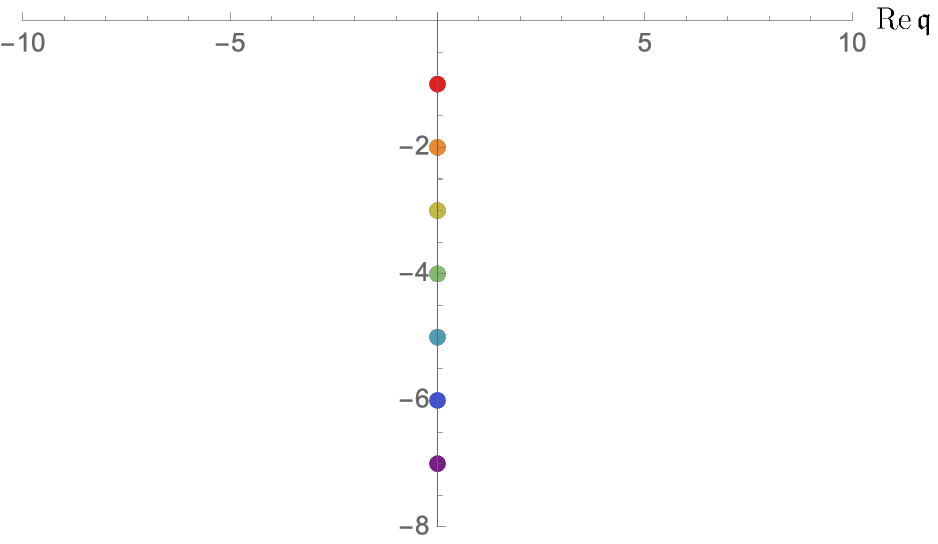}
		\end{subfigure}
	\end{minipage}
}
\fbox{
	\begin{minipage}{0.28\textwidth}
		\centering
		\begin{subfigure}{1.05\linewidth}
			\centering
			\includegraphics[width=\linewidth]{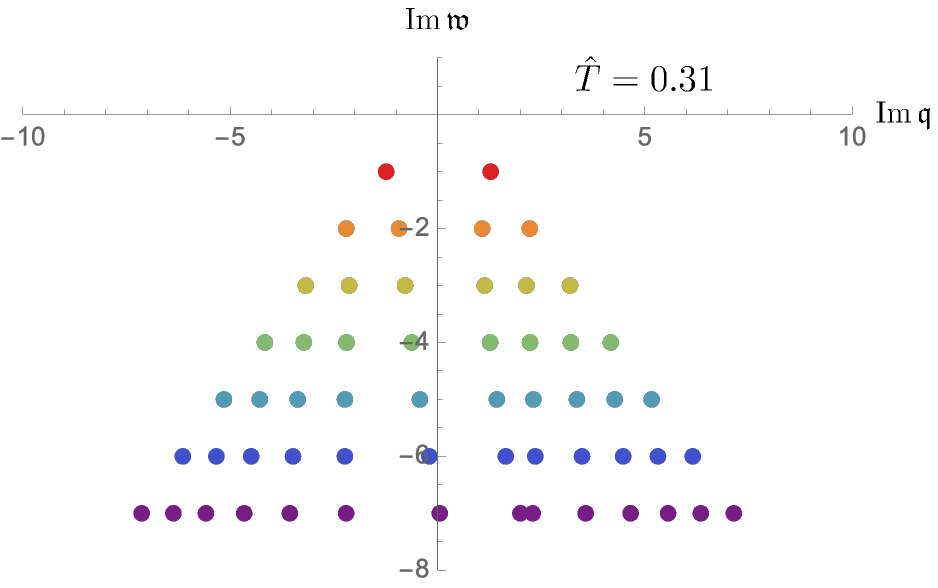}
		\end{subfigure}
		\\
		\begin{subfigure}{1.05\linewidth}
			\centering
			\includegraphics[width=\linewidth]{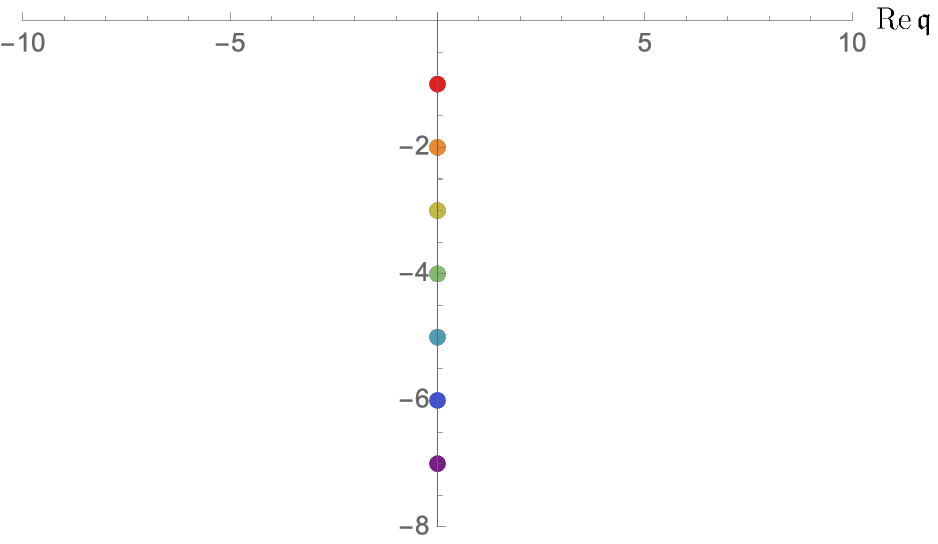}
		\end{subfigure}
	\end{minipage}
}
\fbox{
\begin{minipage}{0.28\textwidth}
	\centering
	\begin{subfigure}{1.05\linewidth}
		\centering
		\includegraphics[width=\linewidth]{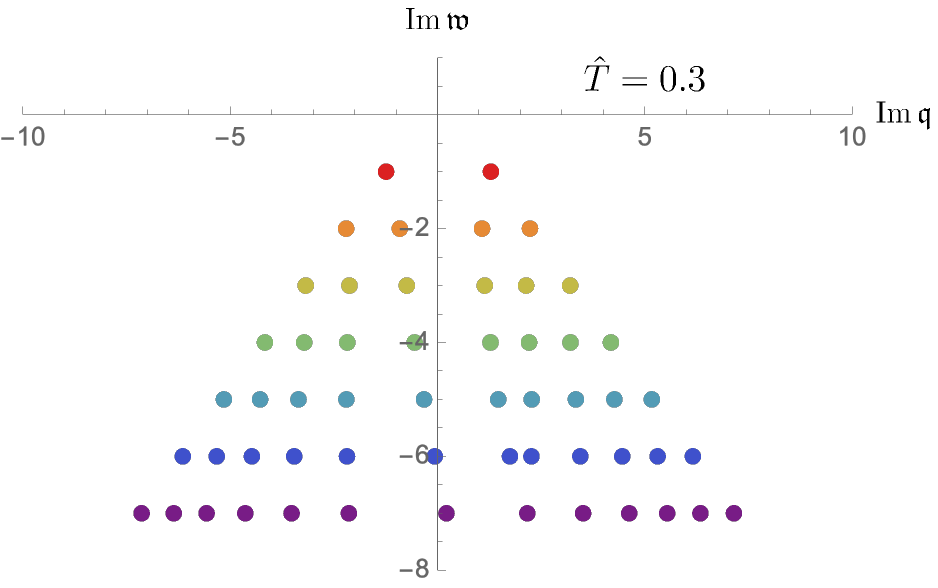}
	\end{subfigure}
	\\
	\begin{subfigure}{1.05\linewidth}
		\centering
		\includegraphics[width=\linewidth]{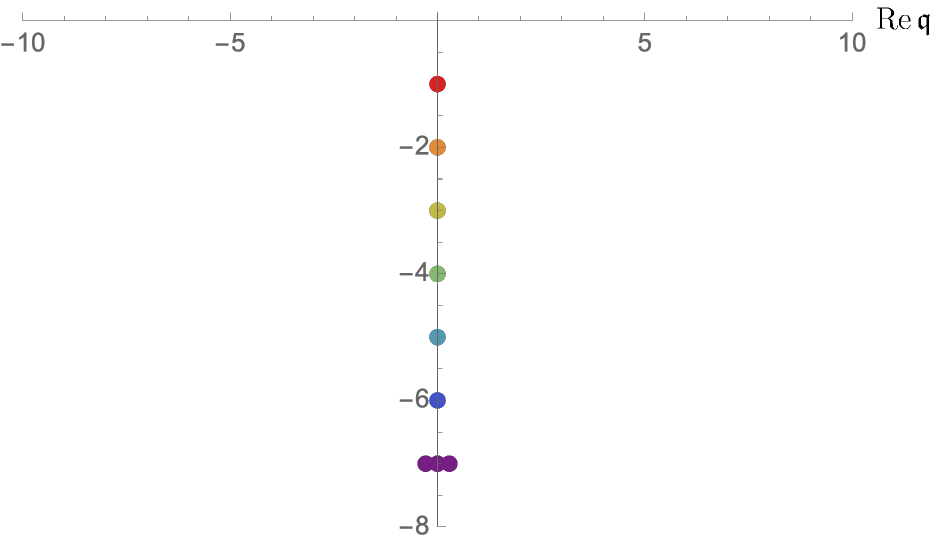}
	\end{subfigure}
\end{minipage}
}
%%%%%%%%%%%%%%%%%%%%%%%%%%%%%%%%%%%%%%%%%%%%%%
\fbox{
	\begin{minipage}{0.28\textwidth}
		\centering
		\begin{subfigure}{1.05\linewidth}
			\centering
			\includegraphics[width=\linewidth]{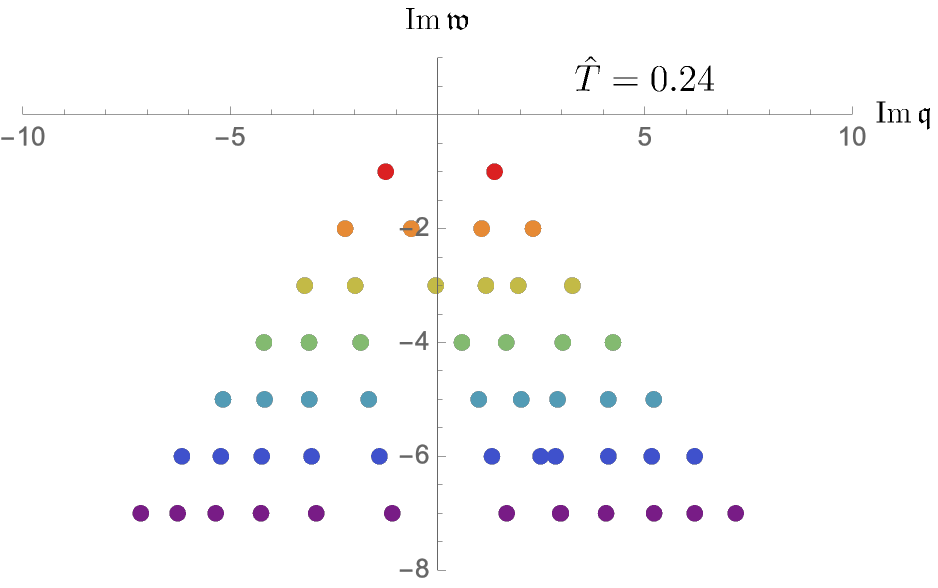}
		\end{subfigure}
		\\
		\begin{subfigure}{1.05\linewidth}
			\centering
			\includegraphics[width=\linewidth]{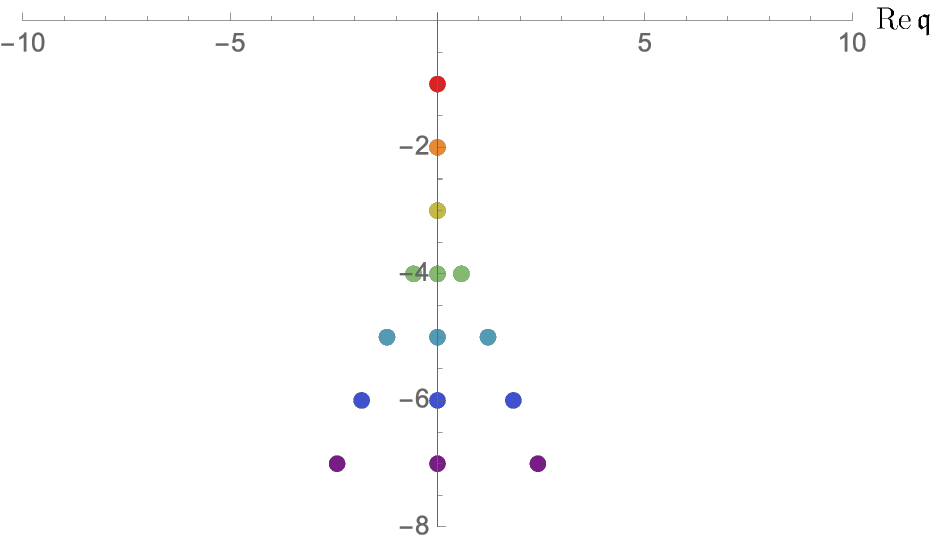}
		\end{subfigure}
	\end{minipage}
}
\fbox{
	\begin{minipage}{0.28\textwidth}
		\centering
		\begin{subfigure}{1.05\linewidth}
			\centering
			\includegraphics[width=\linewidth]{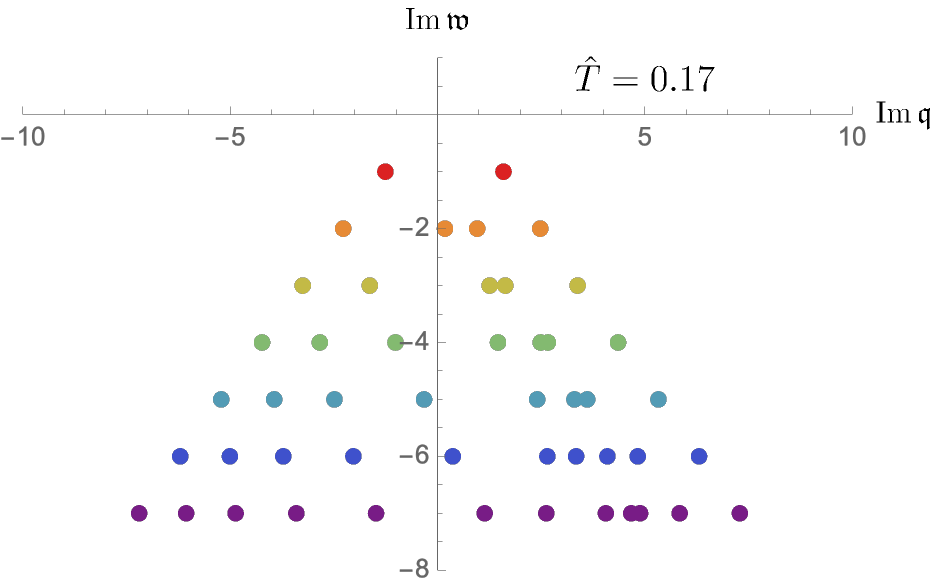}
		\end{subfigure}
		\\
		\begin{subfigure}{1.05\linewidth}
			\centering
			\includegraphics[width=\linewidth]{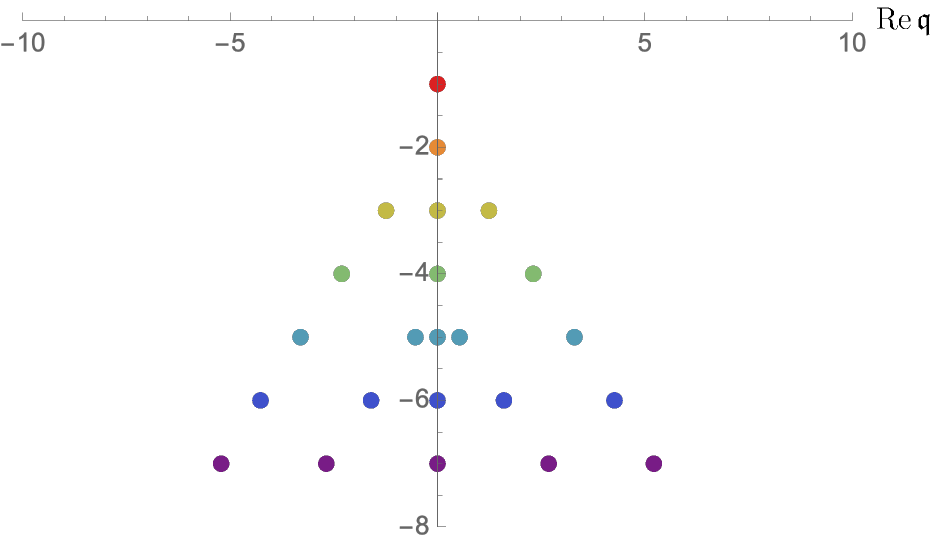}
		\end{subfigure}
	\end{minipage}
}
\fbox{
	\begin{minipage}{0.28\textwidth}
		\centering
		\begin{subfigure}{1.05\linewidth}
			\centering
			\includegraphics[width=\linewidth]{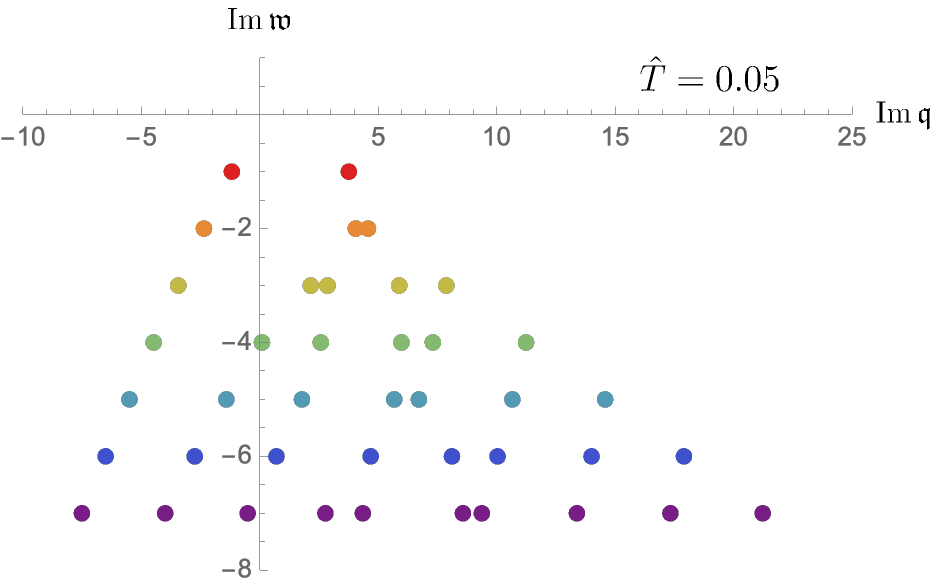}
		\end{subfigure}
		\\
		\begin{subfigure}{1.05\linewidth}
			\centering
			\includegraphics[width=\linewidth]{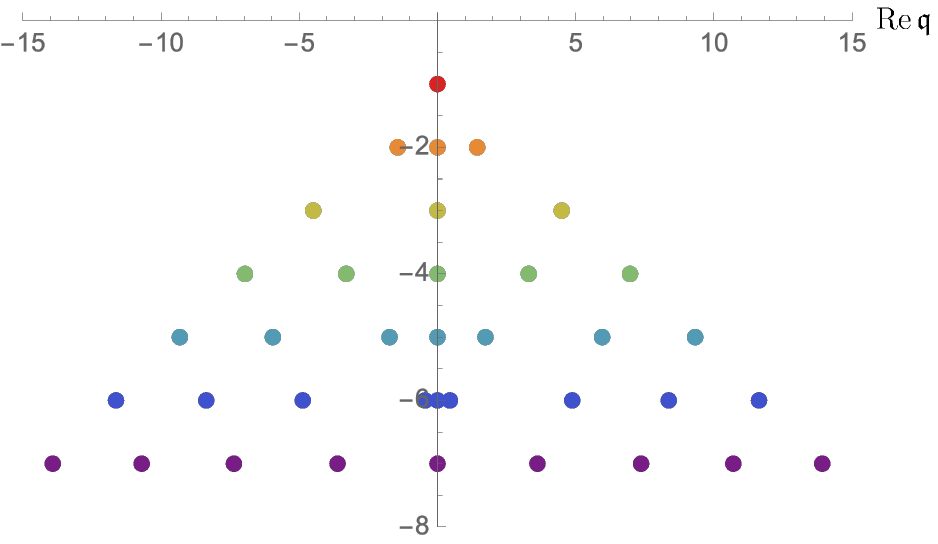}
		\end{subfigure}
	\end{minipage}
}
%%%%%%%%%%%%%%%%%%%%%%%%%%%%%%%%%%%%%%%%%%%%%%
\fbox{
	\begin{minipage}{0.28\textwidth}
		\centering
		\begin{subfigure}{1.05\linewidth}
			\centering
			\includegraphics[width=\linewidth]{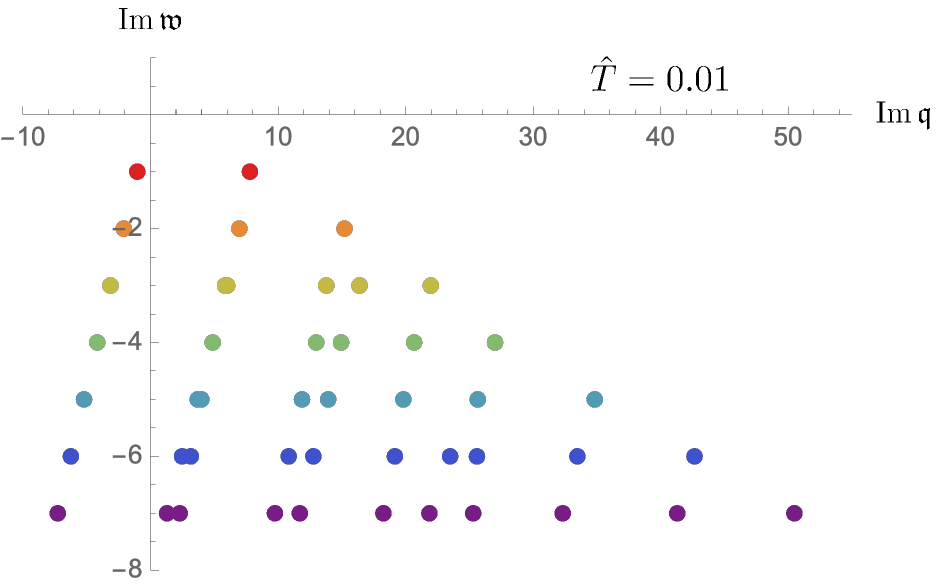}
		\end{subfigure}
		\\
		\begin{subfigure}{1.05\linewidth}
			\centering
			\includegraphics[width=\linewidth]{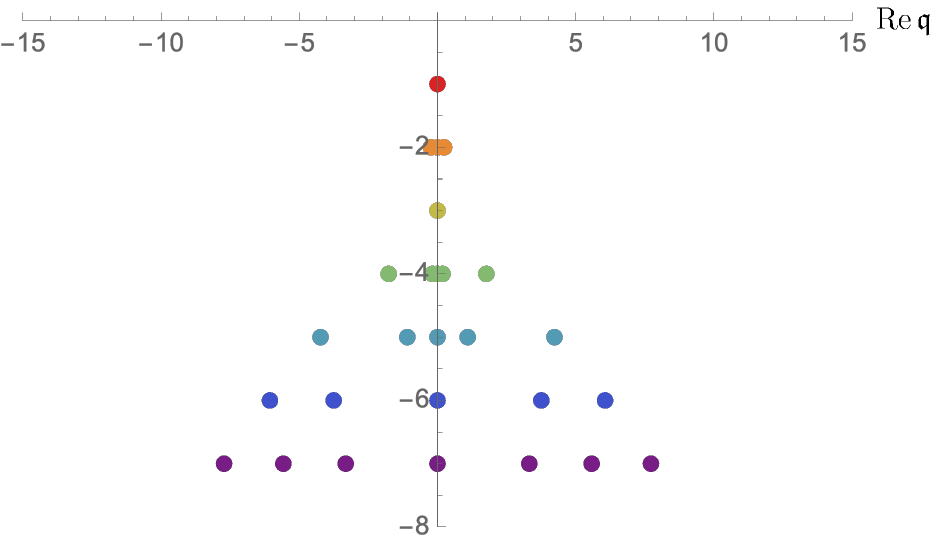}
		\end{subfigure}
	\end{minipage}
}
\fbox{
	\begin{minipage}{0.28\textwidth}
		\centering
		\begin{subfigure}{1.05\linewidth}
			\centering
			\includegraphics[width=\linewidth]{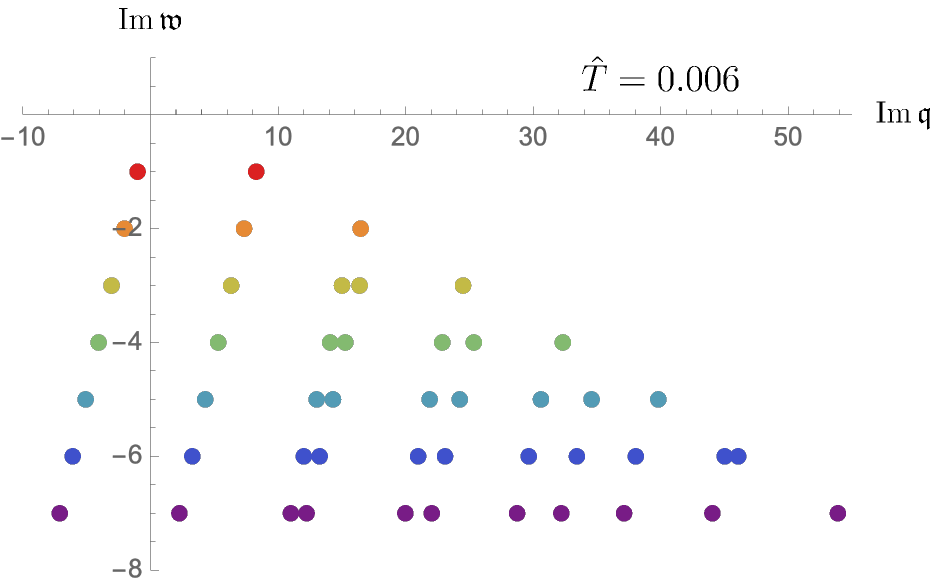}
		\end{subfigure}
		\\
		\begin{subfigure}{1.05\linewidth}
			\centering
			\includegraphics[width=\linewidth]{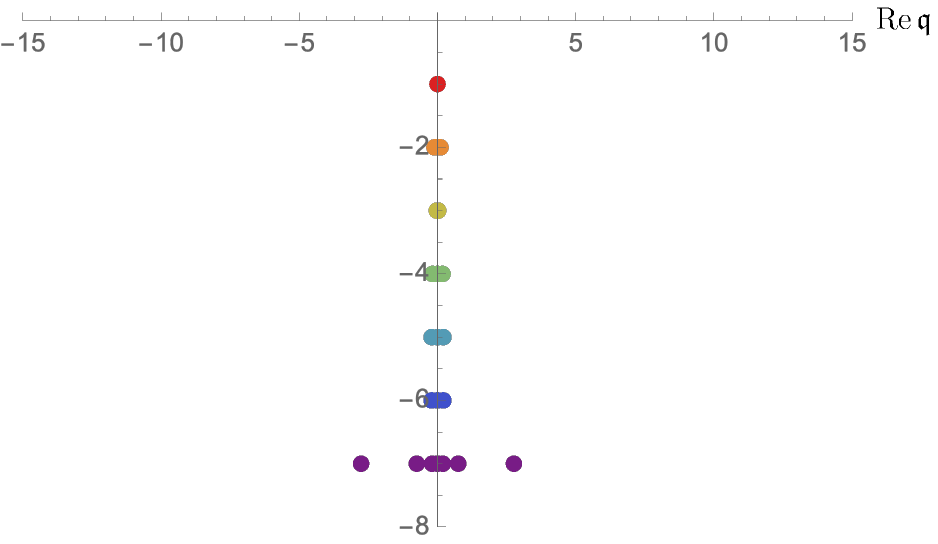}
		\end{subfigure}
	\end{minipage}
}
\fbox{
	\begin{minipage}{0.28\textwidth}
		\centering
		\begin{subfigure}{1.05\linewidth}
			\centering
			\includegraphics[width=\linewidth]{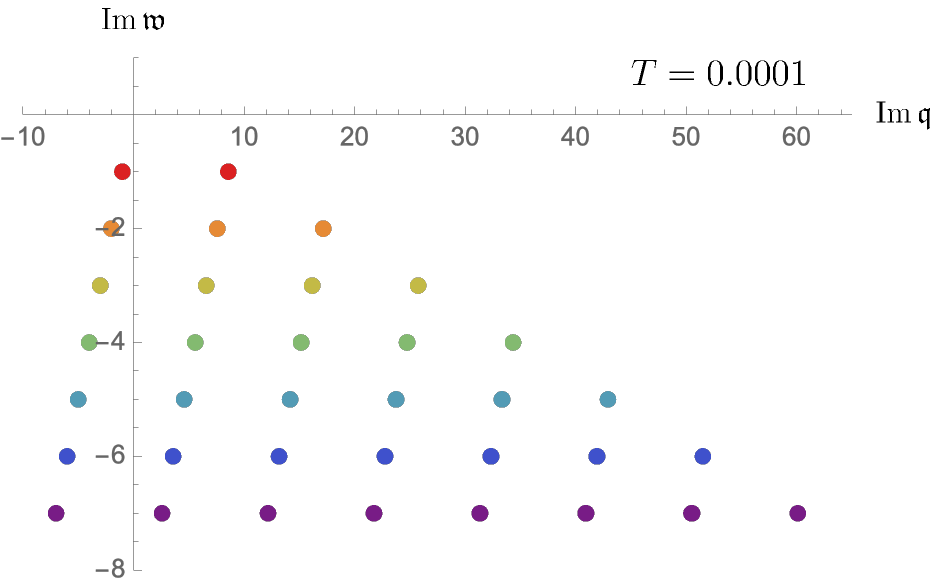}
		\end{subfigure}
		\\
		\begin{subfigure}{1.05\linewidth}
			\centering
			\includegraphics[width=\linewidth]{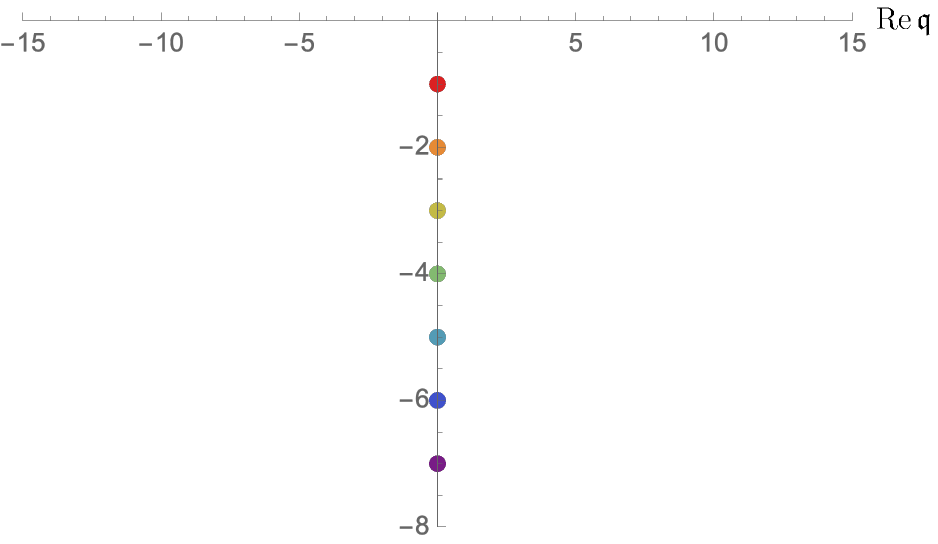}
		\end{subfigure}
	\end{minipage}
}
\caption{Scanning the pole-skipping structure of $G^{R}_{\mathcal{O}\mathcal{O}}$ at $\hat{B}=0.54> \hat{B}_c$. From the above left to the bottom right, temperature decreases from $\hat{T}=0.6$ to $\hat{T}=0.0001$. In each panel, the upper plot shows the pole-skipping points in $\text{Im}\,\wn-\text{Im}\,\qn$ plane, while the lower one the same in $\text{Im}\,\wn-\text{Re}\,\qn$ plane.  }
\label{Scan_above_B_c}
\end{figure}

The first case we consider is $\hat{B}=0.54$, as shown in the figure \ref{Scan_above_B_c}. As one would expect, at high temperature, i.e. $\hat{T}\gtrsim 0.6$, the spectrum is symmetric and lies in the $\text{Im}\,\wn-\text{Im}\,\qn$ plane (see upper left panel). As the temperature decreases, the spectrum deviates from the symmetric form; it actually starts at a higher $\ell$ level. As can be seen from the top middle panel, at $\hat{T}=0.31$, the $\ell=7$ level is more affected than the lower levels. In addition, two of the points belonging to $\ell=7$ are close to each other at this $\hat{T}$ \footnote{For temperatures between $0.31$ and $0.6$, this asymmetry occurs at higher order pole-skipping points, such as $\ell=8,9, \cdots$.}. By decreasing $\hat{T}$, these two points eventually scattter from each other. After scattering, their imaginary parts of momentum become equal, however, they also obtain two non-zero and opposite real momenta. This can be seen at $\hat{T}=0.3$ in the upper right panel.

By reducing $\hat{T}$ further, more and more points from lower levels will collide with each other. At the same time, they will move away from the $\qn=0$ axis on the $\text{Im}\,\wn-\text{Re}\,\qn$ plane. It can be seen from the three figures in the middle row. 

However, the opposite happens when $\hat{T}$ is relatively small; the points start to approach the $\qn=0$ axis on the $\text{Im}\,\wn-\text{Re}\,\qn$ plane. It is illustrated in the bottom panel. Finally, when $\hat{T}$ approaches zero, all real momenta vanish (see bottom right plot). Interestingly, in the latter case, the spectrum at level-$\ell$ consists of $\ell+1$ points. Note that at high $\hat{T}$ (upper left panel), there were $2\ell$ of them; where $\ell$ of them had values of $\text{Im}\,\qn<0$ and the other $\ell$ had values of $\text{Im}\,\qn>0$. We see that by decreasing the temperature, $\ell-1$ points (out of the entire $\ell$ points of $\text{Im}\,\qn<0$) gradually move to the right of $\qn=0$ and eventually coincide with $\ell-1$ points of $\text{Im}\,\qn>0$. Finally, the only pole-skipping points with $\text{Im}\,\qn<0$ at low $\hat{T}$ (i.e., in the lower right panel) are the points corresponding to the leftmost points in the upper left panel.
\\

\begin{figure}[]
	\centering
	\fbox{
		\begin{minipage}{0.28\textwidth}
			\centering
			\begin{subfigure}{1.05\linewidth}
				\centering
				\includegraphics[width=\linewidth]{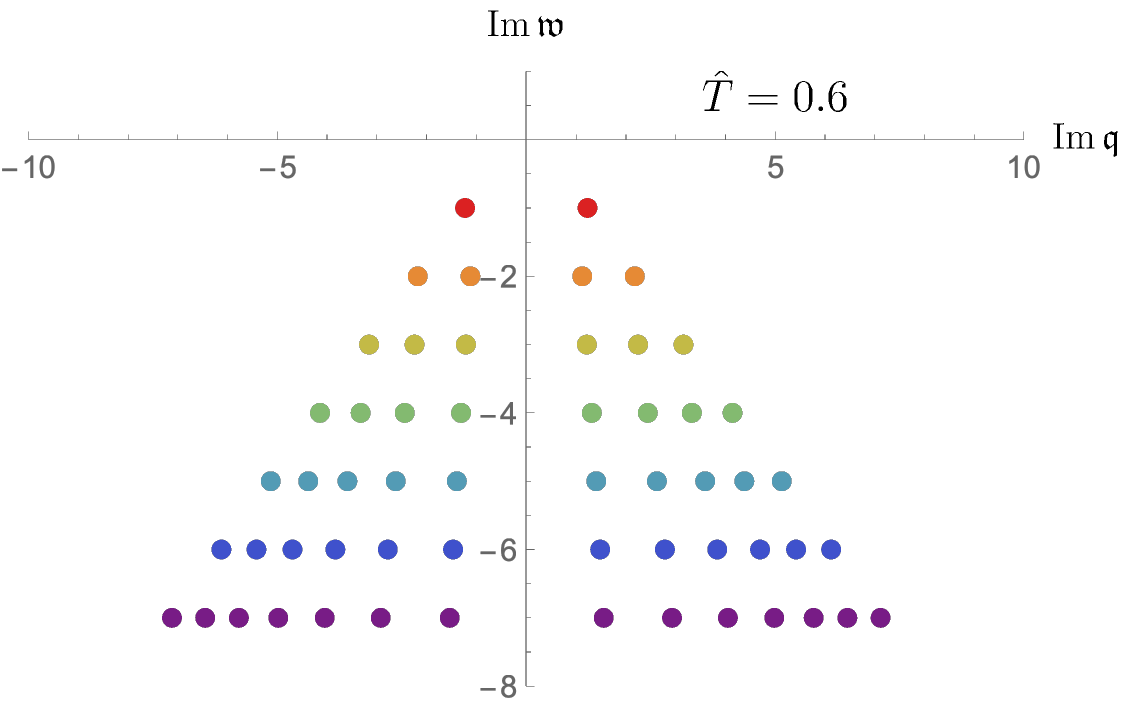}
			\end{subfigure}
			\\
			\begin{subfigure}{1.05\linewidth}
				\centering
				\includegraphics[width=\linewidth]{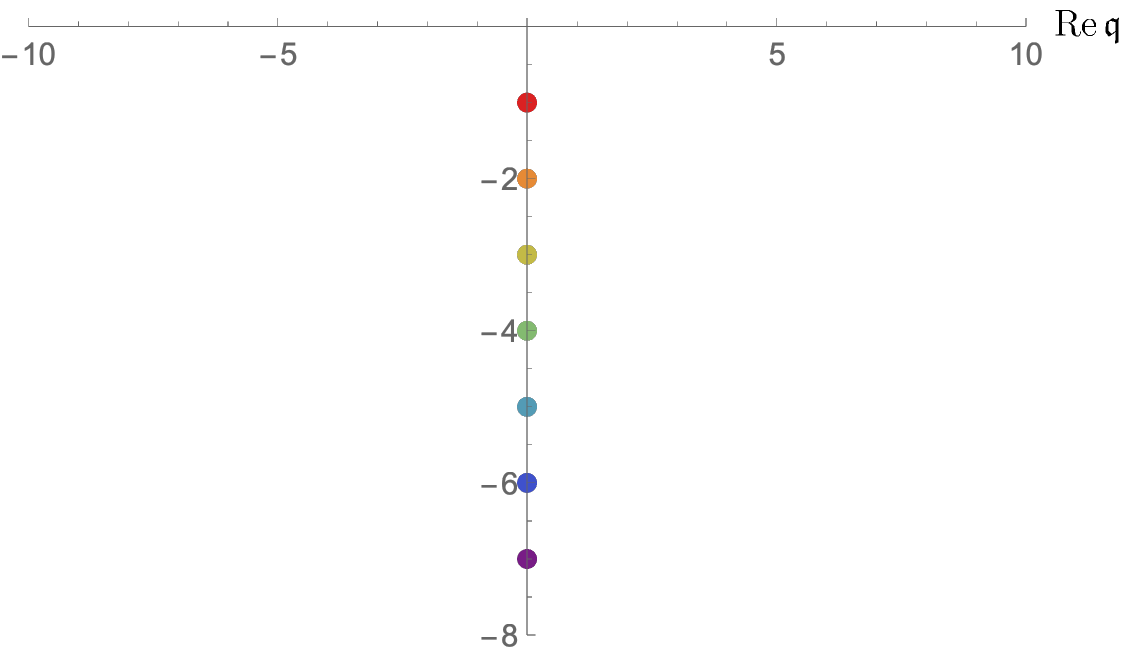}
			\end{subfigure}
		\end{minipage}
	}
	\fbox{
		\begin{minipage}{0.28\textwidth}
			\centering
			\begin{subfigure}{1.05\linewidth}
				\centering
				\includegraphics[width=\linewidth]{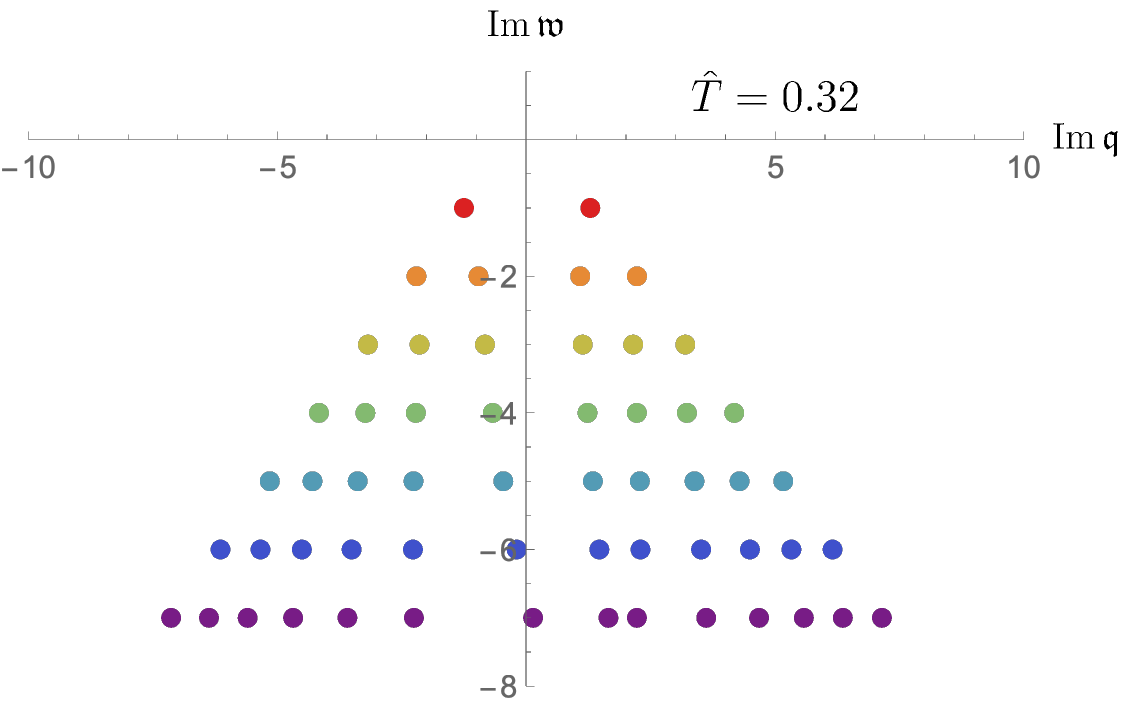}
			\end{subfigure}
			\\
			\begin{subfigure}{1.05\linewidth}
				\centering
				\includegraphics[width=\linewidth]{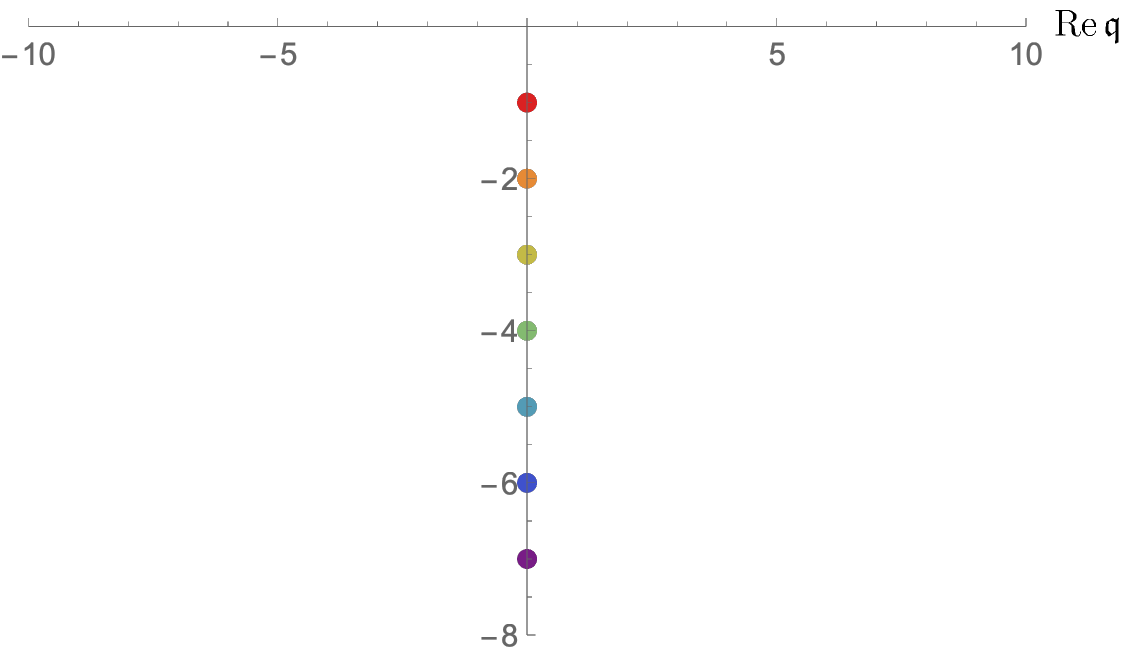}
			\end{subfigure}
		\end{minipage}
	}
	\fbox{
		\begin{minipage}{0.28\textwidth}
			\centering
			\begin{subfigure}{1.05\linewidth}
				\centering
				\includegraphics[width=\linewidth]{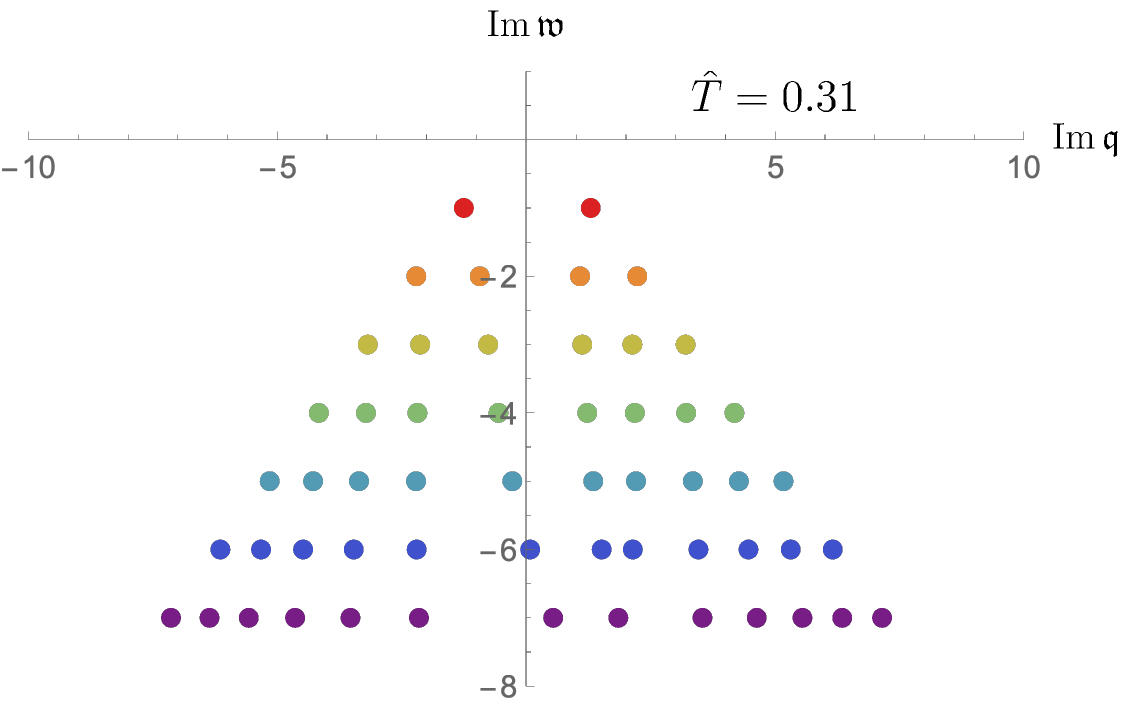}
			\end{subfigure}
			\\
			\begin{subfigure}{1.05\linewidth}
				\centering
				\includegraphics[width=\linewidth]{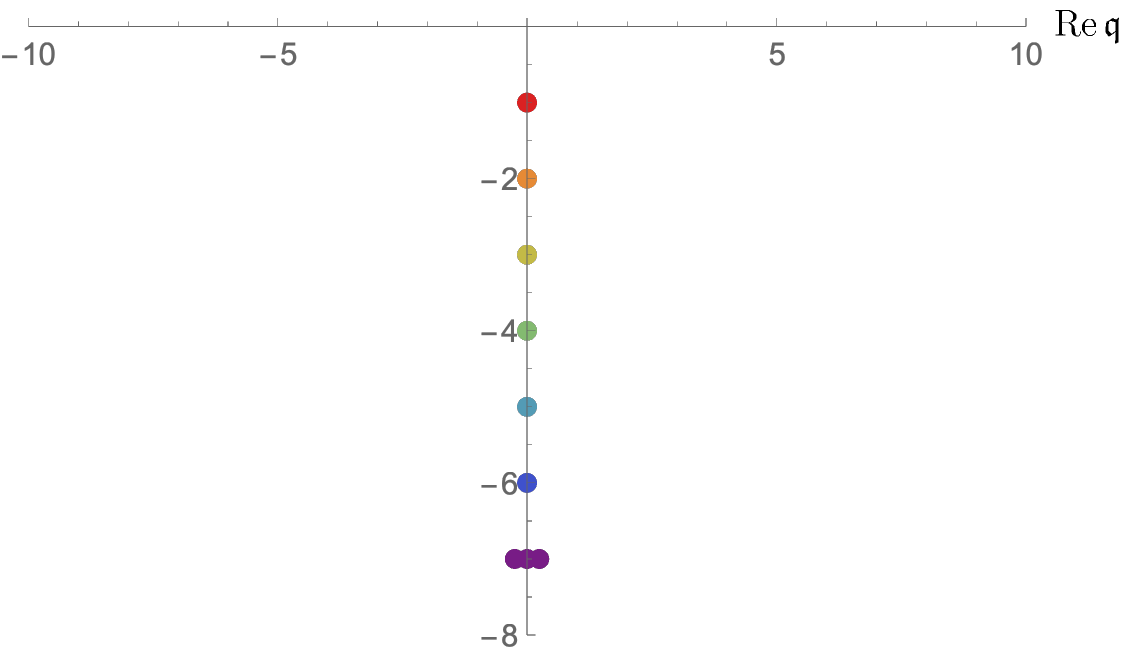}
			\end{subfigure}
		\end{minipage}
	}
	%%%%%%%%%%%%%%%%%%%%%%%%%%%%%%%%%%%%%%%%%%%%%%
	\fbox{
		\begin{minipage}{0.28\textwidth}
			\centering
			\begin{subfigure}{1.05\linewidth}
				\centering
				\includegraphics[width=\linewidth]{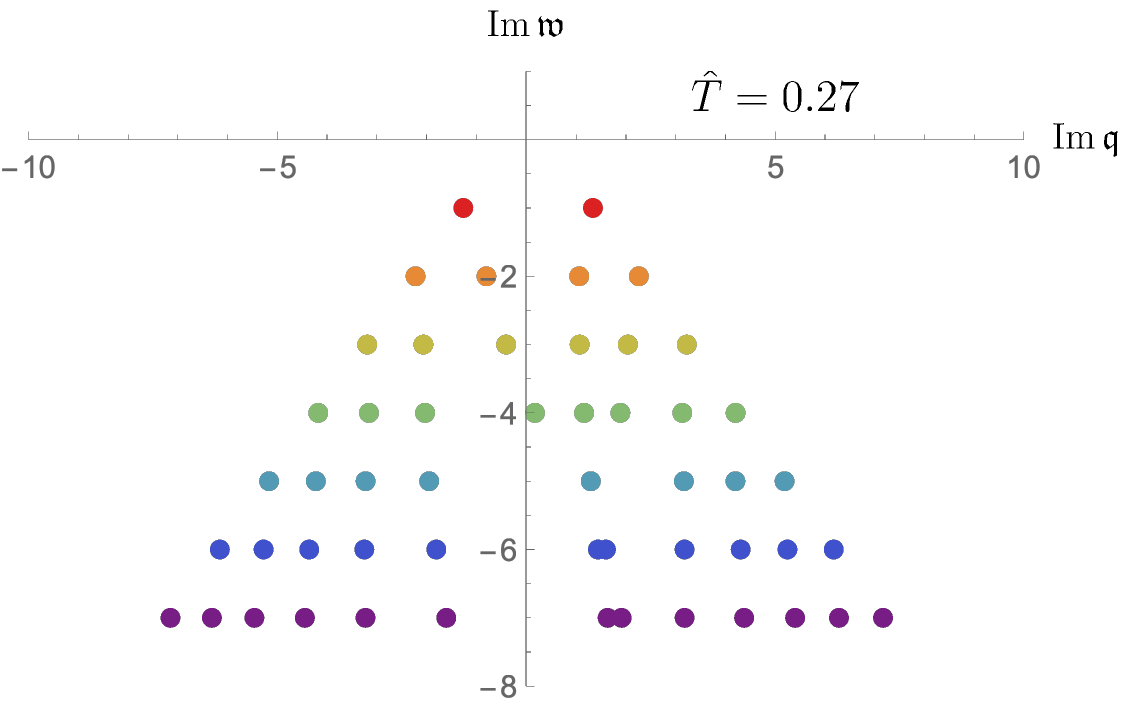}
			\end{subfigure}
			\\
			\begin{subfigure}{1.05\linewidth}
				\centering
				\includegraphics[width=\linewidth]{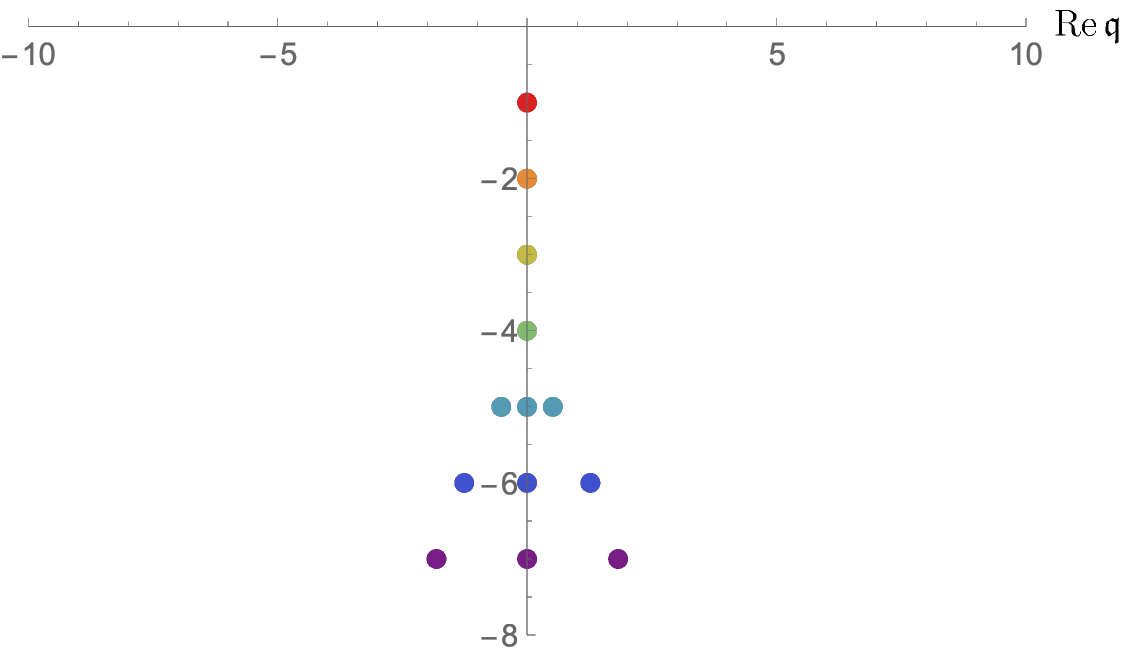}
			\end{subfigure}
		\end{minipage}
	}
	\fbox{
		\begin{minipage}{0.28\textwidth}
			\centering
			\begin{subfigure}{1.05\linewidth}
				\centering
				\includegraphics[width=\linewidth]{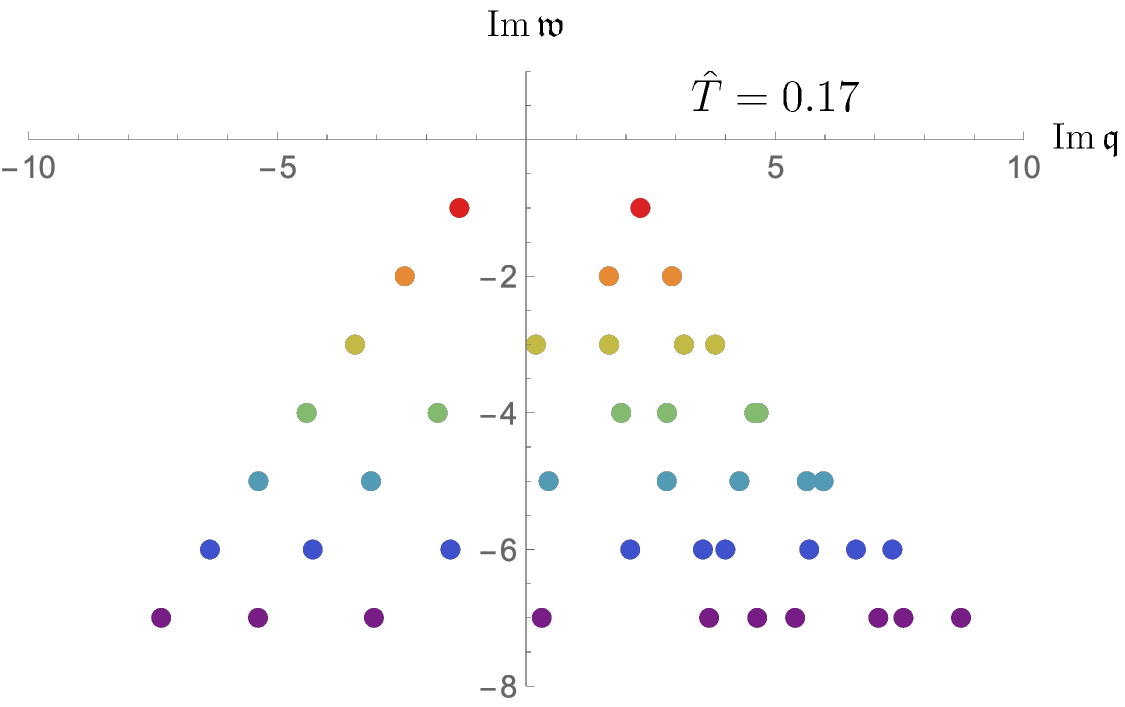}
			\end{subfigure}
			\\
			\begin{subfigure}{1.05\linewidth}
				\centering
				\includegraphics[width=\linewidth]{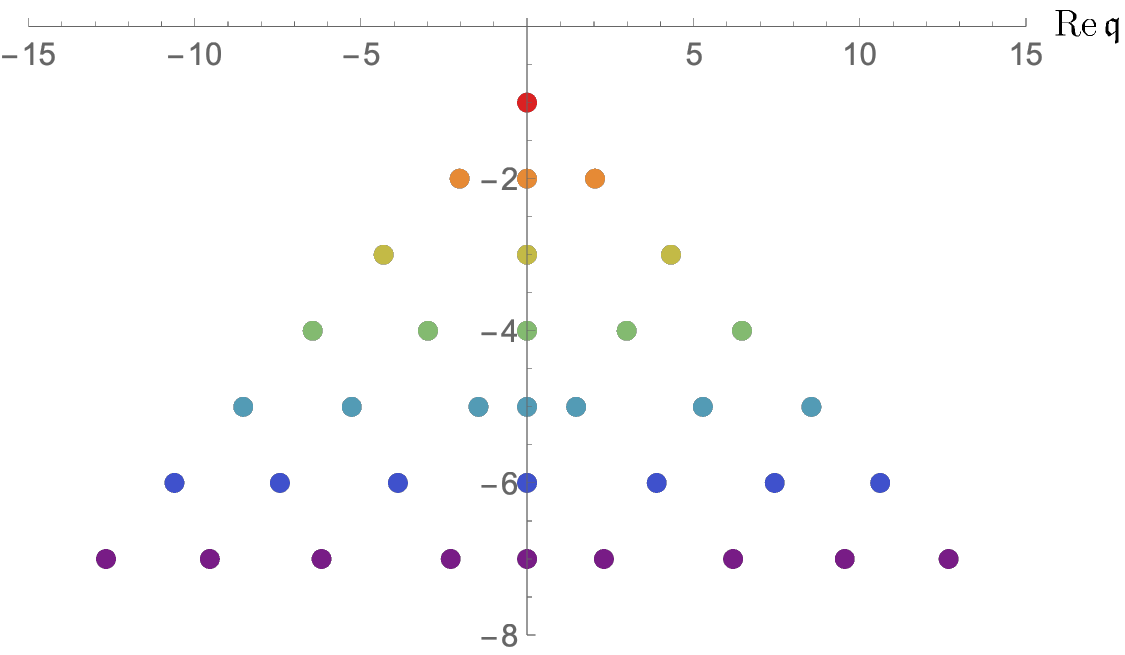}
			\end{subfigure}
		\end{minipage}
	}
	\fbox{
		\begin{minipage}{0.28\textwidth}
			\centering
			\begin{subfigure}{1.05\linewidth}
				\centering
				\includegraphics[width=\linewidth]{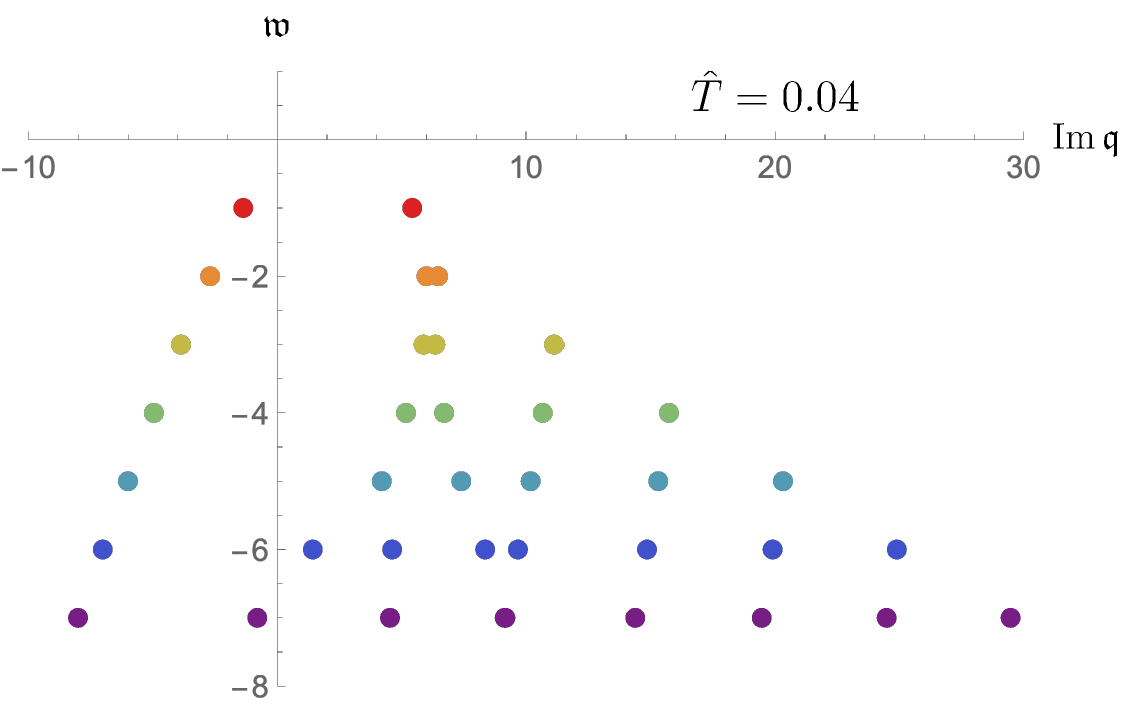}
			\end{subfigure}
			\\
			\begin{subfigure}{1.05\linewidth}
				\centering
				\includegraphics[width=\linewidth]{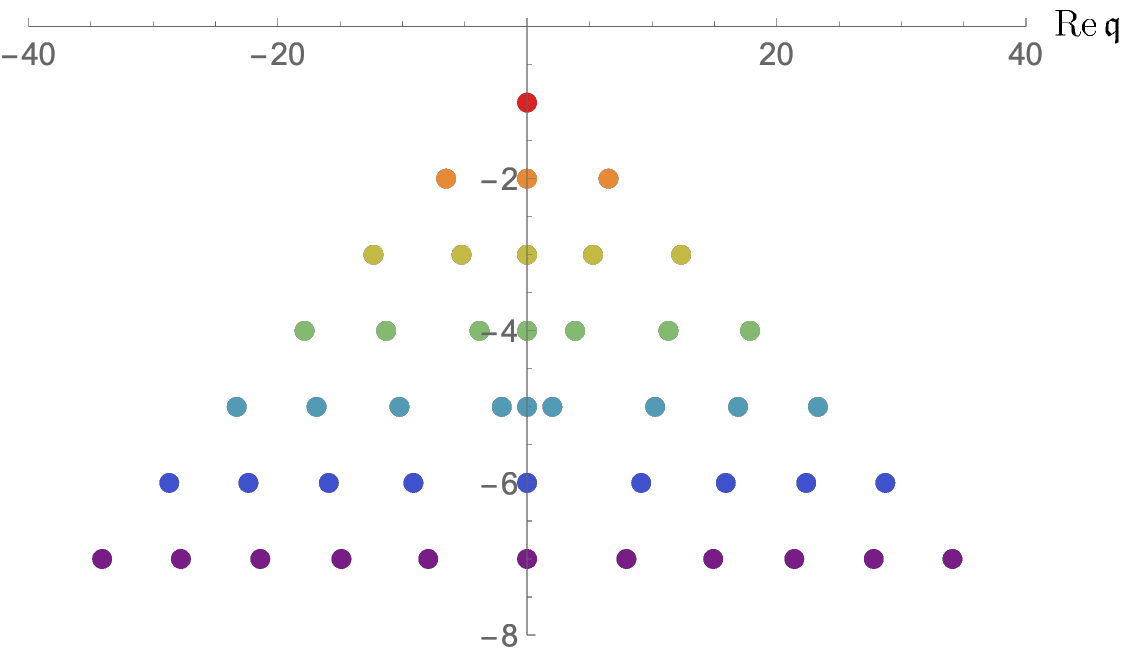}
			\end{subfigure}
		\end{minipage}
	}
	%%%%%%%%%%%%%%%%%%%%%%%%%%%%%%%%%%%%%%%%%%%%%%
	\fbox{
		\begin{minipage}{0.28\textwidth}
			\centering
			\begin{subfigure}{1.05\linewidth}
				\centering
				\includegraphics[width=\linewidth]{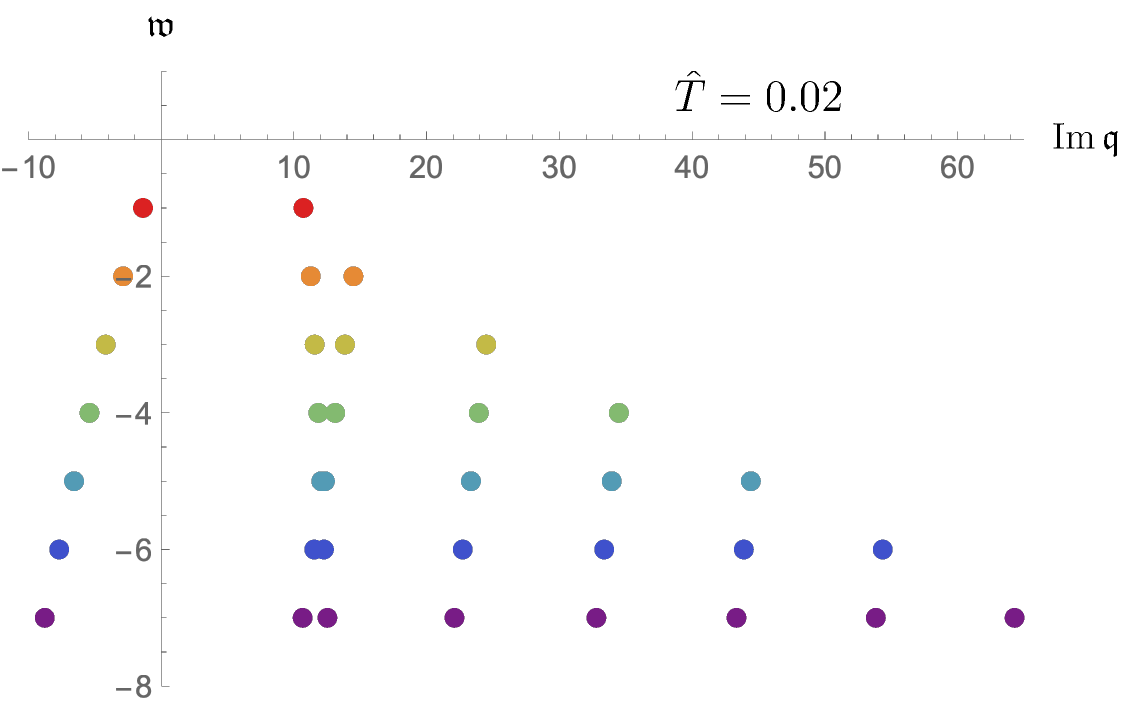}
			\end{subfigure}
			\\
			\begin{subfigure}{1.05\linewidth}
				\centering
				\includegraphics[width=\linewidth]{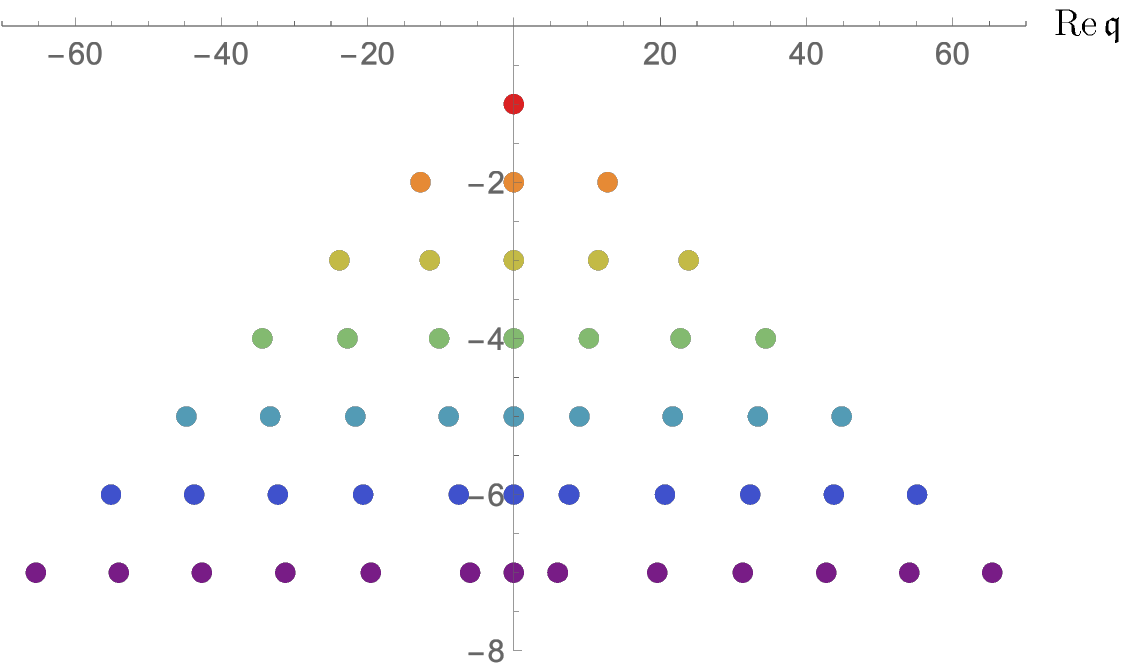}
			\end{subfigure}
		\end{minipage}
	}
	\fbox{
		\begin{minipage}{0.28\textwidth}
			\centering
			\begin{subfigure}{1.05\linewidth}
				\centering
				\includegraphics[width=\linewidth]{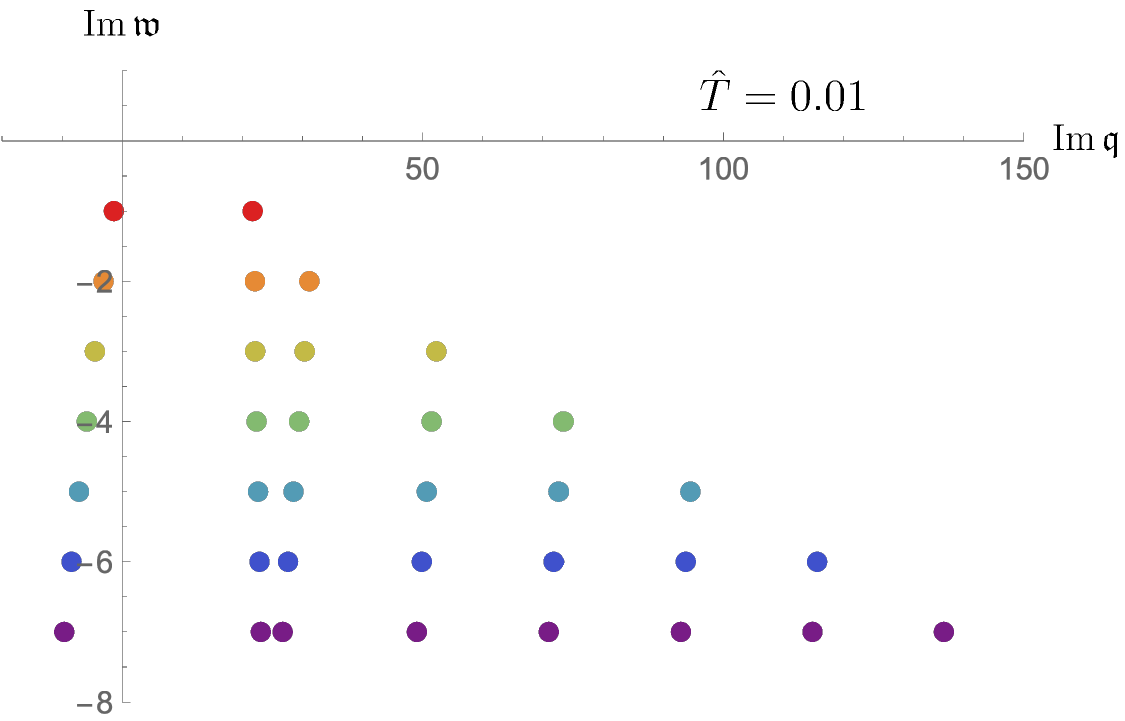}
			\end{subfigure}
			\\
			\begin{subfigure}{1.05\linewidth}
				\centering
				\includegraphics[width=\linewidth]{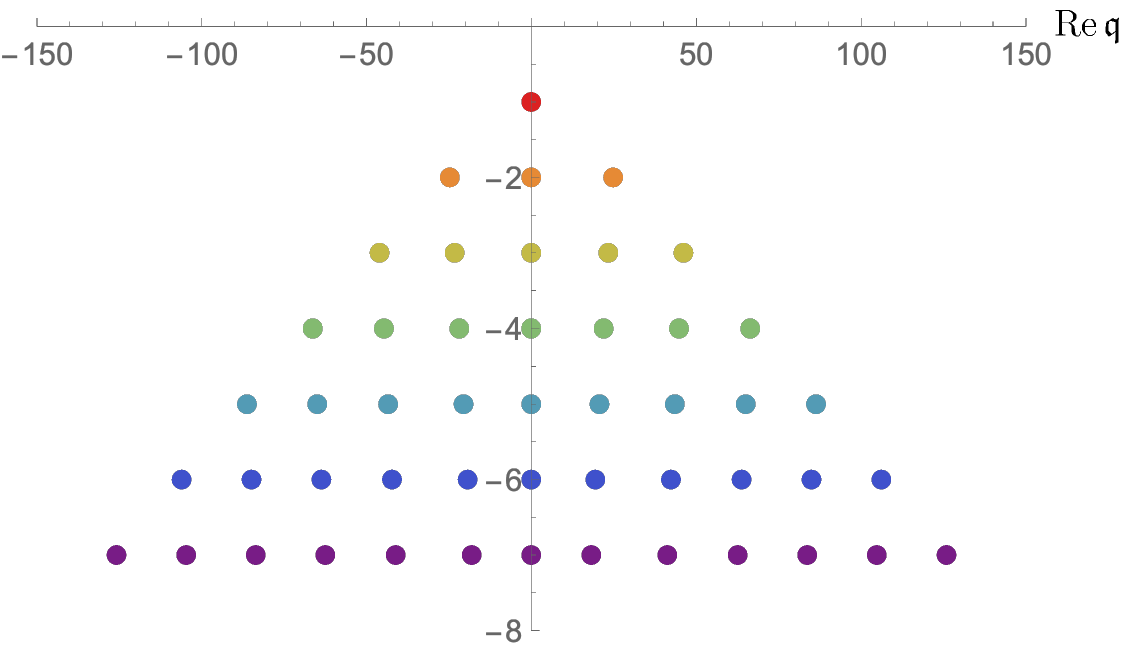}
			\end{subfigure}
		\end{minipage}
	}
	\fbox{
		\begin{minipage}{0.28\textwidth}
			\centering
			\begin{subfigure}{1.05\linewidth}
				\centering
				\includegraphics[width=\linewidth]{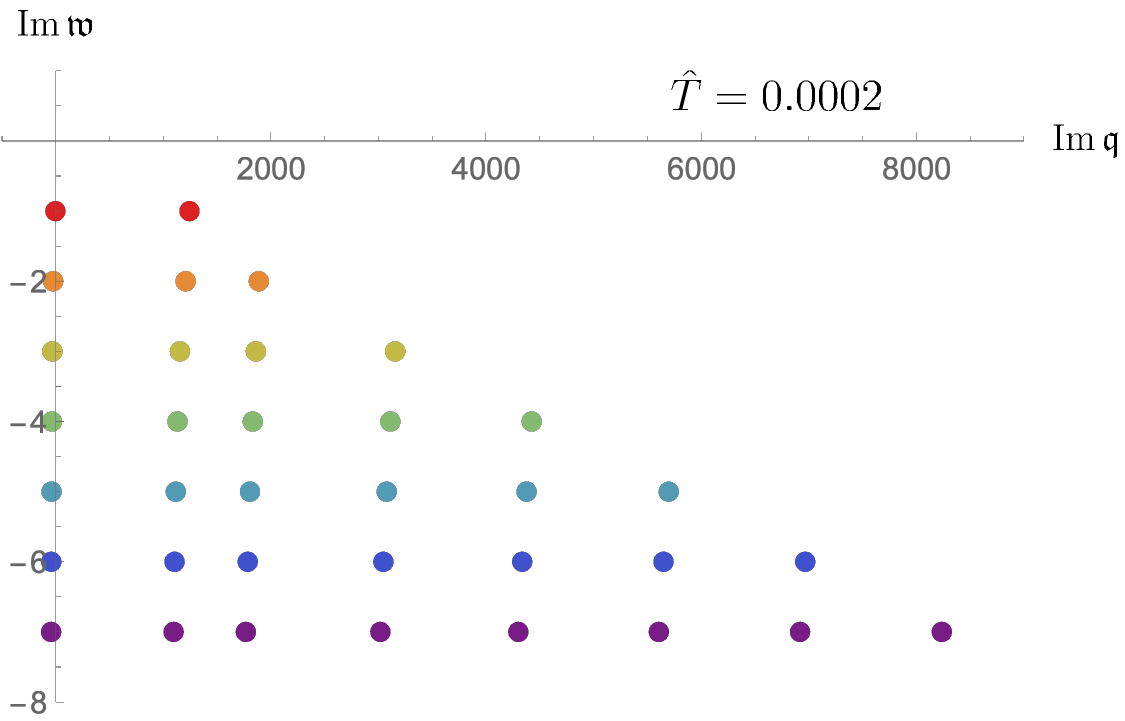}
			\end{subfigure}
			\\
			\begin{subfigure}{1.05\linewidth}
				\centering
				\includegraphics[width=\linewidth]{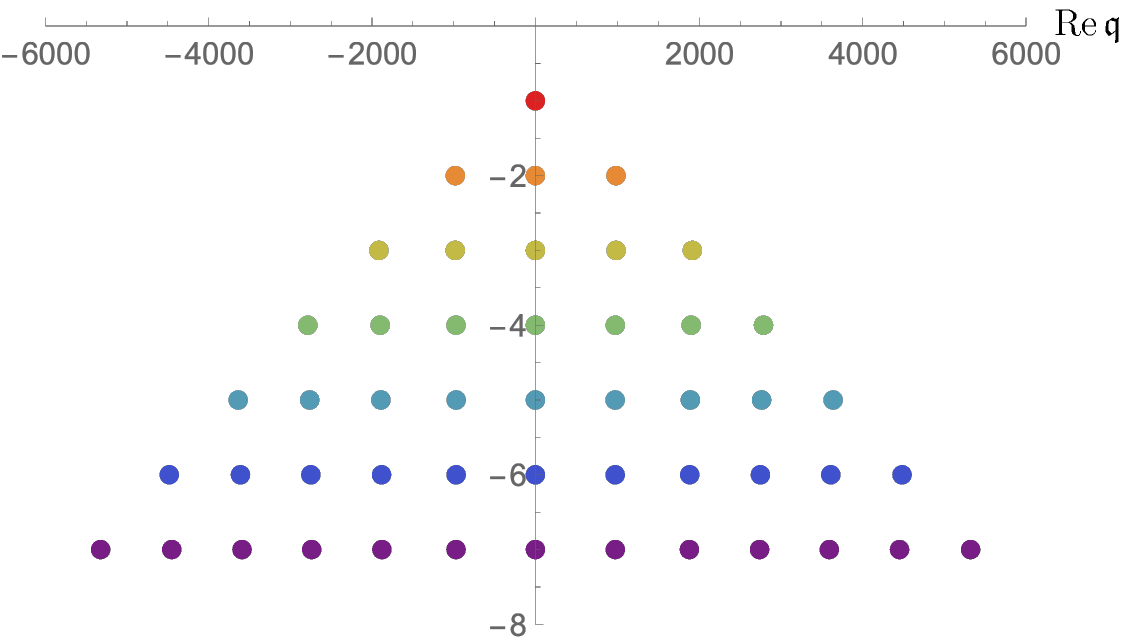}
			\end{subfigure}
		\end{minipage}
	}
	\caption{Scanning the pole-skipping structure of $G^{R}_{\mathcal{O}\mathcal{O}}$ at $\hat{B}=0.44< \hat{B}_c$. From the above left to the bottom right, temperature decreases from $\hat{T}=0.6$ to $\hat{T}=0.0002$. In each panel, the upper plot shows the pole-skipping points in $\text{Im}\,\wn-\text{Im}\,\qn$ plane, while the lower one the same in $\text{Im}\,\wn-\text{Re}\,\qn$ plane.  }
	\label{Scan_below_B_c}
\end{figure}

In figure \ref{Scan_below_B_c}, we scan the spectrum of pole-skipping points at $\hat{B}=0.44$. Similar to the previous case, higher level points start scatttering from each other by lowering the temperature. Lower level points come into play at lower temperatures. However, In contrast to the spectum at $\hat{B}=0.54$, we see here that the real parts of momenta, which is the result of collisins between pole-skipping points,  never vanish.  The lower the temperature, the farther they are from $\qn=0$ on the $\text{Im}\,\wn-\text{Re}\,\qn$ plane.

Finally as $\hat{T}$ approaches zero, we find a spectrum developed on both $\text{Im}\,\wn-\text{Im}\,\qn$ and $\text{Im}\,\wn-\text{Re}\,\qn$ planes. On the lattre plane, it is symmetric and consists of  $\ell-1$ points at level-$\ell$. On the former plane, however, it is assymetric and consists of $\ell+1$ ponits at level-$\ell$.
\\

In summary, we have found two different behaviors for the pole-skipping fates on either side of $\hat{B}_c=0.499$. This simply suggests that there may be some information about the quantum critical point in the pole-skipping spectrum. To explore this, in the next subsection we take a closer look at the spectrum at $\hat{T}=0$.
%%%%%%%%%%%%%%%%%%%%%%%
%%%%%%%%%%%%%%%%%%%%%%%
%%%%%%%%%%%%%%%%%%%%%%%
%%%%%%%%%%%%%%%%%%%%%%%

%______________________________________________________________
\subsection{$\hat{T}\rightarrow 0$: Pole-skipping as the order parameter}
%______________________________________________________________
Based on the results of the previous two subsections, we now scan the spectrum of pole-skipping $\hat{T}\rightarrow0$ by scanning various values of $\hat{B}$. For concreteness, we show the results for this case in a 3D plot consisting of $\text{Im}\,\wn$, $\text{Im}\,\qn$ and $\text{Re}\,\qn$. We show the spectra for six different values of $\hat{B}$ in figure \ref{3D_T_0}.

In the top panel we show the spectra for the three cases $\hat{B}>\hat{B}_c$. The spectrum always lies on the $\text{Im}\,\wn-\text{Im}\,\qn$ plane. In other words, the momentum of a pole-skipping point is purely imaginary. At sufficiently high $\hat{B}$ the spectrum is perfectly symmetric. However, by reducing $\hat{B}$, the spectrum produces an asymmetric behavior. It corresponds to the lower right panel in figure \ref{Scan_above_B_c}.
An important observation here is that ``\textit{the lower the magnetic field, the wider the range of values that $\text{Im}\,\qn$ occupies}".
\\

In the bottom panel of figure \ref{3D_T_0} we show the spectra for the three cases $\hat{B}<\hat{B}_c$. The spectrum no longer lies only on the $\text{Im}\,\wn-\text{Im}\,\qn$ plane \footnote{Pole-skipping points complex momentums were previously observed in \cite{Yuan:2020fvv} in systems with Lifshitz symmetry.}. It finds some real momentum parts: "\textit{The lower the magnetic field, the wider the range of values that $\text{Re}\,\qn$ occupies}".
\\

Considering the above observations, we conclude that there is a direct relationship between the quantum phase transition at $\hat{B}_c$ and the pole-hopping spectrum at $\hat{T}\rightarrow 0$. For states $\hat{B}>\hat{B}_c$, the spectrum does not contain any points with real momentum. Similar to the familiar phase transitions, we can call this state disordered. When $\hat{B}<\hat{B}_c$, the momentum pole-skipping points becomes complex. This state can be viewed as an ordered state. Then to distinguish the ordered state from the disordered state, we introduce the following quantity:

%%%%%%%%%%%%%%%%%%%%%%%
\begin{figure}[]
	\centering
	\includegraphics[width=0.3\textwidth]{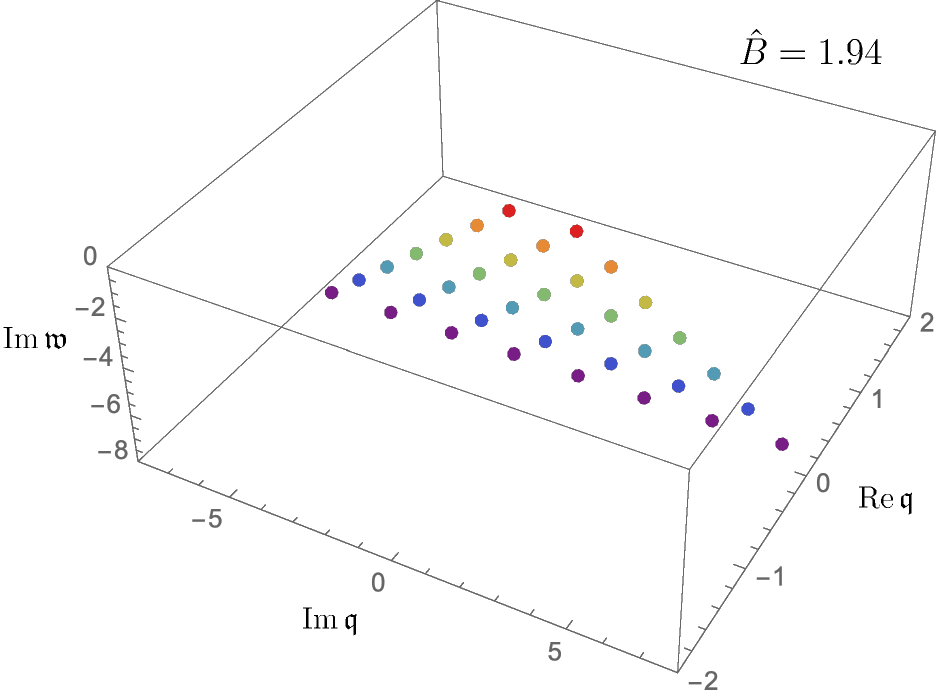}	\includegraphics[width=0.3\textwidth]{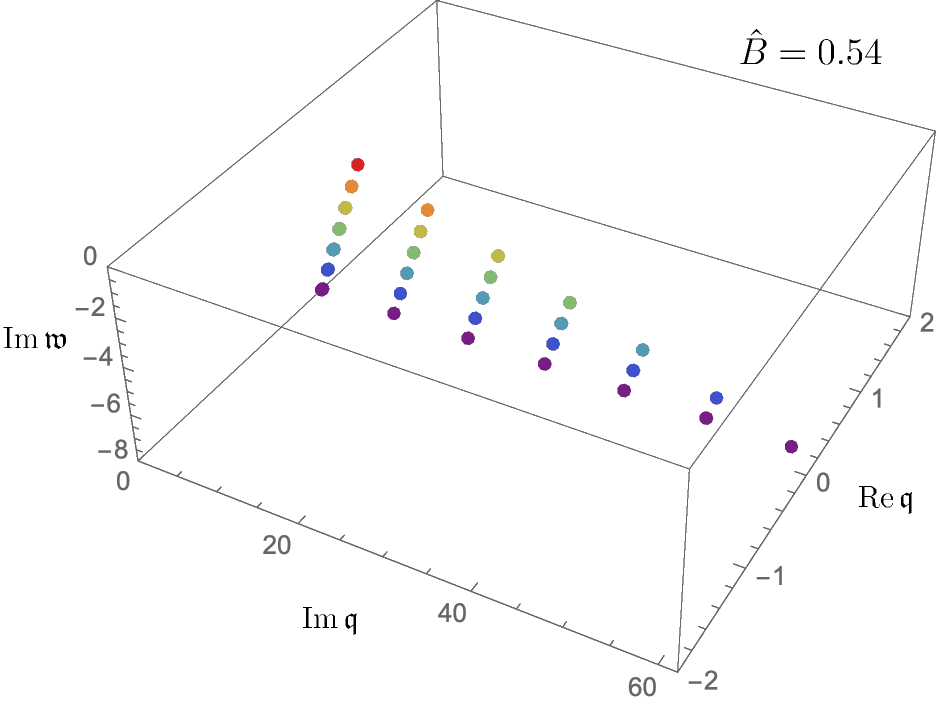}	\includegraphics[width=0.3\textwidth]{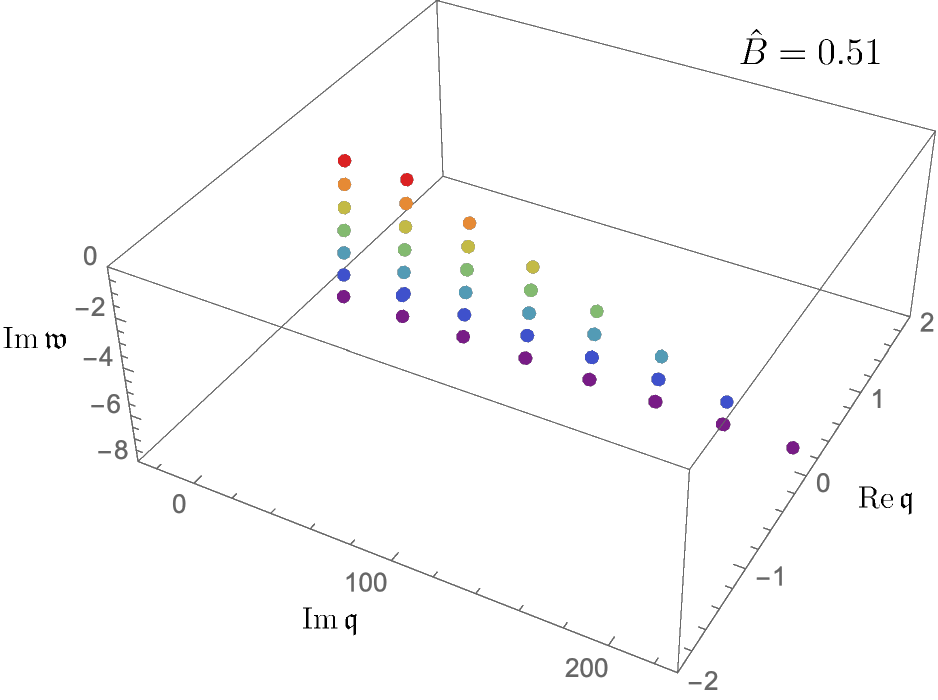}
	\includegraphics[width=0.3\textwidth]{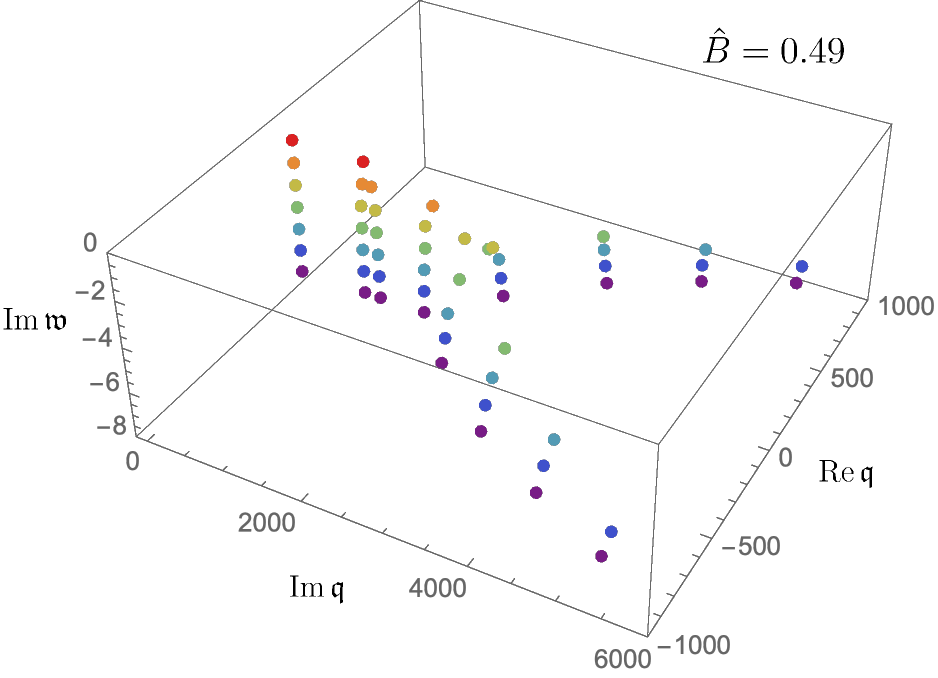}	\includegraphics[width=0.3\textwidth]{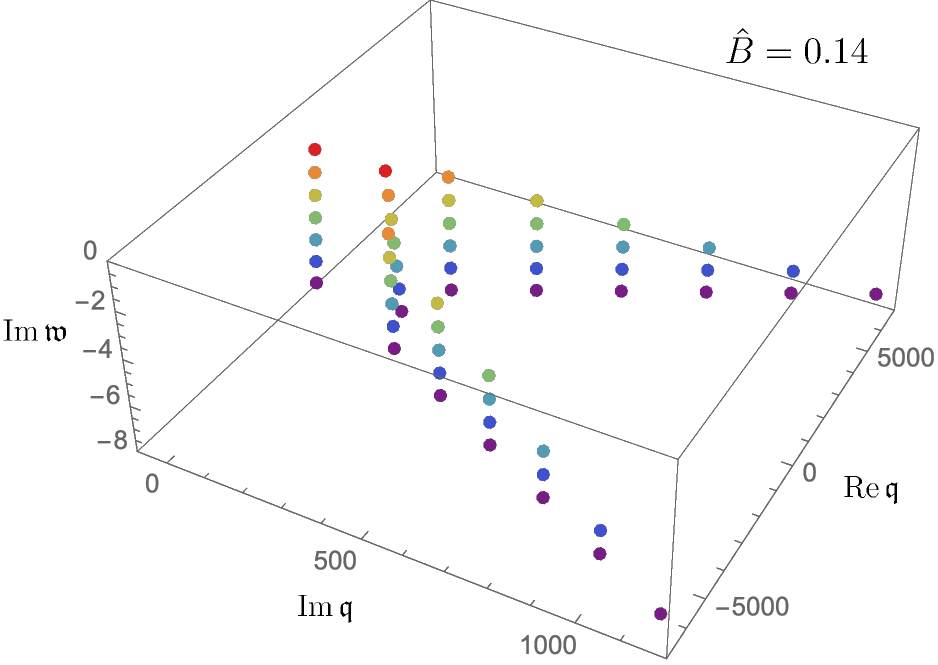} \includegraphics[width=0.3\textwidth]{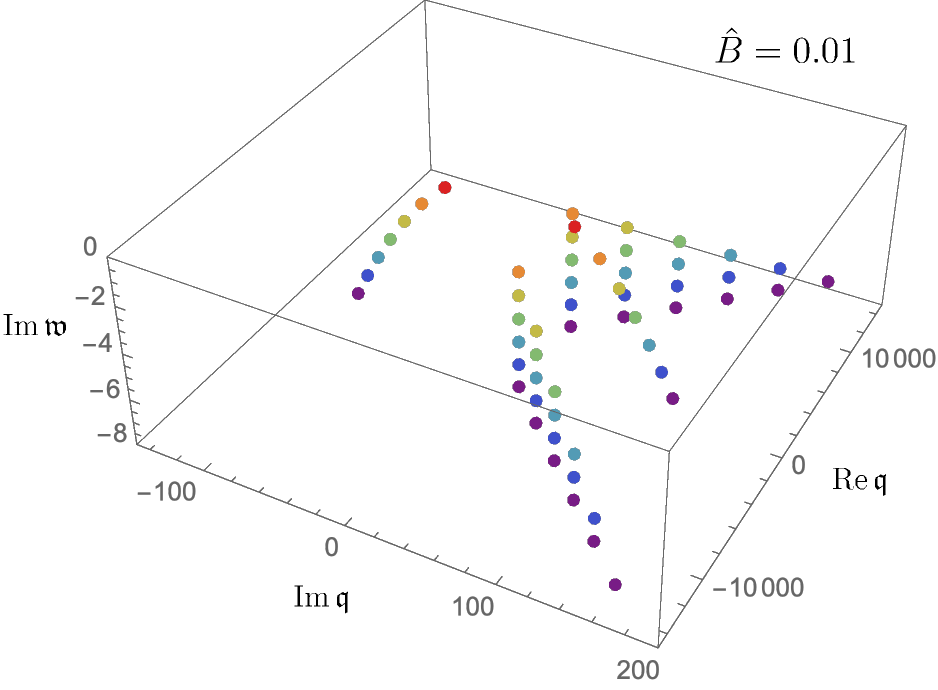}
	%	\,\,\,\includegraphics[width=0.38\textwidth]{Delta_V_B.pdf}
	\caption{Pole-skippng points at low temperature limit $(\hat{T}\rightarrow 0)$: The transition between ordered and disordered states corresponds to a change from the 2D spectrum to the 3D spectrum at $\hat{B}_c \approx 0.499$. }
	\label{3D_T_0}
\end{figure}
%%%%%%%%%%%%%%%%%%%%%%%

	%%%%%%%%%%}%%%%%%%%%%%%%%%%%%%%%%%%
\begin{equation}
\mathcal{M}_{\ell}=\, \sum_{j=1}^{2\ell}\,|\text{Re}\, \tilde{\qn}_{\ell,j}|
\end{equation}
%%%%%%%%%%%%%%%%%%%%%%%%%%%%%%%%%%
Here $\ell$ is the level of pole-skipping point and $j$ counts the the pole-skipping points of this level.
By construction, this quantity is non-negative. We have found that for states well below the critical point, $\mathcal{M}_{\ell}$ becomes positive for all $\ell \ge 2$. For states close to the critical point, positivity starts to emerge at higher $\ell$. On the other hand, $\mathcal{M}_{\ell}$ is zero when none of the pole-skipping points have real momentum. In the latter case, it corresponds to a disordered state of the system. In other words, $\mathcal{M}_{\ell}$ provides us with an infinite set of \textit{order parameters} for distinguishing the two sides of the quantum critical point at $\hat{T}=0$ state.

\begin{figure}
		\centering
	\includegraphics[width=0.55\textwidth]{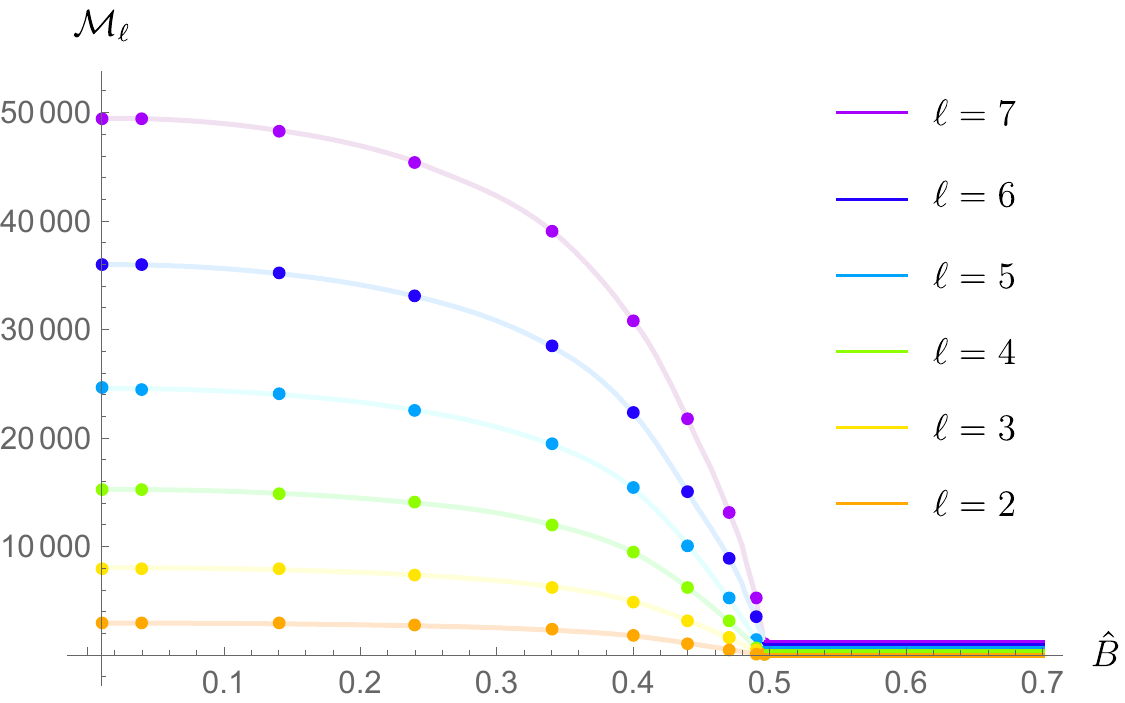}
	%	\,\,\,\includegraphics[width=0.38\textwidth]{Delta_V_B.pdf}
	\caption{Plot of the order parameter $\mathcal{M}_{\ell}$ of $\ell=2,3,\cdots,7$ as a function of the magnetic field. Below the critical magnetic field $\hat{B}_c\approx 0.499$, the order parameter is positive; under a fixed magnetic field, the higher the pole-skipping level, the greater the value of the related order parameter. Above $\hat{B}_c$, all $\mathcal{M}_{\ell}\rightarrow 0$ for $\hat{T}\rightarrow 0$.}
	\label{M_ell}
\end{figure}

In the figure \ref{M_ell} we show $\mathcal{M}_{\ell}$ for $\ell=2,3,\cdots, 7$ as a function of the magnetic field when $\hat{T}=0.0004$.

%%%%%%%%%%%%%%%%%%%%%%%
%%%%%%%%%%%%%%%%%%%%%%%

%______________________________________________________________
\section{Conclusion and outlook}
\label{Conclusion_section}
%______________________________________________________________
%______

We have studied pole-skipping and butterfly velocities in charged and magnetized asymptotically AdS black branes. Our work is a natural continuation of \cite{Abbasi:2019rhy} which initiated this line of research in the limit of weak magnetic fields. The new ingredient here is that we allow strong magnetic fields with significant backreaction onto the geometry. 
These solutions have been studied some time ago in \cite{DHoker:2009ixq,DHoker:2010zpp,DHoker:2012rlj}. One of the main findings there was a quantum phase transition at a critical value of a suitable dimensionless measure of magnetic field strength of $\hat{B}=0.499$. Below that value, the zero temperature limit has non-vanishing entropy whereas above that limit the entropy vanishes as the temperature goes to zero. 

There are several important conclusions that can be drawn from our results. First we confirmed and elaborated further
on the fact that the butterfly velocities are sensitive to the anomaly, which is represented by the Chern-Simons term. Not only the  butterfly velocities are anisotropic but the butterfly velocity in direction of the magnetic field is different from the butterfly velocity in opposite direction to the magnetic field. Again, this can be attributed to the presence of the anomaly. Even more interesting is the fact that the quantum phase transition is visible in the behavior of the butterfly velocities. Approaching the limit $\hat{T}\rightarrow 0$ both butterfly velocities approach the speed of light in the high magnetic field limit. The one in direction against the magnetic field decreases as the magnetic field is lowered
and even vanishes at the quantum critical point. Below that also the forward butterfly velocity decreases with the magnetic field eventually vanishes also. For high temperatures in contrast the butterfly velocities approach the 
usual AdS-Schwarzschild value. Thus is it clear that the quantum phase transition can be detected by the behavior of the butterfly velocities as $\hat{T}\rightarrow 0$.

We also studied that pole-skipping points in the tensor channel. There exists a hierarchy of pole-skipping points indexed by a level $\wn=-i\ell$, $\ell\in\mathbb{N}$ and with momenta $\tilde{\qn}_{\ell, j}$, $j=1, \cdots,2\ell$. Again we find a strong dependence on the magnetic field and temperature. The most important features being that at intermediate temperatures the pole-skipping points arrange asymmetrically around $\text{Im}\, \qn=0$ and they develop also a real part. 
As one approaches $\hat{T}\rightarrow 0$, the behavior depends however drastically upon the field strength being above or below the critical one. Above the critical field strength, the real parts of the complex momenta go to zero such that the pole-skipping spectrum lies again in the plane $\text{Im}\,\wn- \text{Re}\,\qn$ which we can call a $2D$ spectrum. In contrast, for field strengths below the critical value $\text{Re}\, \tilde{\qn}_{\ell,j}\neq0$ and therefore we find a $3D$ spectrum.
This motivates us to suggest the real part of the pole-skipping spectrum to take as an order parameter for the quantum phase transition. Since no symmetry is broken in this quantum phase transition there exists no conventional order parameter. But the real part of the pole-skipping points seem to be a good indicator of of being in a  phase of low field strength, Since no symmetry is broken in this quantum phase transition, no conventional order parameter exists. But the real part of the pole jump seems to be a good indicator of being in a phase of low field strength, so we can call it an ``ordered" phase.

Our study was limited to the holographic field theory with the supersymmetric value of the Chern-Simons coupling, i.e., the holographic dual of the maximally supersymmetric gauge theory in four dimensions. It is natural to ask how our findings translate to the weak coupling limit. While the calculation of butterfly velocities in weakly coupled field theories is complicated there are some hints we can extract from our results. It is tempting to interpret the fact that at large field strength both, the forward and backward, velocities approach the speed of light as a signature of lowest Landau Level physics. A similar behavior has been observed for the anomaly related chiral magnetic wave, which also approached  the speed of light in this limit \cite{Kharzeev:2010gd}. In addition there is a clear relation between the splitting of forward and backward butterfly velocities and the presence of the chiral magnetic effect. On the other hand it is not clear the moment what could correspond to the elaborate pattern of the pole skipping points in the tensor channel at weak coupling. 

It will be interesting to test our results with some experimental setups. One possible place to test our results is the compound $Sr_3Ru_2O_7$. This compound exhibits a series of first-order metamagnetic phase transitions at finite temperatures, ending at a finite temperature critical point \cite{Science}. Similarities to the system studied in this paper are discussed in \cite{DHoker:2010zpp}. This then suggests to experimentally study the speed of the butterfly near the critical point in this compound.

%________________________________________________________
\section*{Acknowledgment}
%______________________________________________________________
We would like to thank Ali Davody, Matthias Kaminski and Dima Kharzeev for fruitful discussions. N.A. thanks the Instituto de F\'\i{}sica Te\'o{}rica, IFT-UAM/CSIC and the Institute for Theoretical Physics of Goethe University for their hospitality when this work was completed.  N.A. was funded by Lanzhou University's ``Double First-Class" start-up fund 561119208. The research of K.L. is supported through the grants CEX2020-001007-S and PID2021-
123017NB-100, PID2021-127726NB-I00 funded by MCIN/AEI/10.13039/501100011033
and by ERDF “A way of making Europe".  

%__________________________________________
\appendix
%__________________________________________
%______________________________________________________________
\section{How to find the upper-half plane pole-skipping point of the energy density correlator?}
\label{Appendix_PS}
%______________________________________________________________
Let us assume a static solution in the bulk in  Eddington-Finkelstein coordinates: $(r,v,x_1,x_2,x_3)$.
In order to find the energy density two-point function, one must consider the perturbations of the $vv$-component of metric $\delta g_{vv}(r,v,x)=\delta g_{vv}(r)e^{-i\omega v+i \vec{k}\cdot \vec{x}}$. 
	 Clearly, the $vv$ component of the Einstein's equation is related to the dynamics of $\delta g_{vv}$: $E_{vv}=0$.
	 There are other perturbations that may be coupled to $\delta g_{vv}$ through the above equation: namely $\delta g_{rr}$, $\delta g_{rv}$, $\delta g_{x^ix^i}$, $\delta g_{x^3x^3}$, $\delta g_{vx^3}$ and $\delta g_{rx^3}$. 
	 We expand all the involved perturbations near the horizon as 
	%%%%%%%%%%%%%%%
	\begin{equation}\label{}
		\delta g_{MN}(r)=	\delta g^{(0)}_{MN}+(r-r_h)	\delta g^{(1)}_{MN}+\cdots
	\end{equation}
	%%%%%%%%%%%%%%%%%

In the absence of perturbations, the equation $E_{vv}=0$ is already satisfied  to \textbf{leading order} of the near horizon expansion; when it comes to the perturbations, however, one finds 
%%%%%%%%%%%%%%%
\begin{eqnarray}\label{}
	\mathcal{G}(\hat{k}, \omega)\, \delta g_{vv}^{(0)}+\,\big(2 \pi T+i \omega\big)\bigg(2 \mathcal
	A\, k \delta g_{vv}^{(0)}+ \mathcal{B}_{T}\,\omega (\delta g_{11}^{(0)}+\delta g_{22}^{(0)})+ \mathcal{B}_{L}\,\omega\delta g_{33}^{(0)})\bigg)=\,0
\end{eqnarray}
%%%%%%%%%%%%%%%%%
At $\omega_p= i 2 \pi T$, the above equation becomes $\mathcal{G}(\hat{k}, \omega)\, g_{vv}^{(0)}=0$. Now there are two possibilities:
\begin{enumerate}
	\item $\delta g_{vv}^{(0)}=0$; we are not interested in this.
	\item $\mathcal{G}(\hat{k}, \omega)=0$ causes $g_{vv}^{(0)}$ to  get decoupled from the rest of $\delta g$ perturbations. Therefore, we cannot solve for the perturbation in the system of equations, since one of the equations ($E_{vv}=0$) is already satisfied, while the number of variables, $\delta g_{MN}$, remains the same as before. This is equivalent to say that the line of poles of energy density correlator on the boundary skips at $(\omega_p, k_p)$, with $k_p$ being the root of  $\mathcal{G}(\hat{k}, \omega)=0$. That is the so-called pole-skipping point. 
\end{enumerate}
It turns out that the condition n $\mathcal{G}(\hat{k}, \omega)=0$ together with $\omega= i 2 \pi T$ precisely identifies the chaos point found from the shock-wave calculations \cite{Blake:2018leo}. The latter refers to the point $(\omega_c, k_c)$ at which OTOC$\,\sim 1- \frac{1}{N^2} e^{-i \omega t + i k x}$ exponentially grows. Then the butterfly velocity is defined as
	%%%%%%%%%%%%%%%
\begin{equation}\label{}
v_B=\,\frac{\omega_c}{k_c}\,.
\end{equation}
%%%%%%%%%%%%%%%%%
However,  the general effective field theory argument shows that, at least in holographic systems \cite{Blake:2017ris} \footnote{This is another way of saying the quantum chaos has hydrodynamic origin is such systems \cite{Blake:2017ris}.}
	%%%%%%%%%%%%%%%
\begin{equation}\label{}
(\omega_p , k_p) \equiv (\omega_c, k_c)\,.
\end{equation}
%%%%%%%%%%%%%%%%%
Then this simply tells us that $v_B$ can also be computed by computing $(\omega_p , k_p)$ .
So all that is to be found is the function $\mathcal{G}(\hat{k}, i 2\pi T)$. It has been found in many cases, including the system \cite{Abbasi:2019rhy} with a single chiral anomaly.

%______________________________________________________________
\section{Relation to the chiral magnetic effect}
%______________________________________________________________
Splitting of longitudinal butterfly velocities found in ref. \cite{Abbasi:2019rhy} (and in this work) suggests that perhaps one can relate this to the chiral magnetic effect. To find rigorous results, we find it convenient to find this relationship by performing analytical calculations.
In the small magnetic field limit, the perturbative solution of the bulk equations is well known \cite{DHoker:2009ixq}. Along the same lines, we parameterize the butterfly velocity in the perturbative expansion of $B$. For the second-order magnetic field, we can write as
%%%%%%%%%%%%%%%%%%%%%%%%%%%%%%%%%%%%%%%% 
\begin{equation}\label{v_B_covariant}\boxed{
		\vec{v}_{B}=\,v_{B}^{(0)}\left(1+ \mathfrak{a}_1\,\mathfrak{b}\cdot\hat{k}+ \mathfrak{a}_2\,\mathfrak{b}^2+ \mathfrak{a}_3\,(\mathfrak{b}\cdot\hat{k}
		)^2\right)\,\hat{k}}\,,\,\,\,\,\,\,\,\mathfrak{b}=\frac{\vec{B}}{T^2}
\end{equation}
%%%%%%%%%%%%%%%%%%%%%%%%%%%%%%%%%%
and $v_B^{(0)}=\sqrt{2/3}$ \cite{Shenker:2013pqa}. Here are a few comments about the above formula:
\begin{itemize}
	\item $\hat{k}$ defines the measurement axis. 
		\item There are  three different structures constructed out of $\mathfrak{b}$ and $\hat{k}$, to second order in $\mathfrak{b}$. The coefficients $\mathfrak{a}_1$, $\mathfrak{a}_2$, and $\mathfrak{a}_3$ can be calculated analytically (see Appendix \ref{U(1)_A}). 
	\item  Due to the term $\mathfrak{b}\cdot\hat{k}$, the speed of the butterfly in two opposite directions relative to a certain axis of measurement is different. The difference depends only on $\mathfrak{a}_1$. As it will be shown in the Appendix \ref{App_C}, when $\mu/T$ and $B/T^2$ are both small
	%%%%%%%%%%%%%%%%%%%%%%%%%%%%%%%%%%%%%%%% 
	\begin{equation}
		\Delta v_B=\, 2 v_B^{(0)}\mathfrak{a}_1 \, |\mathfrak{b}\cdot \hat{k}|\sim \kappa \left(\frac{\mu}{T}\right)^2\left(\frac{B}{T^2}\right)|\cos \theta|
	\end{equation}
	%%%%%%%%%%%%%%%%%%%%%%%%%%%%%%%%%%
	where $\theta$ is the angle between $\hat{k}$ and $\mathfrak{b}$. Note that $B$ is the axial magnetic field.
\end{itemize}
The discussion so far has been for systems with a single $U(1)$ axial current. We can simply extend the discussion to the case of $U_{\text{V}}(1)\times U_{\text{A}}(1)$. As shown in the appendix \ref{sec_U_A_U_V}, in this case we find that
%%%%%%%%%%%%%%%%%%%%%%%%%%%%%%%%%%%%%%%% 
\begin{equation}\label{Delta_v_B}
	\Delta v_B\sim \kappa \left(\frac{\mu_\text{V}}{T}\right)\left(\frac{\mu_\text{A}}{T}\right)\left(\frac{B}{T^2}\right)|\cos \theta|
\end{equation}
%%%%%%%%%%%%%%%%%%%%%%%%%%%%%%%%%%
where $B$ here is the vector magnetic field.

From \eqref{Delta_v_B} it is clear that detecting butterfly velocity difference requires both $\mu_{A}$ and $\mu_V$ to be non-zero.
$\Delta v_B$ is of the form \eqref{Delta_v_B}, which is reminiscent of the chrial magnetic effect \cite{Landsteiner:2012kd} in hydrodynamic energy flow:
%%%%%%%%%%%%%%%%%%%%%%%%%%%%%%%%%%%%%%%% 
\begin{equation}\label{T_mu_nu}
	T^{\mu\nu}\supset\frac{\mu_{\text{V}}\mu_{\text{A}}}{2\pi^2} \big(u^{\mu}B^{\nu}+u^{\nu}B^{\mu}\big)\,.
\end{equation}
%%%%%%%%%%%%%%%%%%%%%%%%%%%%%%%%%%
A comparison between \eqref{Delta_v_B} and \eqref{T_mu_nu} shows that the observation of a non-zero $\Delta v_B$ in experiments may be a sign of the presence of the chiral magnetic effect in the system.
%%%%%%%%%%%%%%%%%%%%%%%
\begin{figure}[h]
	\centering
	\includegraphics[width=0.45\textwidth]{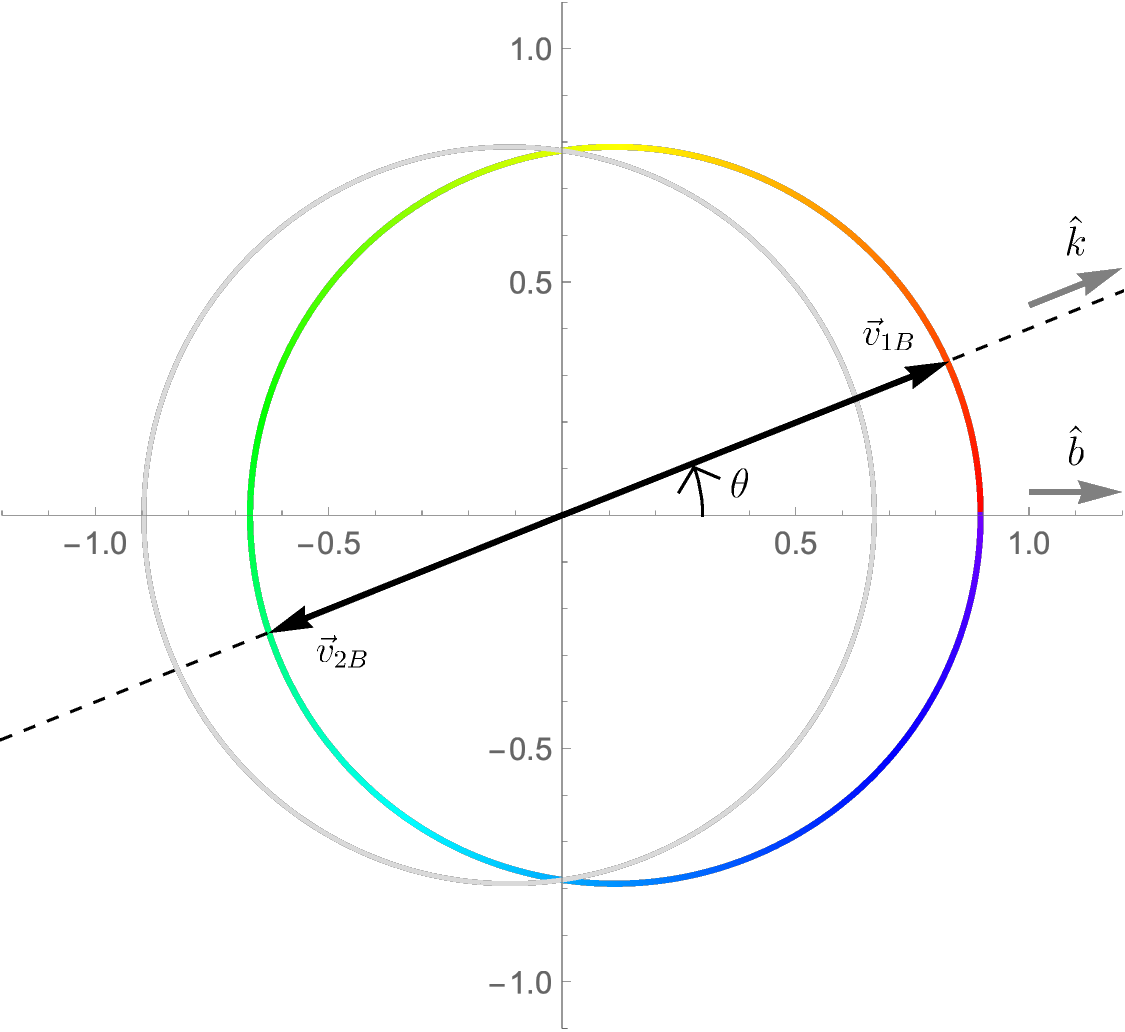}
	%	\,\,\,\includegraphics[width=0.38\textwidth]{Delta_V_B.pdf}
	\caption{Illustration of the two butterfly velocities in an arbitrary direction $\hat{k}$.  The gray curve just shows that the colorful ellipse is not symmetric with respect to the vertical axis. 
		%		Right panel: Illustration of the $|\Delta v_B|$ along differnt directions of measurement, with respect to the magnetic field, in ploar coordinates.  When $\theta$ varies from $-\frac{\pi}{2}$ to $\frac{\pi}{2}$, endpoint of the location vector in polar plane moves on a circle, starting in purpel and ending in green.
	}
	\label{v_B_fig}
\end{figure}
%%%%%%%%%%%%%%%%%%%%%%%

%______________________________________________________________
\section{Magnetized Chirally Charged  RN}
\label{App_C}
%______________________________________________________________
In this Appendix we want to analytically calculate  the butterfly speed in the limit of small magnetic field. We start with the simple case of a $U(1)$ axial current in \ref{U(1)_A} and then extend it to the more realistic case of $U(1)_V\times U(1)_A$ in \ref{sec_U_A_U_V}.

%______________________________________________________________
\subsection{Right-handed case: A single $U(1)$ axial current}
\label{U(1)_A}
%______________________________________________________________
As discussed in the main text, the action in the bulk is given by
%%%%%%%%%%%%%%%%%%%%%%%%%%%%%%%%%%%%%%%% 
\begin{equation}
	\begin{split}
	S = \frac{1}{16 \pi G_5}& \int d^5 x \,\,\sqrt{-g} \left(R + \frac{12}{L^2} - F^{M N} F_{M N}\right)+S_{CS}+ S_{bdy}.\\
	S_{CS} = &\frac{\kappa}{48 \pi G_5} \int d^5 x \,\,\sqrt{-g} \,\epsilon^{\rho \mu \nu \alpha \beta} A_{\rho} F_{\mu \nu}F_{\alpha \beta}
		\end{split}
\end{equation}
%%%%%%%%%%%%%%%%%%%%%%%%%%%%%%%%%%
Let's take the magnetic field in the third direction. The metric and gauge field in the bulk then are parameterized as 
%%%%%%%%%%%%%%%
\begin{eqnarray}\label{final_metric}
	\begin{split}
		ds^2=&-F(r)dv^2+2dr dv+V(r)(dx_1^2+dx_2^2)+W(r)\big(dx_3+C(r) dv\big)^2\\
		A(r)=&\,A_v(r) dv -\frac{1}{2}B x_2 dx_1 +\frac{1}{2} B x_1 dx_2+A_z(r) dz
	\end{split}
\end{eqnarray}
%%%%%%%%%%%%%%%%%
We also take the following ansatz for the metric and gauge field functions:
%%%%%%%%%%%%%%%
\begin{eqnarray}\label{final_ansatz}
	\begin{split}
		F(r)=&f_1(r-r_{h})+\big(f_2+\color{orange}f_{2b}\color{black} B^2\big)(r-r_{h})^2\\
		V(r)=&\,v_0+v_{0b} B^2+\big(v_1+ \color{orange}v_{1b}\color{black} B^2\big)(r-r_{h})+\big(v_2+ \color{orange}v_{2b}\color{black} B^2\big)(r-r_{h})^2\\
		W(r)=&\,v_0+w_{0b} B^2+\big(v_1+\color{orange}w_{1b}\color{black} B^2\big)(r-r_{h})+\big(v_2+w_{2b} B^2\big)(r-r_{h})^2\\
		C(r)=&\,B\big(c_1(r-r_h)+\color{orange}c_2\color{black} (r-r_h)^2\big)\\
		A_{v}(r)=&\,a_{v0}+a_{v1}(r-r_{h})+\big(a_{v2}+\color{orange}a_{v2b}\color{black} B^2\big)(r-r_{h})^2\\
		A_{z}(r)=&\,B\big(a_{z0}+\color{orange}a_{z1}\color{black} (r-r_{h})+\big(a_{z2}+a_{z2b} B^2\big)(r-r_{h})^2\big)
	\end{split}
\end{eqnarray}
%%%%%%%%%%%%%%%%%
	 Compared to the non-chiral case, two new metric and gauge field functions are introduced: $C(r)$ and $A_z(r)$.  This is just due due to the CS term. 
	Note that
	\begin{itemize}
		\item Since CS term breaks the parity,  $C(r)$ and $A_z(r)$ start to contribute at linear order in $B$.
		\item Charge conjugation implies that $C(r)$ and $A_z(r)$ have to be \textbf{even} and \textbf{odd} functions of $\nu$, respectively.
	\item Since we have assumed $F(r_{h})=0$, $C(r)$ must vanish at $r_h$ too. 
		\end{itemize}
Solving the equations, we get (we omit the expression of $v_{2b}$)
%%%%%%%%%%%%%%%
\begin{eqnarray}\label{}
	\begin{split}
	\color{orange}	f_{2b}&=\frac{5}{3 v_0^2}\color{blue}+\frac{3 v_0}{4} c_1^2\color{black}\\
	\color{orange}	v_{1b}&=-\frac{8}{3 f_1 v_0}+8v_0 v_{0b}\big(1-\frac{a_1^2}{6}\big)\\
	\color{orange}	w_{1b}&=\frac{4}{3 f_1 v_0}+8v_0 v_{0b}\big(1-\frac{a_1^2}{6}\big)\color{blue}-\frac{v_0^2}{f_1}c_1^2\color{black}\\
	\color{orange}	a_{v2b}&=\frac{a_{1}}{f_1 v_0^2}\color{blue}-\frac{a_1 v_0}{4 f_1}\,c_1^2+\frac{2 a_1}{f_1 v_0^2} \,\kappa^2\color{black}\\
	\color{orange}c_2&\color{blue}=\left(\frac{11}{3}a_1^2-10\right)\frac{c_1}{f_1}-\frac{4 a_1^2}{f_1 v_0^{3/2}}\,\kappa\\
	\color{orange}a_{z1}&\color{blue}=-\frac{a_1}{f_1}\left(v_0 \,c_1-\frac{2 \kappa}{\sqrt{v_0}}\right)
	\end{split}
\end{eqnarray}
%%%%%%%%%%%%%%%%%
Compared to the non-chiral case, the \color{blue} blue expressions \color{black} have been found to contribute too.

Now let us turn on all the perturbations of the form $\delta g_{MN}e^{- i \omega v+i k x_3}$ that appear in energy dynamics. The $E_{vv}=0$  up to first order in perturbation at $r=r_h$, and to second order in $B$, gives:
%%%%%%%%%%%%%%%
\begin{equation*}\label{}
	\begin{split}
		\left[i k^2\left(1-\frac{w_{0b}}{v_0}B^2\right)+ \color{blue}v_0 c_1 k B \color{black}+\,\frac{12 v_0}{f_1} \omega\big(1-\frac{a_1^2}{6}-\frac{B^2}{6 v_0^2}\color{blue}-\frac{v_0\, c_1^2\,B^2}{24}\color{black}\big)\right]&\delta g_{vv}^{(0)}+\\
		\,\big(\frac{f_1}{2}+i \omega\big)\bigg[2 k \left(1-\frac{w_{0b}}{v_0}B^2\right)\delta g_{vv}^{(0)}+\omega \bigg((\delta g_{11}^{(0)}+\delta g_{22}^{(0)})\left(1-\frac{v_{0b}}{v_0}B^2\right)+&\delta g_{33}^{(0)}\left(1-\frac{w_{0b}}{v_0}B^2\right)\bigg)\bigg]=\,0\,.
	\end{split}
\end{equation*}
%%%%%%%%%%%%%%%%%
Using $f_1=4\pi T$, we find \textbf{two butterfly velocities}
%%%%%%%%%%%%%%%
\begin{equation*}\label{}
	v^{L}_{B, \pm}=\,\pm\frac{2\pi T}{\sqrt{v_0(6-a_1^2)}}\left[1\mp c_1\,\frac{v_0^{1/2}}{2\sqrt{6-a_1^2}}B+\frac{1}{2}\left(\frac{4+v_0^3 c_1^2}{4v_0^2\left(6-a_1^2\right)}-\frac{\color{black}w_{0b}}{v_0}\right)B^2\right]\,.
\end{equation*}
%%%%%%%%%%%%%%%%%
The difference between the two butterfly speeds is given by
%%%%%%%%%%%%%%%
\begin{equation}\label{Delta_analytic}\boxed{
\Delta	v^{L}_{B}=\,c_1\frac{4\pi T}{(6-a_1^2)}B}
\end{equation}
%%%%%%%%%%%%%%%%%
This is actually similar to the result of ref. \cite{Abbasi:2019rhy}. Using symmetry argument and dimensional analysis, we found that $c_1\sim\kappa T^3 a_1^2$ is needed under the small limit of $a_1\sim\mu/T$. Then \eqref{Delta_analytic} simplifies to
%%%%%%%%%%%%%%%
\begin{equation}\label{}
\Delta v^{L}_{B}\sim \kappa \left(\frac{\mu}{T}\right)^2\left(\frac{B}{T^2}\right)\,.
\end{equation}
%%%%%%%%%%%%%%%%%
Notably, ref. \cite{Abbasi:2019rhy} can specify the numerical coefficient on the right-hand side of the above equation as $\frac{8 \pi^4}{3^{3/2}}(\log 4-1)$ . The reason is that the above reference uses a complete solution for metric and gauge fields. However, here we only solve the bulk equation near the horizon.

It is easy to show that if we take the perturbation as $\delta g_{MN}e^{- i \omega v+i k x_1}$, which is equivalent to measuring the butterfly effect on the axis perpendicular to the magnetic field, we find two butterfly velocities same size
%%%%%%%%%%%%%%%
\begin{equation*}\label{}
	v^{T}_{B, \pm}=\,\pm\frac{2\pi T}{\sqrt{v_0(6-a_1^2)}}\left[1+\frac{1}{2}\left(\frac{4+v_0^3 c_1^2}{4v_0^2\left(6-a_1^2\right)}-\frac{\color{black}v_{0b}}{v_0}\right)B^2\right]\,.
\end{equation*}
%%%%%%%%%%%%%%%%%

%______________________________________________________________
\subsection{A $U(1)$ axial current together with a $U(1)$ vector current}
\label{sec_U_A_U_V}
%______________________________________________________________
In this case we consider two types of fermions on the boundary; right-handed and left handed. Then for each of them we consider a distinct gauge field in the bulk, say $A^{R}_{M}(r,x^{\mu})$ and $A^L_M(r,x^{\mu})$. By definition, the CS term has opposite sign for left and right handed fermions; therefore we write:
%%%%%%%%%%%%%%%%%%%%%%%%%%%%%%%%%%%%%%%% 
\begin{equation}\label{}
	S_{CS} = \frac{\kappa}{48 \pi G_5} \int d^5 x \,\,\sqrt{-g} \,\epsilon^{\rho \mu \nu \alpha \beta} A^R_{\rho} F^R_{\mu \nu}F^R_{\alpha \beta}-\frac{\kappa}{48 \pi G_5} \int d^5 x \,\,\sqrt{-g} \,\epsilon^{\rho \mu \nu \alpha \beta} A^L_{\rho} F^L_{\mu \nu}F^L_{\alpha \beta}
\end{equation}
%%%%%%%%%%%%%%%%%%%%%%%%%%%%%%%%%%
To make the ansatz we have to be careful with the\textbf{ magnetic field}. In practice, instead of \textbf{(right-handed, left-handed)} we work with \textbf{(axial A, vector V)} currents. Then in an experimental setup, we would have 
%%%%%%%%%%%%%%%%%%%%%%%%%%%%%%%%%%%%%%%% 
\begin{equation}\label{}
	B\equiv B_{\text{V}}\,,\,\,\,\,\,\,\,B_{\text{A}}=0\,,
\end{equation}
%%%%%%%%%%%%%%%%%%%%%%%%%%%%%%%%%%
or equivalently we write in terms of gauge fields
%%%%%%%%%%%%%%%%%%%%%%%%%%%%%%%%%%%%%%%% 
\begin{equation}\label{A_V_A_A}
	A_{\text{V}}=-\frac{1}{2}B_{\text{V}}\,x_2 dx_1 +\frac{1}{2} B_{\text{V}}\,x_1 dx_2\,,\,\,\,\,\,\,\,\,\,A_{\text{A}}=\,0\,.
\end{equation}
%%%%%%%%%%%%%%%%%%%%%%%%%%%%%%%%%%
Then by using $A^{R,L}=A_{\text{V}}\pm A_{\text{A}}$, we can write \eqref{A_V_A_A} in terms of $R$ and $L$ gauge fields:
%%%%%%%%%%%%%%%%%%%%%%%%%%%%%%%%%%%%%%%% 
\begin{equation}\label{}
	A^{R}=A^{L}=-\frac{1}{2}B_{\text{V}}\,x_2 dx_1 +\frac{1}{2} B_{\text{V}}\,x_1 dx_2\,.
\end{equation}
%%%%%%%%%%%%%%%%%%%%%%%%%%%%%%%%%%
Using this, we can simply parameterize the metric and gauge fields in the bulk
%%%%%%%%%%%%%%%
\begin{eqnarray}\label{}
	\begin{split}
		ds^2=&-F(r)dv^2+2dr dv+V(r)(dx_1^2+dx_2^2)+W(r)\big(dx_3+C(r) dv\big)^2\\
		A^R(r)=&\,A^R_v(r) dv -\frac{1}{2}B x_2 dx_1 +\frac{1}{2} B x_1 dx_2+A^R_z(r) dz\\
		A^L(r)=&\,A^L_v(r) dv -\frac{1}{2}B x_2 dx_1 +\frac{1}{2} B x_1 dx_2+A^L_z(r) dz
	\end{split}
\end{eqnarray}
%%%%%%%%%%%%%%%%%
We also take the following ansatz for the metric and gauge field functions:
%%%%%%%%%%%%%%%
\begin{eqnarray}\label{}
	\begin{split}
		F(r)=&f_1(r-r_{h})+\big(\color{blue}f_2 \color{black} +\color{orange}f_{2b}\color{black} B^2\big)(r-r_{h})^2\\
		V(r)=&\,v_0+v_{0b} B^2+\big(\color{blue}v_1 \color{black}+ \color{orange}v_{1b}\color{black} B^2\big)(r-r_{h})+\big(\color{blue}v_2 \color{black}+ \color{orange}v_{2b}\color{black} B^2\big)(r-r_{h})^2\\
		W(r)=&\,v_0+w_{0b} B^2+\big(\color{blue} v_1 \color{black}+w_{1b} B^2\big)(r-r_{h})+\big(\color{blue}v_2 \color{black}+w_{2b} B^2\big)(r-r_{h})^2\\
		C(r)=&\,B\big(c_1(r-r_h)+\color{orange}c_2\color{black} (r-r_h)^2\big)\\
		A^R_{v}(r)=&\,a^R_{v0}+a^R_{v1}(r-r_{h})+\big(\color{blue}a^R_{v2} \color{black}+\color{orange}a^R_{v2b}\color{black} B^2\big)(r-r_{h})^2\\
		A^R_{z}(r)=&\,B\big(a^R_{z0}+\color{orange}a_{z1}\color{black} (r-r_{h})+\big(a^R_{z2}+a^R_{z2b} B^2\big)(r-r_{h})^2\big)\\
		A^L_{v}(r)=&\,a^L_{v0}+a^L_{v1}(r-r_{h})+\big(\color{blue}a^L_{v2} \color{black}+\color{orange}a^L_{v2b}\color{black} B^2\big)(r-r_{h})^2\\
		A^L_{z}(r)=&\,B\big(a^L_{z0}+\color{orange}a^L_{z1}\color{black} (r-r_{h})+\big(a^L_{z2}+a^L_{z2b} B^2\big)(r-r_{h})^2\big)
	\end{split}
\end{eqnarray}
%%%%%%%%%%%%%%%%%
Equations of motion at $r=r_h$ and at $O(B^0)$ give the \color{blue} blue coefficients \color{black}
%%%%%%%%%%%%%%%
\begin{eqnarray}\label{}
	\begin{split}
		v_1=&\frac{8v_0}{f_1}-\frac{4  \bigg((a^R_{v1})^2+(a^L_{v1})^2\bigg) v_0}{f_1}\\
		v_2=&\frac{v_1^2}{4 v_0}=\frac{16v_0}{f_1^2}\left(1-\frac{(a^R_{v1})^2+(a^L_{v1})^2}{6}\right)^2\\
		f_2=&-2+\frac{7}{3}\bigg((a^R_{v1})^2+(a^L_{v1})^2\bigg)\\
		a^{R,L}_{v2}=&-6\frac{a^{R,L}_{v1}}{f_1}\left(1-\frac{(a^R_{v1})^2+(a^L_{v1})^2}{6}\right)
	\end{split}
\end{eqnarray}
%%%%%%%%%%%%%%%%%
And at $O(B^2)$, solving the equations at $r=r_h$ give the \color{orange} orange coefficients \color{black}
%%%%%%%%%%%%%%%
\begin{eqnarray}\label{}
	\begin{split}
		f_{2b}&=\frac{5}{3 v_0^2}+\frac{3 v_0}{4} c_1^2\,,\,\,\,\,\,\,\,\,\,\,\,\,v_{2b}= \text{complicated}\\
		v_{1b}&=-\frac{8}{3 f_1 v_0}+8v_0 v_{0b}\bigg(1-\frac{(a^R_{v1})^2+(a^L_{v1})^2}{6}\bigg)\\
		w_{1b}&=\frac{4}{3 f_1 v_0}+8v_0 v_{0b}\bigg(1-\frac{(a^R_{v1})^2+(a^L_{v1})^2}{6}\bigg)-\frac{v_0^2}{f_1}c_1^2\\
		a^{R,L}_{2b}&=\frac{a^{R,L}_{v1}}{f_1 v_0^2}-\frac{a^{R,L}_{v1} v_0}{4 f_1}\,c_1^2+\frac{2 a^{R,L}_{v1}}{f_1 v_0^2} \,\kappa^2\\
		c_2&=\left(\frac{11}{3}\big((a^R_{v1})^2+(a^L_{v1})^2\big)-10\right)\frac{c_1}{f_1}-4\frac{(a^R_{v1})^2+(a^L_{v1})^2}{f_1 v_0^{3/2}}\,\kappa\\
		a^{R,L}_{z1}&=-\frac{a^{R,L}_{v1}}{f_1}\left(v_0 \,c_1-\frac{2 \kappa}{\sqrt{v_0}}\right)\,.
	\end{split}
\end{eqnarray}
%%%%%%%%%%%%%%%%%
The other coefficients either come from the input data or do not contribute to the final result.

Now let us turn on all the perturbations of the form $\delta g_{MN}e^{- i \omega v+i k x_3}$ that arise in energy dynamics. The $E_{vv}=0$,  up to first order in perturbation at $r=r_h$, and to second order in $B$, gives:
%%%%%%%%%%%%%%%
\begin{equation*}\label{}
	\begin{split}
		\left[i k^2\left(1-\frac{w_{0b}}{v_0}B^2\right)+ v_0 c_1 k B +\,\frac{12 v_0}{f_1} \omega\big(1-\frac{(a^R_{v1})^2+(a^L_{v1})^2}{6}-\frac{B^2}{6 v_0^2}-\frac{v_0\, c_1^2\,B^2}{24}\big)\right]&\delta g_{vv}^{(0)}+\\
		\,\big(\frac{f_1}{2}+i \omega\big)\bigg[2 k \left(1-\frac{w_{0b}}{v_0}B^2\right)\delta g_{vv}^{(0)}+\omega \bigg((\delta g_{11}^{(0)}+\delta g_{22}^{(0)})\left(1-\frac{v_{0b}}{v_0}B^2\right)+&\delta g_{33}^{(0)}\left(1-\frac{w_{0b}}{v_0}B^2\right)\bigg)\bigg]=\,0\\
	\end{split}
\end{equation*}
%%%%%%%%%%%%%%%%%
Using $f_1=4\pi T$, we find \textbf{two butterfly velocities}:
%%%%%%%%%%%%%%%
\begin{equation*}\label{}
	\begin{split}
		v^{L}_{B, \pm}=\,&\pm\frac{2\pi T}{\sqrt{v_0(6-\big((a^R_{v1})^2+(a^L_{v1})^2\big)}}\\
		&\,\,\,\,\times \,\left[1\mp c_1\,\frac{v_0^{1/2}}{2\sqrt{6-((a^R_{v1})^2+(a^L_{v1})^2}}B+\frac{1}{2}\left(\frac{4+v_0^3 c_1^2}{4v_0^2\big(6-((a^R_{v1})^2+(a^L_{v1})^2\big)}-\frac{\color{black}w_{0b}}{v_0}\right)B^2\right]\,.
	\end{split}
\end{equation*}
%%%%%%%%%%%%%%%%%
\textbf{A comment:}
\\It can be seen that, in general, the two longitudinal butterfly velocities have different magnitudes. This is due to the linear term in $B$. This term has a factor of $c_1$ in the coefficient, which cannot be determined by near-horizon analysis. However, as we discuss \eqref{final_ansatz} below, it is proportional to $\kappa$ and must be an even function of $\nu^R$ and $\nu^L$. Also, it must disappear when $\nu^R=\nu^L$. Taking these points into account, we can parameterize the general form of this quantity:
%%%%%%%%%%%%%%%%%%%%%%%%%%%%%%%%%%%%%%%% 
\begin{equation}\label{c_1}
	\begin{split}
		c_1= &\,\frac{\kappa}{v_0^{3/2}}\, \big((\nu^R)^2-(\nu^L)^2\big)\,\mathcal{J}\big(\nu^R,\nu^L\big)\\
		=& \, \frac{4\,\kappa}{v_0^{3/2}}\, \nu_{\text{A}}\,\nu_{\text{V}}\,\mathcal{J}\big(\nu_{\text{A}},\nu_{\text{V}}\big)
	\end{split}
\end{equation}
%%%%%%%%%%%%%%%%%%%%%%%%%%%%%%%%%%
Where $\mathcal{J}$ is an even function of its two parameters, and $\mathcal{J}(0,0)=1$.

It is easy to show that if we take the perturbations as $\delta g_{MN}e^{- i \omega v+i k x_1}$, which is equivalent to measuring the butterfly effect on the axis perpendicular to the magnetic field, we find two butterfly velocities with the same magnitude
%%%%%%%%%%%%%%%
\begin{equation*}\label{}
	v^{T}_{B, \pm}=\,\pm\frac{2\pi T}{\sqrt{v_0\big(6-((a^R_{v1})^2+(a^L_{v1})^2))}}\left[1+\frac{1}{2}\left(\frac{4+v_0^3 c_1^2}{4v_0^2\big(6-((a^R_{v1})^2+(a^L_{v1})^2)\big)}-\frac{\color{black}v_{0b}}{v_0}\right)B^2\right]\,.
\end{equation*}
%%%%%%%%%%%%%%%%%

%______________________________________________________________
\section{Near horizon data  in the tensor channel}
\label{Appendix_PS_H_xy}
%______________________________________________________________
The near horizon information about the dynamics of $H_{xy}$ is encoded in equations \eqref{M_11}-\eqref{M_41}. The first three coefficients in these equations is as the following:

\begin{equation}
	\begin{split}
M_{11}=&\,\bigg[k-\left(\frac{p}{q}+ 2 \kappa b\right)\omega\bigg]^2+ 2 i \omega \big(6 - b^2 - q^2\big)\,,\\
M_{21}=&\,8 k^2 \left(6-5 b^2-q^2\right)+4 i \omega  \left[10 b^4+20 b^2 \left(q^2-6\right)-3 \left(p^2-2 \left(q^2-6\right)^2\right)\right]-24  \,p\, q\,k\, \omega\\
&+\,8 \bigg[  \left(10 b^2+5 q^2-12\right)\,k\, \omega+p\, q \left(3 \omega ^2+3 i \omega -2\right)\bigg]\left(\frac{p}{q}+ 2 \kappa b\right)\,,\\
&+\,4 \omega  \bigg[2 \left(6-5 b^2- 4 q^2\right) \omega - 3 i  q^2\bigg]\left(\frac{p}{q}+ 2 \kappa b\right)^2\\
M_{22}=&\,6\, k^2+4 i \omega\left(11 \left(6-q^2\right)-19b^2\right)+\,6 \left(4 b^2-24\right)\\
&\,-12 k \omega \,\left(\frac{p}{q}+ 2 \kappa b\right)+\,6 \left(\omega ^2+i \omega -1\right)\,\left(\frac{p}{q}+ 2 \kappa b\right)^2\,.
	\end{split}
\end{equation}

\bibliographystyle{utphys}
%\bibliography{nlhydro}

\begin{thebibliography}{100}

\bibitem{Shenker:2013pqa} S.~H.~Shenker and D.~Stanford,
	``Black holes and the butterfly effect,''
	JHEP \textbf{03} (2014), 067
	%doi:10.1007/JHEP03(2014)067
	[arXiv:1306.0622 [hep-th]].
	%1076 citations counted in INSPIRE as of 29 Apr 2023

\bibitem{Shenker:2014cwa} S.~H.~Shenker and D.~Stanford,
	``Stringy effects in scrambling,''
	JHEP \textbf{05} (2015), 132
	%doi:10.1007/JHEP05(2015)132
	[arXiv:1412.6087 [hep-th]].
	%457 citations counted in INSPIRE as of 29 Apr 2023

\bibitem{Maldacena:2015waa} J.~Maldacena, S.~H.~Shenker and D.~Stanford,
	``A bound on chaos,''
	JHEP \textbf{08} (2016), 106
	%doi:10.1007/JHEP08(2016)106
	[arXiv:1503.01409 [hep-th]].
	%1515 citations counted in INSPIRE as of 29 Apr 2023

\bibitem{Roberts:2014ifa} D.~A.~Roberts and D.~Stanford,
	``Two-dimensional conformal field theory and the butterfly effect,''
	Phys. Rev. Lett. \textbf{115} (2015) no.13, 131603
	%doi:10.1103/PhysRevLett.115.131603
	[arXiv:1412.5123 [hep-th]].
	%355 citations counted in INSPIRE as of 29 Apr 2023

\bibitem{Maldacena:2016hyu} J.~Maldacena and D.~Stanford,
	``Remarks on the Sachdev-Ye-Kitaev model,''
	Phys. Rev. D \textbf{94} (2016) no.10, 106002
	%doi:10.1103/PhysRevD.94.106002
	[arXiv:1604.07818 [hep-th]].
	%1355 citations counted in INSPIRE as of 29 Apr 2023

\bibitem{Stanford:2015owe} D.~Stanford,
	``Many-body chaos at weak coupling,''
	JHEP \textbf{10} (2016), 009
	%doi:10.1007/JHEP10(2016)009
	[arXiv:1512.07687 [hep-th]].
	%140 citations counted in INSPIRE as of 29 Apr 2023

\bibitem{Blake:2017ris} M.~Blake, H.~Lee and H.~Liu,
``A quantum hydrodynamical description for scrambling and many-body chaos,''
JHEP \textbf{10} (2018), 127
%doi:10.1007/JHEP10(2018)127
[arXiv:1801.00010 [hep-th]].
%48 citations counted in INSPIRE as of 28 May 2020

\bibitem{Grozdanov:2017ajz} S.~Grozdanov, K.~Schalm and V.~Scopelliti,
``Black hole scrambling from hydrodynamics,''
Phys. Rev. Lett. \textbf{120} (2018) no.23, 231601
%doi:10.1103/PhysRevLett.120.231601
[arXiv:1710.00921 [hep-th]].
%43 citations counted in INSPIRE as of 28 May 2020

\bibitem{Amado:2007yr} I.~Amado, C.~Hoyos-Badajoz, K.~Landsteiner and S.~Montero,
``Residues of correlators in the strongly coupled N=4 plasma,''
Phys. Rev. D \textbf{77} (2008), 065004
%doi:10.1103/PhysRevD.77.065004
[arXiv:0710.4458 [hep-th]].
%29 citations counted in INSPIRE as of 26 Jul 2023

\bibitem{Amado:2008ji} I.~Amado, C.~Hoyos-Badajoz, K.~Landsteiner and S.~Montero,
``Hydrodynamics and beyond in the strongly coupled N=4 plasma,''
JHEP \textbf{07} (2008), 133
%doi:10.1088/1126-6708/2008/07/133
[arXiv:0805.2570 [hep-th]].
%51 citations counted in INSPIRE as of 26 Jul 2023

\bibitem{Blake:2018leo} M.~Blake, R.~A.~Davison, S.~Grozdanov and H.~Liu,
``Many-body chaos and energy dynamics in holography,''
JHEP \textbf{10} (2018), 035
%doi:10.1007/JHEP10(2018)035
[arXiv:1809.01169 [hep-th]].
%94 citations counted in INSPIRE as of 29 Apr 2023

\bibitem{DHoker:2009ixq} E.~D'Hoker and P.~Kraus,
``Charged Magnetic Brane Solutions in AdS (5) and the fate of the third law of thermodynamics,''
JHEP {\bf 1003}, 095 (2010)
[arXiv:0911.4518 [hep-th]].
%%CITATION = doi:10.1007/JHEP03(2010)095;%%
%91 citations counted in INSPIRE as of 15 Oct 2018

\bibitem{DHoker:2010zpp} E.~D'Hoker and P.~Kraus,
``Holographic Metamagnetism, Quantum Criticality, and Crossover Behavior,''
JHEP \textbf{05} (2010), 083
%doi:10.1007/JHEP05(2010)083
[arXiv:1003.1302 [hep-th]].
%62 citations counted in INSPIRE as of 23 Jul 2023

\bibitem{DHoker:2012rlj} E.~D'Hoker and P.~Kraus,
%``Quantum Criticality via Magnetic Branes,''
Lect. Notes Phys. \textbf{871} (2013), 469-502
doi:10.1007/978-3-642-37305-3\_18
[arXiv:1208.1925 [hep-th]].
%30 citations counted in INSPIRE as of 27 Jul 2023

\bibitem{Abbasi:2019rhy} N.~Abbasi and J.~Tabatabaei,
``Quantum chaos, pole-skipping and hydrodynamics in a holographic system with chiral anomaly,''
JHEP \textbf{03} (2020), 050
%doi:10.1007/JHEP03(2020)050
[arXiv:1910.13696 [hep-th]].
%22 citations counted in INSPIRE as of 29 Apr 2023

\bibitem{Fukushima:2008xe} K.~Fukushima, D.~E.~Kharzeev and H.~J.~Warringa,
%``The Chiral Magnetic Effect,''
Phys. Rev. D \textbf{78} (2008), 074033
doi:10.1103/PhysRevD.78.074033
[arXiv:0808.3382 [hep-ph]].
%1822 citations counted in INSPIRE as of 27 Jul 2023

\bibitem{Ling:2016ibq} Y.~Ling, P.~Liu and J.~P.~Wu,
``Holographic Butterfly Effect at Quantum Critical Points,''
JHEP \textbf{10} (2017), 025
%doi:10.1007/JHEP10(2017)025
[arXiv:1610.02669 [hep-th]].
%50 citations counted in INSPIRE as of 24 Jul 2023

\bibitem{Ammon:2017ded} M.~Ammon, M.~Kaminski, R.~Koirala, J.~Leiber and J.~Wu,
``Quasinormal modes of charged magnetic black branes \& chiral magnetic transport,''
JHEP \textbf{04} (2017), 067
%doi:10.1007/JHEP04(2017)067
[arXiv:1701.05565 [hep-th]].
%41 citations counted in INSPIRE as of 23 Jul 2023

\bibitem{Blake:2019otz} M.~Blake, R.~A.~Davison and D.~Vegh,
``Horizon constraints on holographic Green\textquoteright{}s functions,''
JHEP \textbf{01} (2020), 077
%doi:10.1007/JHEP01(2020)077
[arXiv:1904.12883 [hep-th]].
%56 citations counted in INSPIRE as of 23 Jul 2023

\bibitem{Natsuume:2019sfp} M.~Natsuume and T.~Okamura,
``Holographic chaos, pole-skipping, and regularity,''
arXiv:1905.12014 [hep-th].
%%CITATION = ARXIV:1905.12014;%%
%4 citations counted in INSPIRE as of 27 Oct 2019

\bibitem{Natsuume:2019vcv} M.~Natsuume and T.~Okamura,
``Pole-skipping with finite-coupling corrections,''
arXiv:1909.09168 [hep-th].
%%CITATION = ARXIV:1909.09168;%%
%1 citations counted in INSPIRE as of 08 Oct 2019

\bibitem{Wu:2019esr} X.~Wu,
``Higher curvature corrections to pole-skipping,''
arXiv:1909.10223 [hep-th].
%%CITATION = ARXIV:1909.10223;%%

\bibitem{Ahn:2019rnq} Y.~Ahn, V.~Jahnke, H.~S.~Jeong and K.~Y.~Kim,
``Scrambling in Hyperbolic Black Holes: shock waves and pole-skipping,''
arXiv:1907.08030 [hep-th].
%%CITATION = ARXIV:1907.08030;%%
%3 citations counted in INSPIRE as of 08 Oct 2019

%\cite{Grozdanov:2019uhi}
\bibitem{Grozdanov:2019uhi}
S.~Grozdanov, P.~K.~Kovtun, A.~O.~Starinets and P.~Tadi\'c,
``The complex life of hydrodynamic modes,''
JHEP \textbf{11}, 097 (2019)
%doi:10.1007/JHEP11(2019)097
[arXiv:1904.12862 [hep-th]].
%100 citations counted in INSPIRE as of 30 Jul 2023

\bibitem{Li:2019bgc} W.~Li, S.~Lin and J.~Mei,
``Thermal diffusion and quantum chaos in neutral magnetized plasma,''
Phys.\ Rev.\ D {\bf 100}, no. 4, 046012 (2019)
%doi:10.1103/PhysRevD.100.046012
[arXiv:1905.07684 [hep-th]].
%%CITATION = doi:10.1103/PhysRevD.100.046012;%%

\bibitem{Ceplak:2019ymw} N.~Ceplak, K.~Ramdial and D.~Vegh,
``Fermionic pole-skipping in holography,''
arXiv:1910.02975 [hep-th].
%%CITATION = ARXIV:1910.02975;%%
%1 citations counted in INSPIRE as of 15 Oct 2019

\bibitem{Das:2019tga} S.~Das, B.~Ezhuthachan and A.~Kundu,
``Real Time Dynamics in Low Point Correlators,''
arXiv:1907.08763 [hep-th].
%%CITATION = ARXIV:1907.08763;%%
%1 citations counted in INSPIRE as of 15 Oct 2019

%\cite{Abbasi:2020ykq}
\bibitem{Abbasi:2020ykq}
N.~Abbasi and S.~Tahery,
``Complexified quasinormal modes and the pole-skipping in a holographic system at finite chemical potential,''
JHEP \textbf{10} (2020), 076
%doi:10.1007/JHEP10(2020)076
[arXiv:2007.10024 [hep-th]].
%50 citations counted in INSPIRE as of 30 Jul 2023


%\cite{Jansen:2020hfd}
\bibitem{Jansen:2020hfd}
A.~Jansen and C.~Pantelidou,
``Quasinormal modes in charged fluids at complex momentum,''
JHEP \textbf{10} (2020), 121
%doi:10.1007/JHEP10(2020)121
[arXiv:2007.14418 [hep-th]].
%41 citations counted in INSPIRE as of 30 Jul 2023

%\cite{Choi:2020tdj}
\bibitem{Choi:2020tdj}
C.~Choi, M.~Mezei and G.~S\'arosi,
``Pole skipping away from maximal chaos,''
%doi:10.1007/JHEP02(2021)207
[arXiv:2010.08558 [hep-th]].
%44 citations counted in INSPIRE as of 30 Jul 2023

\bibitem{Liu:2020yaf} Y.~Liu and A.~Raju,
``Quantum Chaos in Topologically Massive Gravity,''
[arXiv:2005.08508 [hep-th]].
%1 citations counted in INSPIRE as of 03 Jul 2020

\bibitem{Ahn:2020baf} Y.~Ahn, V.~Jahnke, H.~S.~Jeong, K.~S.~Lee, M.~Nishida and K.~Y.~Kim,
``Classifying pole-skipping points,''
JHEP \textbf{03} (2021), 175
%doi:10.1007/JHEP03(2021)175
[arXiv:2010.16166 [hep-th]].
%13 citations counted in INSPIRE as of 18 Dec 2021

\bibitem{Kim:2020url} K.~Y.~Kim, K.~S.~Lee and M.~Nishida,
``Holographic scalar and vector exchange in OTOCs and pole-skipping phenomena,''
JHEP \textbf{04} (2021), 092
[erratum: JHEP \textbf{04} (2021), 229]
%doi:10.1007/JHEP04(2021)092
[arXiv:2011.13716 [hep-th]].
%7 citations counted in INSPIRE as of 18 Dec 2021

\bibitem{Sil:2020jhr} K.~Sil,
``Pole skipping and chaos in anisotropic plasma: a holographic study,''
JHEP \textbf{03} (2021), 232
%doi:10.1007/JHEP03(2021)232
[arXiv:2012.07710 [hep-th]].
%6 citations counted in INSPIRE as of 18 Dec 2021

\bibitem{Yuan:2020fvv} H.~Yuan and X.~H.~Ge,
``Pole-skipping and hydrodynamic analysis in Lifshitz, AdS$_{2}$ and Rindler geometries,''
JHEP \textbf{06} (2021), 165
%doi:10.1007/JHEP06(2021)165
[arXiv:2012.15396 [hep-th]].
%3 citations counted in INSPIRE as of 18 Dec 2021

\bibitem{Abbasi:2020xli} N.~Abbasi and M.~Kaminski,
``Constraints on quasinormal modes and bounds for critical points from pole-skipping,''
JHEP \textbf{03} (2021), 265
%doi:10.1007/JHEP03(2021)265
[arXiv:2012.15820 [hep-th]].
%6 citations counted in INSPIRE as of 18 Dec 2021

\bibitem{Ceplak:2021efc} N.~Ceplak and D.~Vegh,
``Pole-skipping and Rarita-Schwinger fields,''
Phys. Rev. D \textbf{103} (2021) no.10, 106009
%doi:10.1103/PhysRevD.103.106009
[arXiv:2101.01490 [hep-th]].
%5 citations counted in INSPIRE as of 18 Dec 2021

\bibitem{Yuan:2021ets} H.~Yuan and X.~H.~Ge,
``Analogue of the pole-skipping phenomenon in acoustic black holes,''
[arXiv:2110.08074 [hep-th]].
%0 citations counted in INSPIRE as of 18 Dec 2021

\bibitem{Haehl:2018izb} F.~M.~Haehl and M.~Rozali,
``Effective Field Theory for Chaotic CFTs,''
JHEP {\bf 1810}, 118 (2018)
%doi:10.1007/JHEP10(2018)118
[arXiv:1808.02898 [hep-th]].
%%CITATION = doi:10.1007/JHEP10(2018)118;%%
%25 citations counted in INSPIRE as of 25 Sep 2019

\bibitem{Ramirez:2020qer} D.~M.~Ramirez,
``Chaos and pole skipping in CFT$_{2}$,''
JHEP \textbf{12} (2021), 006
%doi:10.1007/JHEP12(2021)006
[arXiv:2009.00500 [hep-th]].
%14 citations counted in INSPIRE as of 18 Dec 2021

\bibitem{Blake:2021hjj} M.~Blake and R.~A.~Davison,
``Chaos and pole-skipping in rotating black holes,''
[arXiv:2111.11093 [hep-th]].
%0 citations counted in INSPIRE as of 18 Dec 2021

\bibitem{Mahish:2022xjz} S.~Mahish and K.~Sil,
``Quantum information scrambling and quantum chaos in little string theory,''
JHEP \textbf{08} (2022), 041
%doi:10.1007/JHEP08(2022)041
[arXiv:2202.05865 [hep-th]].
%2 citations counted in INSPIRE as of 23 Jul 2023

\bibitem{Wang:2022mcq} D.~Wang and Z.~Y.~Wang,
``Pole Skipping in Holographic Theories with Bosonic Fields,''
Phys. Rev. Lett. \textbf{129} (2022) no.23, 231603
%doi:10.1103/PhysRevLett.129.231603
[arXiv:2208.01047 [hep-th]].
%7 citations counted in INSPIRE as of 23 Jul 2023

\bibitem{Amano:2022mlu} M.~A.~G.~Amano, M.~Blake, C.~Cartwright, M.~Kaminski and A.~P.~Thompson,
``Chaos and pole-skipping in a simply spinning plasma,''
JHEP \textbf{02} (2023), 253
%doi:10.1007/JHEP02(2023)253
[arXiv:2211.00016 [hep-th]].
%5 citations counted in INSPIRE as of 23 Jul 2023

\bibitem{Baishya:2023nsb} B.~Baishya and K.~Nayek,
``Probing Pole Skipping through Scalar-Gauss-Bonnet coupling,''
[arXiv:2301.03984 [hep-th]].
%1 citations counted in INSPIRE as of 23 Jul 2023

\bibitem{Yuan:2023tft} H.~Yuan, X.~H.~Ge, K.~Y.~Kim, C.~W.~Ji and Y.~Ahn,
``Pole-skipping points in 2D gravity and SYK model,''
[arXiv:2303.04801 [hep-th]].
%4 citations counted in INSPIRE as of 23 Jul 2023

\bibitem{Grozdanov:2023txs} S.~Grozdanov and M.~Vrbica,
``Pole-skipping of gravitational waves in the backgrounds of four-dimensional massive black holes,''
[arXiv:2303.15921 [hep-th]].
%4 citations counted in INSPIRE as of 23 Jul 2023

\bibitem{Natsuume:2023lzy} M.~Natsuume and T.~Okamura,
``Pole-skipping in a non-black hole geometry,''
[arXiv:2306.03930 [hep-th]].
%1 citations counted in INSPIRE as of 23 Jul 2023

\bibitem{Jeong:2023zkf} H.~S.~Jeong, C.~W.~Ji and K.~Y.~Kim,
``Pole-Skipping in Rotating BTZ Black Holes,''
[arXiv:2306.14805 [hep-th]].
%0 citations counted in INSPIRE as of 23 Jul 2023

\bibitem{Natsuume:2023hsz} M.~Natsuume and T.~Okamura,
``Pole-skipping as ''missing states'',''
[arXiv:2307.11178 [hep-th]].
%0 citations counted in INSPIRE as of 28 Jul 2023

\bibitem{Cvetic:1999ne} M.~Cvetic and S.~S.~Gubser,
``Phases of R charged black holes, spinning branes and strongly coupled gauge theories,''
JHEP {\bf 9904}, 024 (1999)
%doi:10.1088/1126-6708/1999/04/024
[hep-th/9902195].
%%CITATION = doi:10.1088/1126-6708/1999/04/024;%%
%361 citations counted in INSPIRE as of 15 Oct 2019

\bibitem{Chamblin:1999tk} A.~Chamblin, R.~Emparan, C.~V.~Johnson and R.~C.~Myers,
``Charged AdS black holes and catastrophic holography,''
Phys.\ Rev.\ D {\bf 60}, 064018 (1999)
%doi:10.1103/PhysRevD.60.064018
[hep-th/9902170].
%%CITATION = doi:10.1103/PhysRevD.60.064018;%%
%865 citations counted in INSPIRE as of 15 Oct 2019

\bibitem{Kharzeev:2010gd} D.~E.~Kharzeev and H.~U.~Yee,
``Chiral Magnetic Wave,''
Phys. Rev. D \textbf{83} (2011), 085007
%doi:10.1103/PhysRevD.83.085007
[arXiv:1012.6026 [hep-th]].
%332 citations counted in INSPIRE as of 28 Jul 2023

\bibitem{Science} A. ~W. ~Rost, R. ~S. ~Perry, J. ~F. ~Mercure, A. ~P. ~Mackenzie, and S. ~A. ~Grigera,
 ``Entropy Landscape of Phase Formation Associated with Quantum Criticality in $Sr_3Ru_2O_7$'',
  Science Vol. 325. no. 5946, pp. 1360 - 1363 (2009).


\bibitem{Landsteiner:2012kd} K.~Landsteiner, E.~Megias and F.~Pena-Benitez,
``Anomalous Transport from Kubo Formulae,''
Lect. Notes Phys. \textbf{871} (2013), 433-468
%doi:10.1007/978-3-642-37305-3\_17
[arXiv:1207.5808 [hep-th]].
%135 citations counted in INSPIRE as of 29 Apr 2023


\end{thebibliography}
\providecommand{\href}[2]{#2}\begingroup\raggedright

%%%%% CLEAR DOUBLE PAGE!
\newpage{\pagestyle{empty}\cleardoublepage}

\end{document}